\documentclass{jfm}

\usepackage{graphicx}
\usepackage{natbib}
\usepackage{color}
\usepackage{amsmath}
\usepackage{amssymb}

\ifCUPmtlplainloaded \else
  \checkfont{eurm10}
  \iffontfound
    \IfFileExists{upmath.sty}
      {\typeout{^^JFound AMS Euler Roman fonts on the system,
                   using the 'upmath' package.^^J}%
       \usepackage{upmath}}
      {\typeout{^^JFound AMS Euler Roman fonts on the system, but you
                   dont seem to have the}%
       \typeout{'upmath' package installed. JFM.cls can take advantage
                 of these fonts,^^Jif you use 'upmath' package.^^J}%
      }
  \else
  \fi
\fi

\ifCUPmtlplainloaded \else
  \checkfont{msam10}
  \iffontfound
    \IfFileExists{amssymb.sty}
      {\typeout{^^JFound AMS Symbol fonts on the system, using the
                'amssymb' package.^^J}%
       \usepackage{amssymb}%

      }{}
  \fi
\fi

\ifCUPmtlplainloaded \else
  \IfFileExists{amsbsy.sty}
    {\typeout{^^JFound the 'amsbsy' package on the system, using it.^^J}%
     \usepackage{amsbsy}}
    {}
\fi

\newsavebox{\astrutbox}
\sbox{\astrutbox}{\rule[-5pt]{0pt}{20pt}}

\title[Worthington Jets After Cavity Collapse.]{Generation and Breakup of Worthington Jets After Cavity Collapse.}

\author[Stephan Gekle and Jos\'e Manuel Gordillo]
{S\ls T\ls E\ls P\ls H\ls A\ls N \ns G\ls E\ls K\ls L\ls E\ls$^1$
\and J.\ns M.\ns G\ls O\ls R\ls D\ls I\ls L\ls L\ls
O\ls$^2$\break}
\affiliation{$^1$ Department of Applied Physics and J.M. Burgers
Centre for Fluid Dynamics, University of Twente, P.O. Box 217,
7500 AE Enschede, The Netherlands\\
$^2$ \'Area de Mec\'anica de Fluidos, Departamento de Ingener\'ia
Aeroespacial y Mec\'anica de Fluidos, Universidad de Sevilla, Avda.
de los Descubrimientos s/n 41092, Sevilla, Spain.\\}

\date{?? and in revised form ??}

\begin{document}

\maketitle

\begin{abstract}
Helped by the careful analysis of their experimental data,
\cite{Worthington97,Worthington} described roughly the mechanism
underlying the formation of high-speed jets ejected after the
impact of an axisymmetric solid on a liquid-air interface. They
made the fundamental observation that the intensity of these sharp
jets was intimately related to the formation of an axisymmetric
air cavity in the wake of the impactor. In this work we combine
detailed boundary-integral simulations with analytical modeling to
describe the formation and break-up of such Worthington jets in
two common physical systems: the impact of a circular disc on a
liquid surface and the release of air bubbles from an underwater
nozzle. We first show that the jet base dynamics can be predicted
for both systems using our earlier model in Gekle, Gordillo, van
der Meer and Lohse. Phys. Rev. Lett. 102 (2009). Nevertheless, our
main point here is to present a model which allows us to
accurately predict the shape of the entire jet. In our model, the
flow structure inside the jet is divided into three different
regions: The \emph{axial acceleration region}, where the radial
momentum of the incoming liquid is converted into axial momentum,
the \emph{ballistic region}, where fluid particles experience no
further acceleration and move constantly with the velocity
obtained at the end of the acceleration region and the \emph{jet
tip region} where the jet eventually breaks into droplets. Good
agreement with numerics and some experimental data is found.
Moreover, we find that, contrarily to the capillary breakup of
liquid cylinders in vacuum studied by \cite{Rayleigh}, the breakup
of stretched liquid jets at high values of both Weber and Reynolds
numbers is not triggered by the growth of perturbations coming
from an external source of noise. Instead, the jet breaks up due
to the capillary deceleration of the liquid at the tip which
produces a corrugation to the jet shape. This perturbation, which
is self-induced by the flow, will grow in time promoted by a
capillary mechanism. Combining these three regions for the base,
the jet, and the tip we are able to predict the exact shape
evolution of Worthington jets ejected after the impact of a solid
object - including the size of small droplets ejected from the tip
due to a surface-tension driven instability - using as the single
input parameters the minimum radius of the cavity and 
the flow field \textit{before} the jet emerges.
\end{abstract}

%
%
%
%
%
%

\section{Introduction}

\begin{figure}
  \centerline{\includegraphics[width=0.5\textwidth]{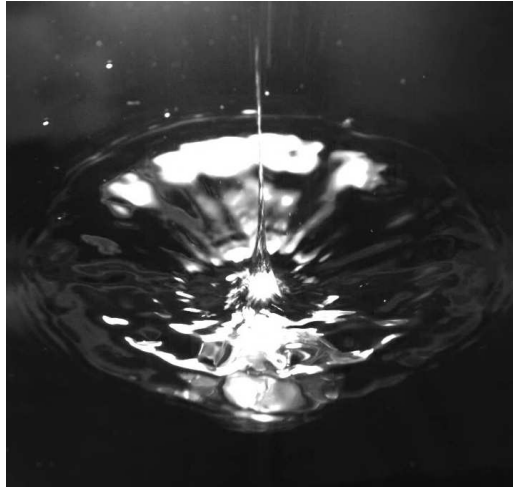} }
  \caption{Image of the high-speed jet ejected after the impact of 2cm disc with 1 m/s on a quiescent water surface.}\label{discExp}
\end{figure}

The impact of a solid object against a liquid interface is
frequently accompanied by the ejection of a high speed jet emerging
out of the liquid bulk into the air. Figure \ref{discExp}, which
shows the effect of a horizontal disc that impacts on a
pool of water, illustrates a liquid jet which flows $\sim 20$ times
faster than the disc impact speed. The qualitative description of
this common and striking phenomenon was firstly elucidated at the
beginning of the twentieth century by \cite{Worthington97,Worthington}. Through the careful analysis of the photographs taken after a solid sphere was dropped into water, \cite{Worthington97,Worthington} realized that these type of liquid
threads emerge as a consequence of the hydrostatic collapse of the
air-filled cavity which is created at the wake of the impacting solid.
\cite{Worthington97,Worthington} also made the remarkable observation that the generation of such cavities was very much influenced by the surface properties of the spherical solid. One century after their original observations,
\cite{Bocquet} quantified the conditions that determine the
existence of the air cavity in terms of the surface properties of
the solid and the material properties of the liquid.

High speed jets emerging out of a liquid interface are also
frequently observed in many other situations. For instance, it is
very usual to perceive that the liquid ``jumps'' out of the
surface of sparkling drinks, a fact which is known to happen as a
consequence of bubbles bursting at the liquid interface
[\cite{BoultonStoneBlake_JFM_1993, DucheminEtAl_PhysFluids_2002,
LigerBelairPolidoriJeandet_ChemSocRev_2008}]. Similarly, the
impact of a drop on a liquid interface or solid surface
[\cite{OguzProsperetti_JFM_1990, ShinMcMahon_PhysFluids_1990,
Rein_FluidDynRes_1993, MortonRudmanLiow_PhysFluids_2000,
DengAnilkumarWang_JFM_2007, BartoloJosserandBonn_PRL_2006}], is
commonly accompanied by the ejection of liquid jets whose
velocities can be substantially larger than that of the impacting
drop. Less familiar situations such as those related to the
focussing of capillary [\cite{MacIntyre_JPhysChem_1968,
ThoroddsenEtohTakehara_PhysFluids_2007}] or Faraday waves
[\cite{HogrefeEtAl_PhysicaD_1998, ZeffEtAl_Nature_2000}] also give
rise to the same type of phenomenon. Nevertheless, in spite of the
clear analogies, the main difference between the situations
enumerated above and the case of jet formation after cavity
collapse is that, in the latter case, surface tension does not
play any role in the jet ejection process [see \cite*{PRL09} for
details]. Indeed, the type of Worthington jets to be described
here depend on a purely inertial mechanism, namely the radial
energy focussing along the narrow cavity wall right before the
cavity pinches-off. This fact makes our process also somewhat
different from situations in which jets are induced by pressure
waves [\cite{OhlIkink_PRL_2003, TjanPhillips_JFM_2007,
AntkowiakEtAl_JFM_2007,BlakeEtAl_JFM_1993}].

Moreover, contrarily to what could be expected from the analogy with
other related physical situations [\cite{LonguetHiggins_JFM_1983,
LonguetHigginsOguz_JFM_1995}], \cite{PRL09} pointed out that jets
formed after cavity collapse are not significantly influenced by
the hyperbolic type of flow existing at the pinch-off location.
Instead, the description of this type of jets shares many
similarities with the very violent jets of fluidized metal which are
ejected after the explosion of lined cavities [e.g.
\cite{BirkhoffEtAl_JApplPhys_1948}], with those formed when an
axisymmetric bubble collapses inside a stagnant liquid pool
[\cite{Manasseh1,PoFRocioI}] or possibly even with the granular jets
observed when an object impacts a fluidized granular material
[\cite{Thoroddsengranular,Lohsegranular}].

Most of the results presented here refer to the perpendicular
impact of a circular disc with radius $R_D$ and \emph{constant}
velocity $V_D$ against a liquid surface. The fact that the solid
is a disc instead of a sphere leads to the formation of an air
cavity which is attached at the disc periphery, independent of the
surface properties. Thus, this choice for the solid geometry
avoids the additional difficulty of determining the position of
the void attachment line on the solid surface. The differences
pointed out above set our system somewhat apart from similar
studies [\cite{DuclauxEtAl_JFM_2007,
GlasheenMcMahon_PhysFluids_1996}]. The experimental realization of
the setup to which the numerical simulations presented are
referred, is described by
\cite{BergmannEtAl_PRL_2006,BergmannEtAl_preprint_2008,GekleEtAl_PRL_2008,PRL09},
who show that boundary-integral simulations are in excellent
agreement with experiments. In addition, potential flow numerical
simulations to study of the type of Worthington jets ejected after
bubble pinch-off from an underwater nozzle sticking into a
quiescent pool of water [\cite{Manasseh1,
LonguetHigginsKermanLunde_JFM_1991, OguzProsperetti_JFM_1993,
Burton2005, Keim, Thoroddsen07,ThoroddsenAnnuRev,
GordilloSevillaMartinezBazan_PhysFluids_2007,
BurtonTaborek_PRL_2008, Gordillo_PhysFluids_2008, PoFRocioI,
SchmidtEtAl_NaturePhys_2009}] are also reported in this paper. As
in the case of Worthington jets ejected after solid body impact,
similar boundary-integral simulations have been shown to be in
excellent agreement with experiments [see
\cite{OguzProsperetti_JFM_1993,PoFRocioI}].

This paper is organized as follows: In section~\ref{sec:description}
we present the three different numerical methods used.
Section~\ref{sec:results} presents the results from the simulations
which are compared to the analytical model in
section~\ref{sec:model}. Conclusions are drawn in
section~\ref{sec:conclusion}.


\section{Numerical methods}\label{sec:description}

In this paper we have used three types of boundary-integral
simulations. The first two model, respectively, the normal impact
of a disc on a free surface and the pinch-off of a bubble from an
underwater nozzle. With the purpose of simulating the capillary
breakup of the jets formed in the first two situations, the third
type of simulation represents a jet issued from a
constant-diameter nozzle with an imposed axial strain rate. The
latter type of numerical simulations have the advantage of
allowing us to directly impose the values of both the strain rate
and the Weber number, which are the parameters controlling the
breakup of the jet, as will become clear from the discussion
below.

\subsection{Disc impact simulations}

\begin{figure}
\begin{picture}(400,266)(0,0)
 \put(0,133) {
 \begin{picture}(0,0)(0,0)
   \put(0,0){\includegraphics[width=.33\textwidth]{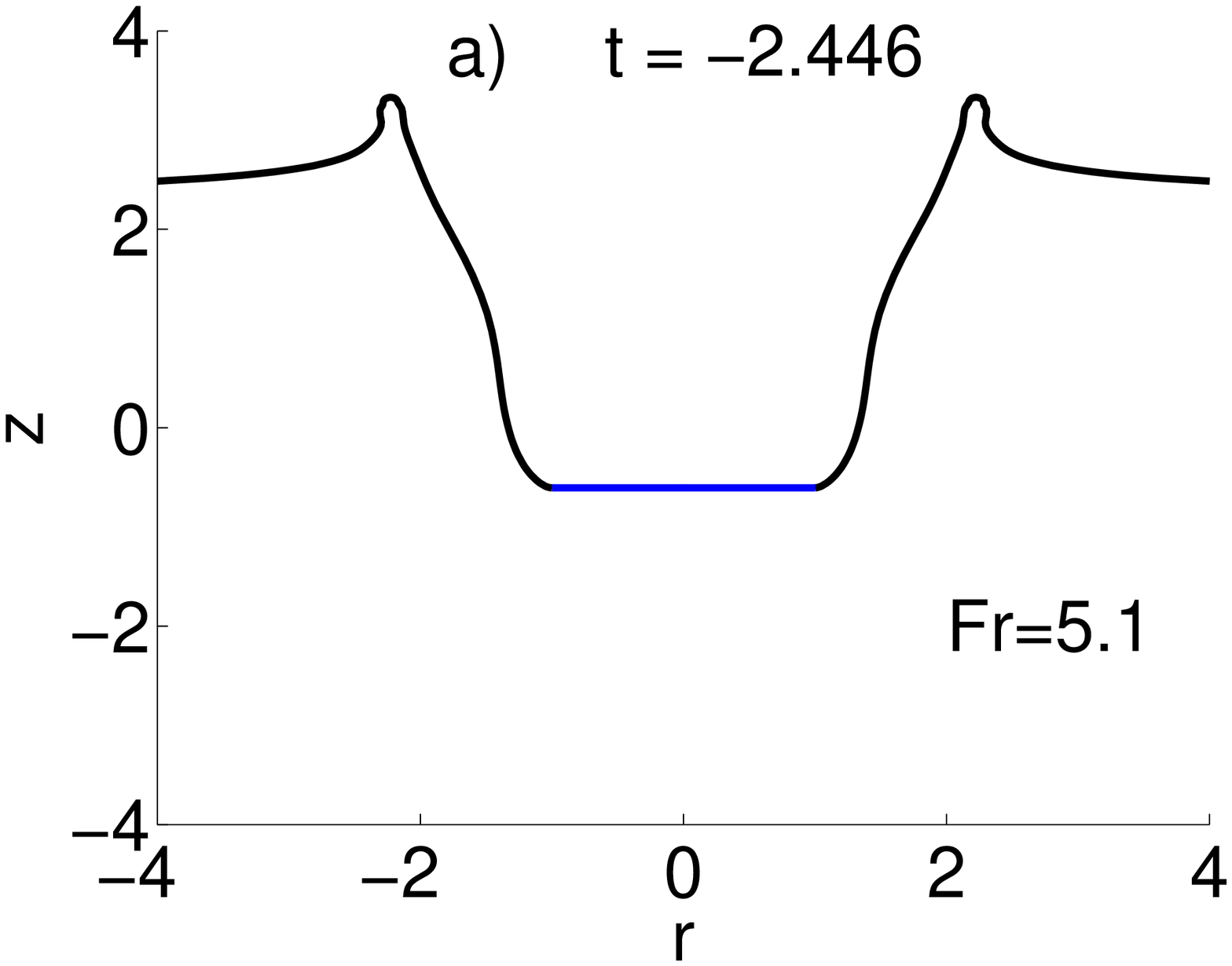}}
 \end{picture}
}
\put(133,133){
\begin{picture}(0,0)(0,0)
  \put(0,0){\includegraphics[width=.33\textwidth]{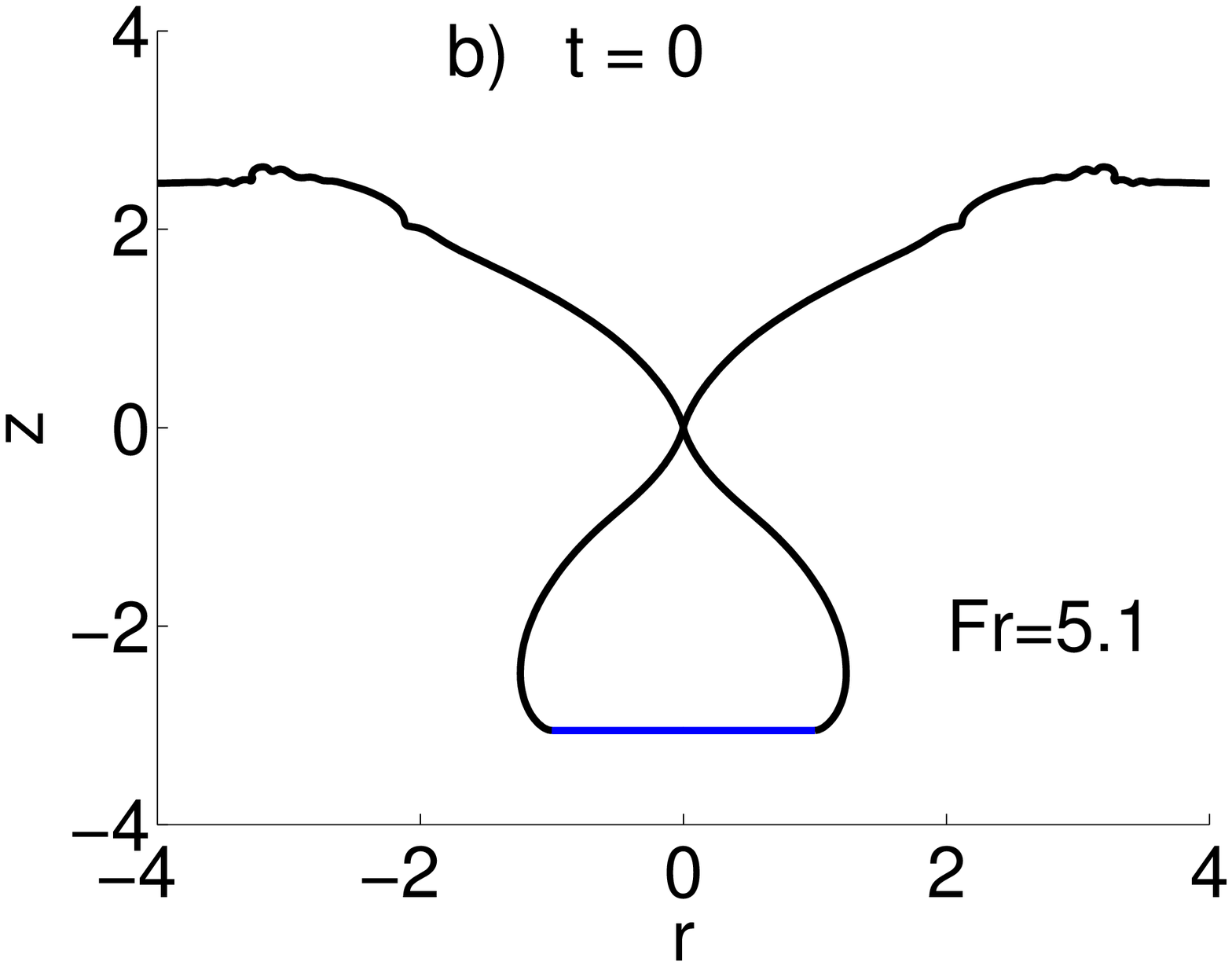}}
 \end{picture}
}
\put(266,133) {
\begin{picture}(0,0)(0,0)
 \put(0,0){\includegraphics[width=.33\textwidth]{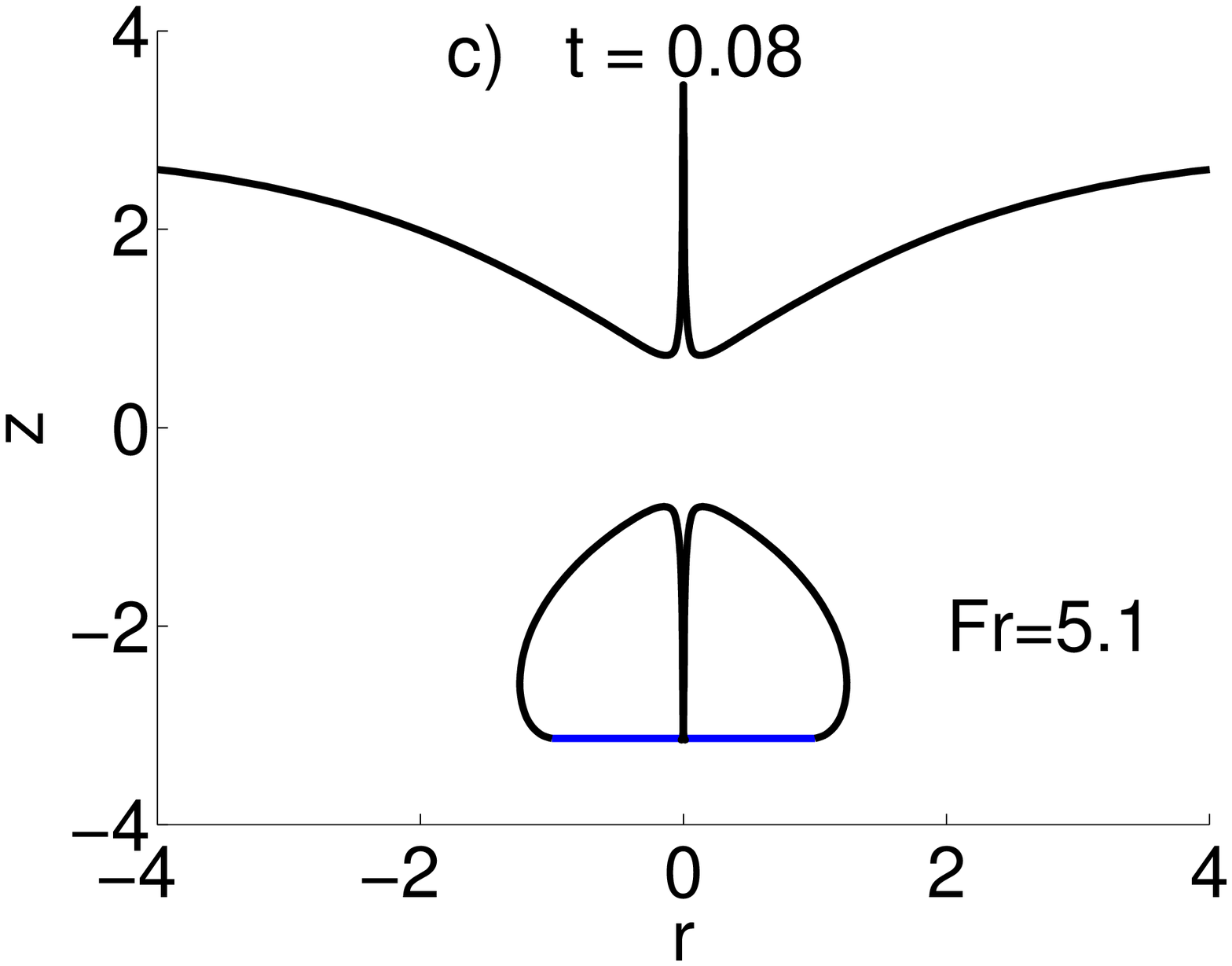}}
\end{picture}
}
\put(0,0) {
 \begin{picture}(0,0)(0,0)
   \put(0,0){\includegraphics[width=.33\textwidth]{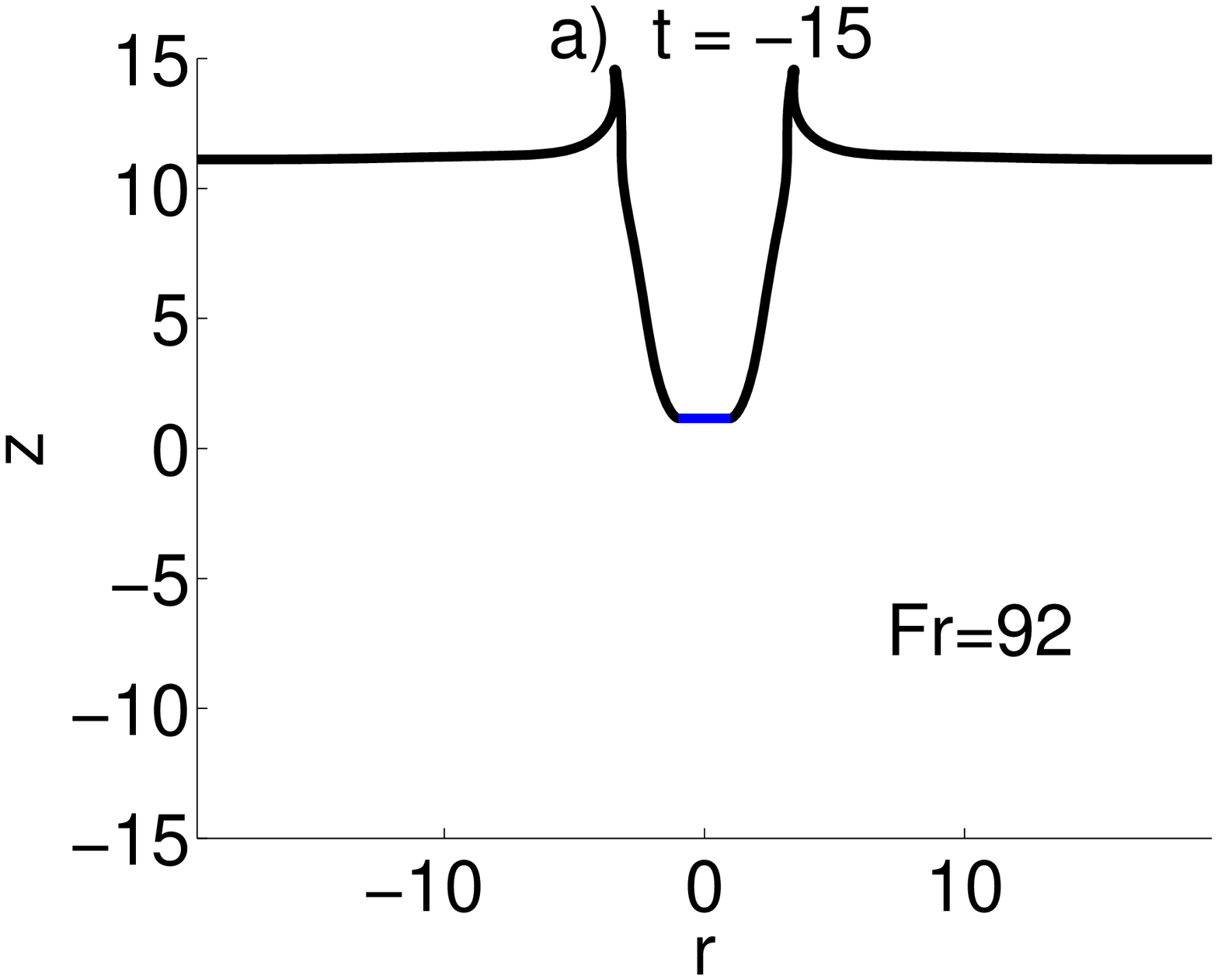}}
 \end{picture}
}
\put(133,0){
\begin{picture}(0,0)(0,0)
  \put(0,0){\includegraphics[width=.33\textwidth]{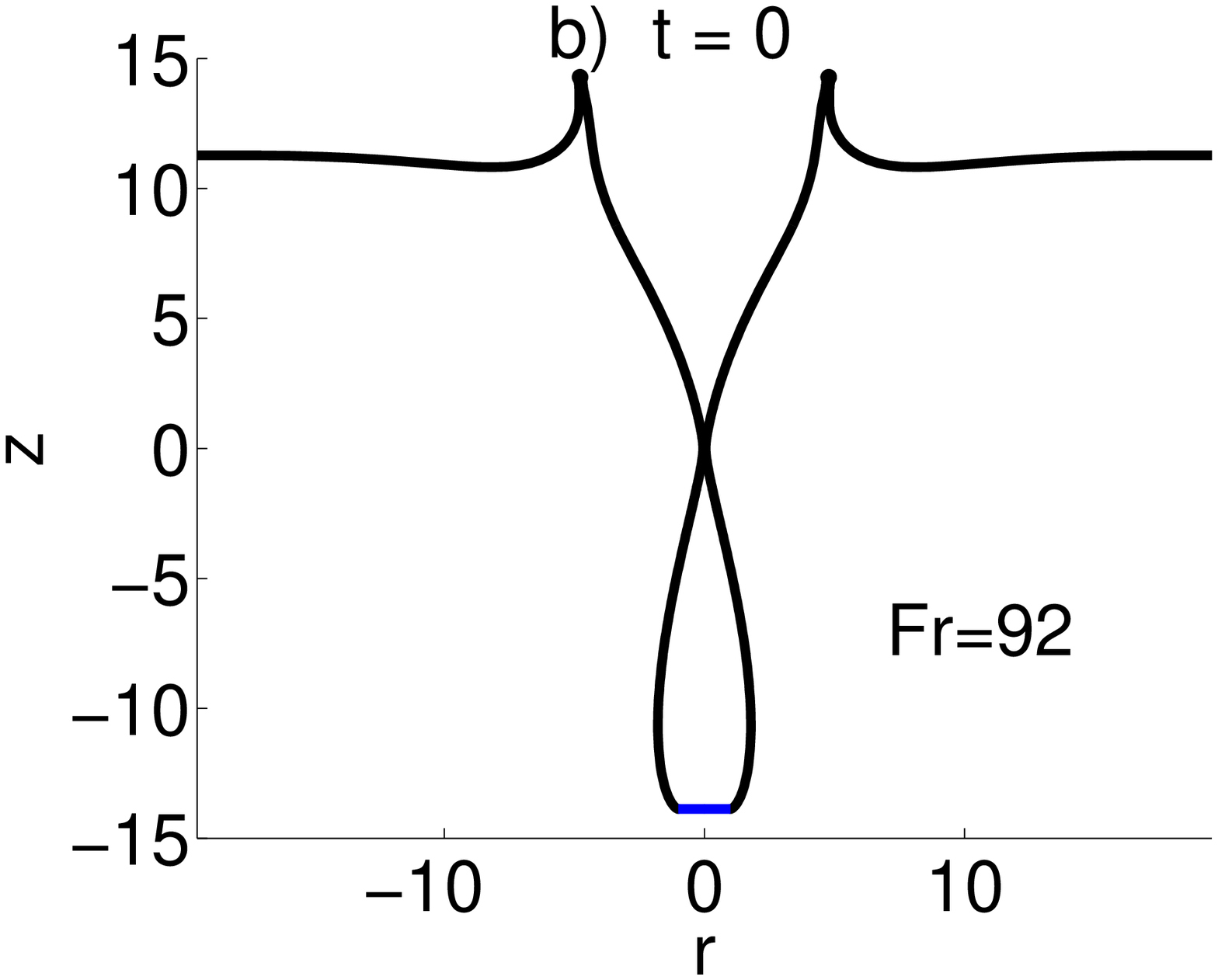}}
 \end{picture}
}
\put(266,0) {
\begin{picture}(0,0)(0,0)
 \put(0,0){\includegraphics[width=.33\textwidth]{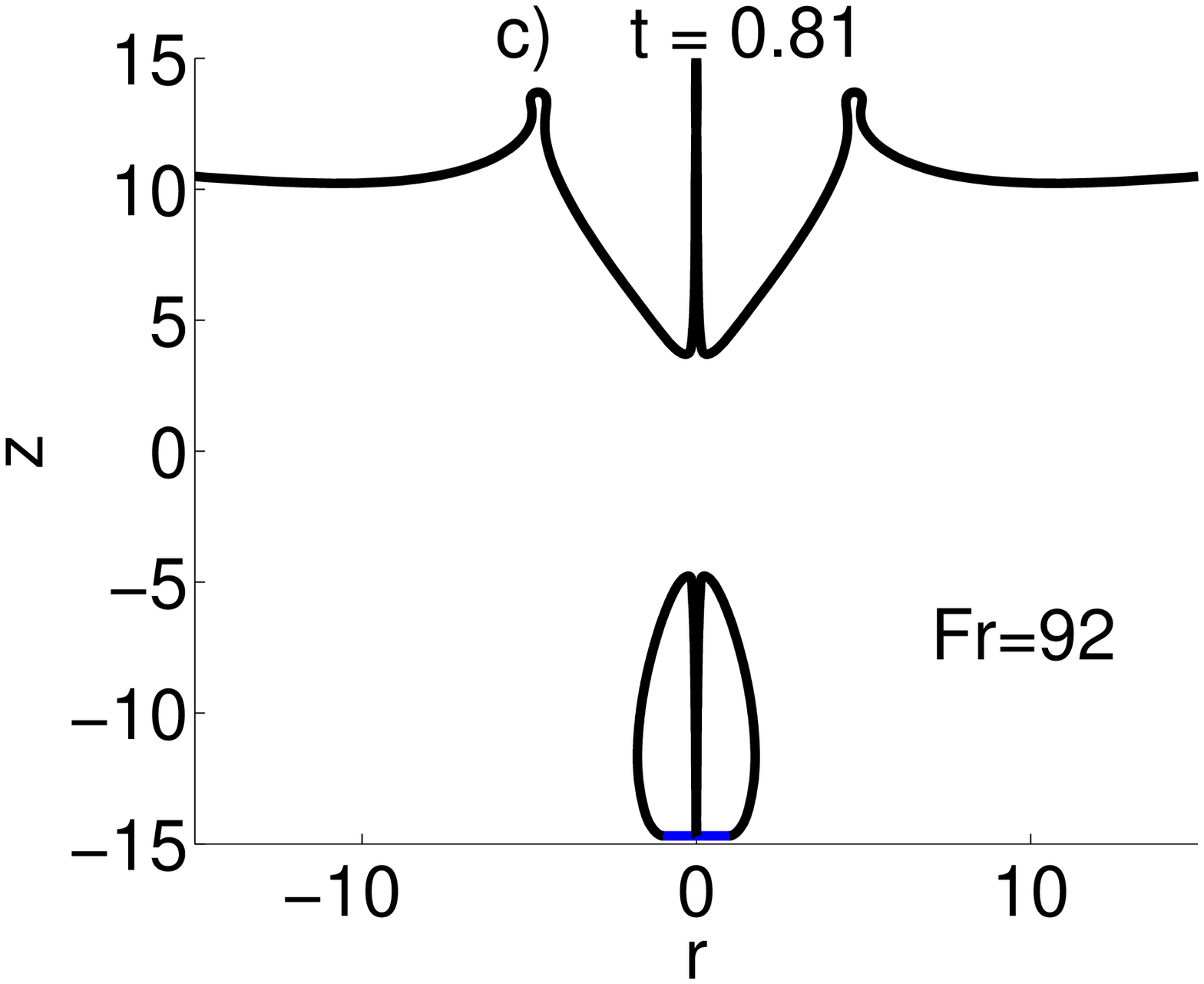}}
\end{picture}
}
\end{picture}
\caption{Numerical results obtained when a circular disc (blue line)
impacts perpendicularly and at constant velocity on a flat liquid
interface. Upon impact a cavity attached at the disc periphery is
created in the liquid (a) which collapses under the influence of
hydrostatic pressure (b). As a consequence of the cavity collapse,
two jets with velocities much larger than that of the impact solid,
are ejected upwards and downwards. The influence of increasing the
impact Froude number from $\mathrm{Fr}=5.1$ -- top row -- to $\mathrm{Fr}=92$ - bottom
row - is that the cavity becomes more slender.}\label{Cavities}
\end{figure}

The process of disc impact [see also \cite{BergmannEtAl_PRL_2006,
BergmannEtAl_preprint_2008, PRL09}] is illustrated in figure
\ref{Cavities}: after impact a large cavity is created beneath the
surface which subsequently collapses about halfway due to the
hydrostatic pressure from the liquid bulk. From the closure location
two high-speed jets are ejected up- and downwards. Here positions,
velocities and time are made dimensionless using as characteristic
quantities the disc radius $R_D$, the impact velocity $V_D$, and
$T_D=R_D/V_D$, respectively. (Variables in capital letters will be
used to denote dimensional quantities whereas their lower case
analogs will indicate the corresponding dimensionless variable).
Moreover, it will be assumed that axisymmetry is preserved and,
thus, a polar coordinate system $(r,z)$ will be used. The origins of
both the axial polar coordinate $z$ and of time $t$ are set at the
cavity pinch-off height and at the pinch-off instant, respectively.

Since global and local Reynolds numbers are large and the generation
of vorticity is negligible [\cite{BergmannEtAl_preprint_2008,PRL09}]
we can make use of a flow potential to describe the liquid flow
field. The numerical details, including the ``surface surgery''
needed to accurately capture the transition from the cavity collapse
process to the jet ejection, are given elsewhere [see
\cite{PRL09,BergmannEtAl_preprint_2008}]. These simulations have
shown excellent agreement with experimental high-speed recordings
and particle image velocimetry measurements
[\cite{BergmannEtAl_PRL_2006, GekleEtAl_PRL_2008, PRL09,
BergmannEtAl_preprint_2008}]. The simulation stops when the downward
jet hits the disc surface.

Since the Reynolds number is large, the dimensionless parameters
controlling the jet ejection process are the Froude number,
$\mathrm{Fr}=V_D^2/(R_D\,g)$, and the Weber number, $\mathrm{We}=\rho V^2_D
R_D/\sigma$ where $g$, $\rho$ and $\sigma$ indicate the
gravitational acceleration, the liquid density and the interfacial
tension, respectively. Since $\mathrm{We} \gtrsim O(10^2)$ in all
cases considered here, the jet ejection is not promoted by surface
tension [\cite{PRL09}]. Nonetheless, capillarity is essential to
describe the jet breakup process, as will become clear from the
discussion below. Air effects, which play an essential role during
the latest stages of cavity collapse
[\cite*{GordilloEtAl_PRL_2005,Gordillo_PhysFluids_2008,PRL09Air}],
are not taken into explicit consideration here. Instead the cut-off
radius at which the cavity geometry is changed into the jet geometry
is fixed manually verifying carefully that the exact value of this
parameter does not influence our results. The only consequence of
this simplification is that a tiny fraction of the jet - the jet tip
- may not be accurately described neither by our numerical
simulations nor by our theory as will be discussed in section
\ref{sec:gas}.

\subsection{Bubble pinch-off from an underwater nozzle}

\begin{figure}
  \centerline{\includegraphics[width=0.5\textwidth]{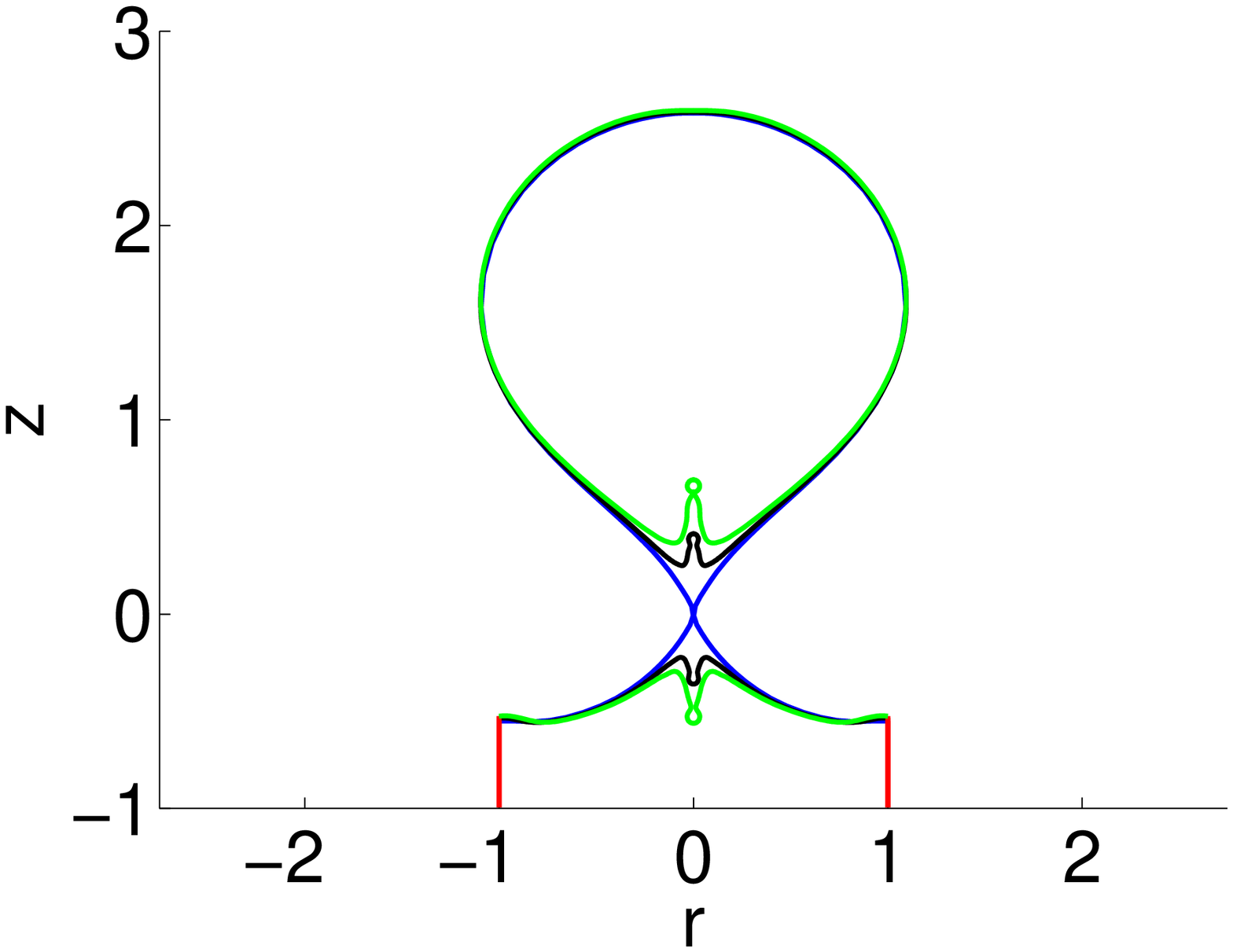}
\includegraphics[width=0.5\textwidth]{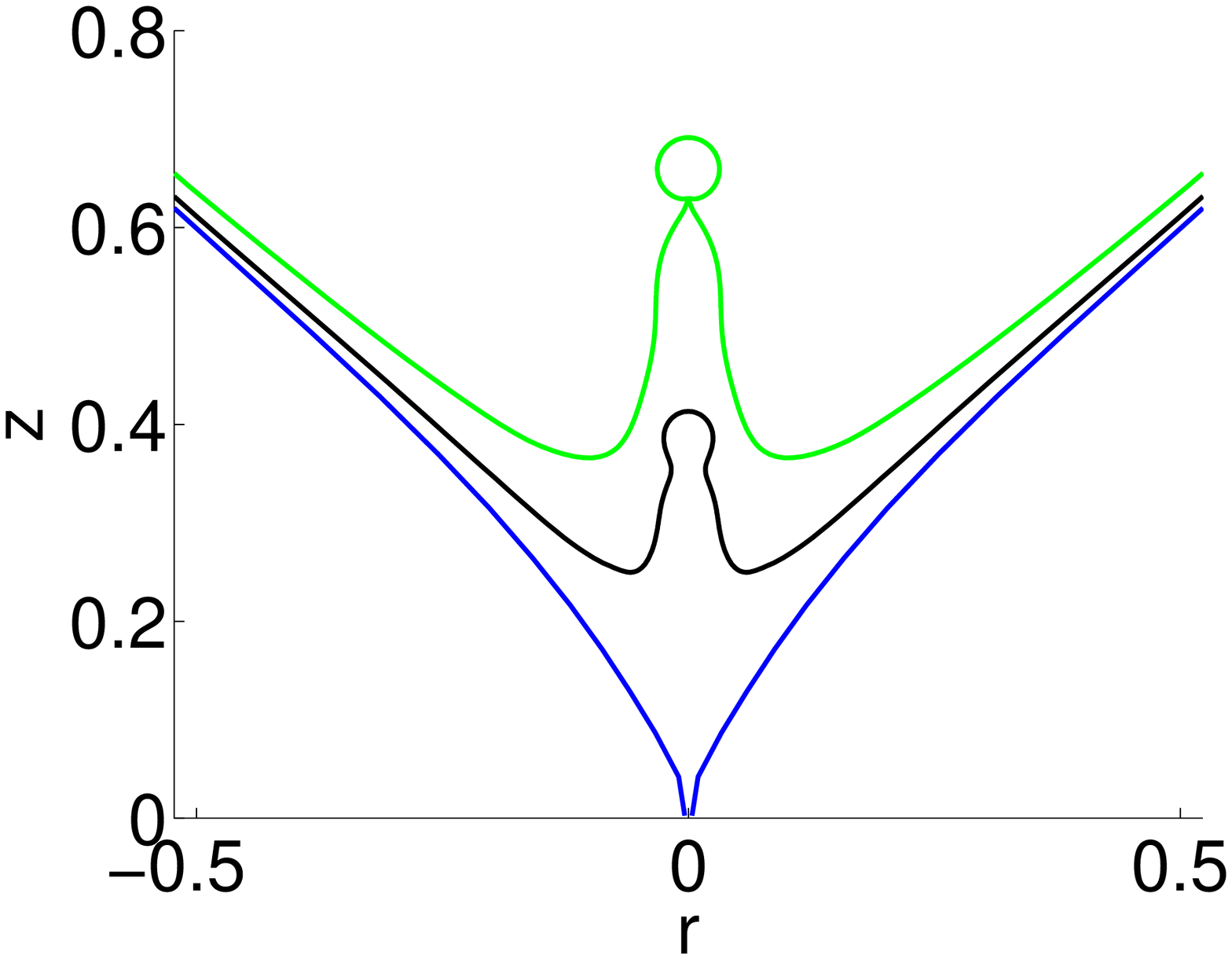} }
  \caption{(a) Time evolution of jets formed after the collapse of gas bubbles injected into a quiescent liquid pool through a nozzle (red line), showing the ejection of the first drop, for $\mathrm{Bo}=2.1$. b) Closeup view of the jet region in (a). The colors correspond to different dimensionless times: $t=0$ (blue), $t=0.0014$ (black) and $t=0.0027$ (green)}. \label{surfaceProfilesNeedle}
\end{figure}

In the second type of simulations a bubble grows and detaches when
a constant gas flow rate is injected from an underwater nozzle
into a quiescent pool of liquid. \cite{Manasseh1} and
\cite{PoFRocioI} experimentally showed that this process also
creates high speed jets. Indeed, as the bubble grows in size, the
neck becomes more and more elongated and, eventually, surface
tension triggers the pinch-off of the bubble, leading to the
formation of two fast and small jets as illustrated in figure
\ref{surfaceProfilesNeedle}. Surface tension also leads to the
pinch-off of a small droplet at the jet tip, which is precisely the
instant when the simulation stops.

Here, distances are made non-dimensional using the nozzle radius $R_N$ as the characteristic length scale;
moreover, the prescribed gas flow rate $Q$ is used to derive the
typical time scale $T_N=(\pi R_N^3)/Q$. For the quasi-static
injection conditions considered here, the relevant dimensionless
parameter characterizing this physical situation is the Bond
number $\mathrm{Bo}=\rho R_N^2 g/\sigma$
[\cite{LonguetHigginsKermanLunde_JFM_1991, PoFRocioI}], which in
the case presented here equals $2.1$. More details of our
simulation method are given in \cite{OguzProsperetti_JFM_1993,
GekleSnoeijerEtAl_preprint}. Note that the present numerical
simulations [\cite{OguzProsperetti_JFM_1993}] as well as those
reported using a very similar numerical method [\cite{PoFRocioI}],
are in excellent agreement with experiments.

\subsection{Simulations of a jet ejected at constant diameter}

\begin{figure}
  \centerline{\includegraphics[width=0.7\textwidth]{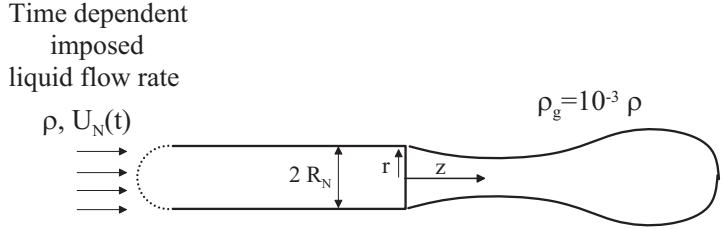}   }
  \caption{Sketch defining the geometry of the numerical simulations used to describe the capillary breakup of a stretched liquid jet of density $\rho$ injected into a gaseous atmosphere of density $\rho_g=10^{-3}\rho$.
  The liquid velocity profile imposed at the boundary which delimits the nozzle on the left is uniform and decreases linearly with time.}\label{Geometry}
\end{figure}

As will be shown by our theoretical analysis below, the jet
breakup process can be described in terms of two dimensionless
parameters evaluated nearby the base of the jet, namely, the local
Weber number and the dimensionless axial strain rate. These
quantities depend non-trivially on the input parameters of our
physical simulations (disc speed, nozzle size etc.). In order to
obtain a way of systematically varying both the local Weber number
and strain rate we conducted a third type of simulation by
adapting the axisymmetric (two-fluid) boundary integral method
described in \cite{GordilloSevillaMartinezBazan_PhysFluids_2007}
to a situation that retains the essential ingredients to describe
the capillary breakup process in the first two types of
simulations. For this purpose, we have simulated the discharge of
a liquid injected through a constant radius needle
with a length of 20 times its radius into a gaseous atmosphere.
The density ratio of the inner and outer fluids is $10^3$ and a
uniform velocity profile linearly decreasing with
time is imposed on the boundary that delimits the
computational domain on the left (see figure \ref{Geometry}).
Initially, the liquid interface is assumed to be a hemisphere
attached at the nozzle tip. The uniform velocity with which the
liquid is injected varies in time according to
\begin{equation}
U_N(t_N)=U_N(0)(1-\alpha\,t_N)
\end{equation}
with the dimensionless strain rate $\alpha$ and the initial
velocity $U_N(0)$ determined by the physical situation which one
intends to imitate (jets formed either after the disc impact or
from the underwater nozzle). For these type of simulations positions,
velocities and time will be made non dimensional using, as
characteristic dimensional quantities, the injection needle radius
$R_N$, the initial velocity $U_N(0)$, and $T_N=R_N/U_N(0)$
respectively.

In section \ref{sec:breakup} we demonstrate very good agreement
between the results of these type of simulations and those related
to the formation of jets after bubble pinch-off from an underwater
nozzle. Unfortunately, the extremely large values of the Weber
number reached at the tip of the liquid jets formed after the
impact of a disc on a free surface ($\sim O(10^3)$) unavoidably
lead to the development of numerical instabilities
[\cite{TjanPhillips_JFM_2007}]. This fact makes a direct
comparison between the simulations of  the axial
strain system sketched in figure \ref{Geometry} and those
corresponding to the impacting disc impossible.

%
%
%
%
%
%

\section{Analysis of numerical results}\label{sec:results}

\subsection{Effects of azimuthal asymmetries in the determination of the cut-off radius}\label{sec:gas}

The value of $r_{min}$ (the minimum radius of the cavity before
the jet emerges) would be zero under the ideal conditions of our
simulations, which do not take into account gas effects
[\cite{GordilloEtAl_PRL_2005,Gordillo_PhysFluids_2008,PRL09Air,
BurtonTaborek_PRL_2008}], liquid viscosity
[\cite{Burton2005,Thoroddsen07,PoFrocioII}] or small azimuthal
asymmetries that may be present in the flow [\cite{Keim,
SchmidtEtAl_NaturePhys_2009}]. This would imply that the initial
jet velocity would be infinity. However, all the effects
enumerated above are known to strongly influence the spatial
region surrounding the cavity neck during the very last stages of
bubble pinch-off and, therefore, are essential to determine the
real value of $r_{min}$ (\cite{Gordillo_PhysFluids_2008,
PRL09Air}).

Note first that, the larger $r_{min}$ is, the smaller will be the
maximum liquid velocity at the tip of the jet.
\begin{figure}
  \centerline{\includegraphics[width=0.8 \textwidth]{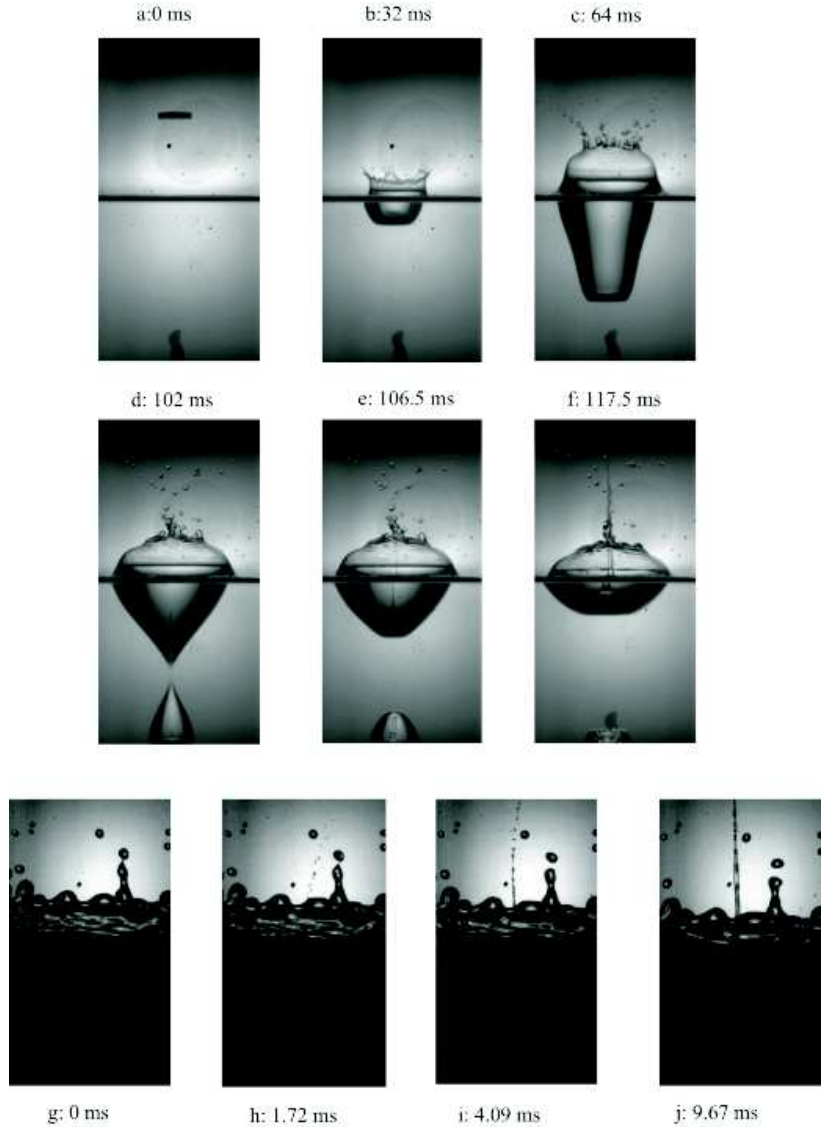}   }
  \caption{Pictures (a)-(f) show the smooth cavity formed after the normal impact of a brass disc against a water interface. The disc dimensions are 22~mm in diameter and 4.7~mm in height. The disc falls by gravity and the impact velocity is $V_\mathrm{impact}=1.85$~m/s. Note that, while the time between impact and cavity closure is roughly 70~ms, the upwards jet reaches the free surface in less than 4~ms, indicating that the jet velocity is much larger than the impactor's velocity. Indeed, the initial velocity of the tip of the jet,
  measured from detailed images of the type (g)-(j), is larger than -- since drops might not be in a plane perpendicular to the free surface -- 22.71~m/s and thus larger than 12.28 times the disc velocity. The huge velocities reached by the liquid jet can also be visually appreciated by comparison with the velocity of the drops formed in the corona splash which hardly change their position between images (g) and (j). Let us also remark that, initially, the jet is not axisymmetric ((h) and (i)). Nevertheless, after a few milliseconds, picture (j) shows that the jet becomes approximately axisymmetric.}\label{ExperimentosI}
\end{figure}
\begin{figure}
  \centerline{\includegraphics[width=0.6\textwidth]{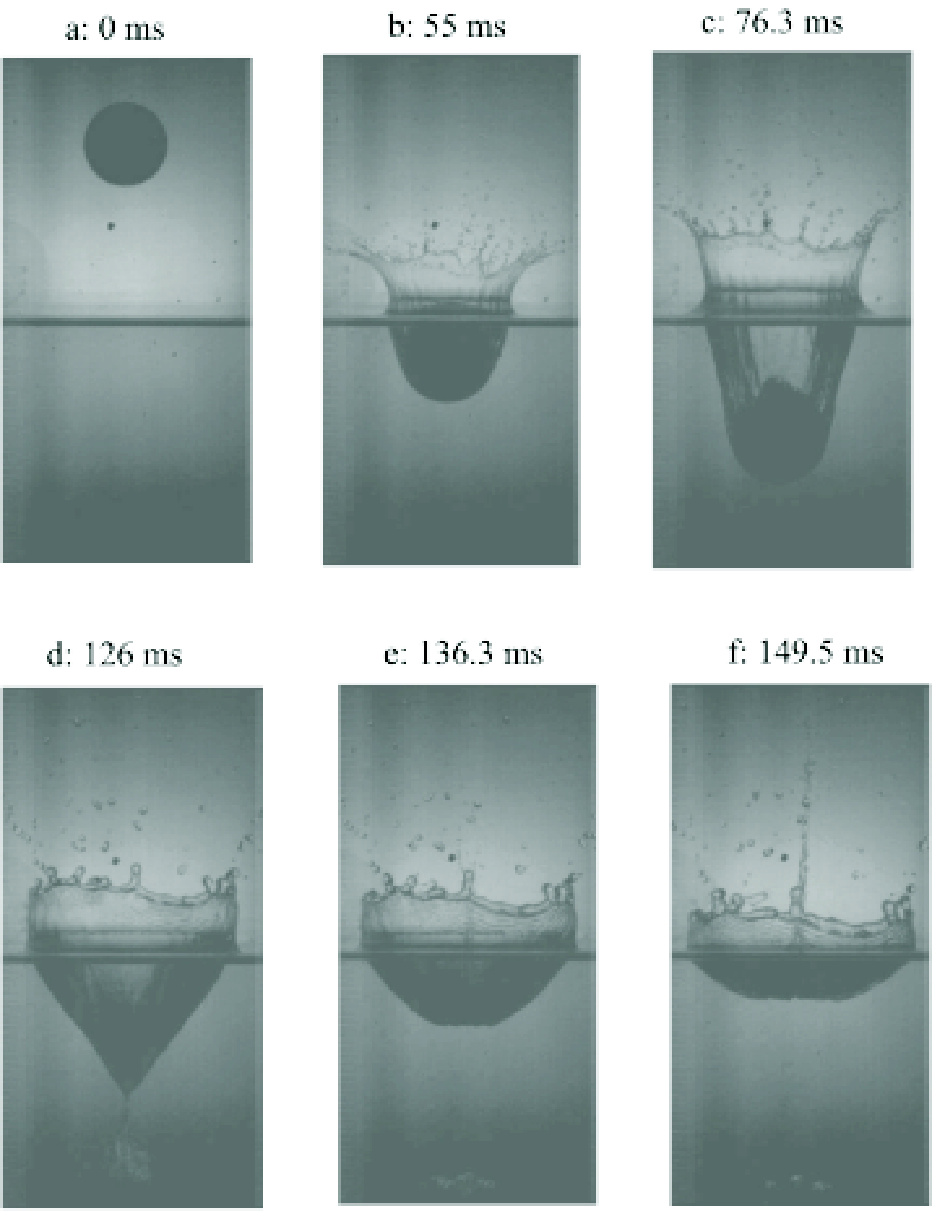}   }
  \caption{Pictures (a)-(f) show the cavity formation caused by a golf ball with a diameter of 42.75~mm impacting with a velocity of 2.03~m/s. Compared to figure \ref{ExperimentosI} the surface shape is visibly distorted (c) due to the rough surface structure of the ball. Nevertheless, it can be inferred from a detailed image analysis that the jet velocity is again much larger than the ball's velocity. However, in spite of both the impact velocity and the ball diameter being larger than those of the disc, the maximum velocity of the jet is only $V_\mathrm{impact}\simeq 20$~m/s and thus smaller than for the impacting disc.}\label{ExperimentosII}
\end{figure}
Here we will provide experimental evidence showing that
non-axisymmetric perturbations are of crucial importance to fix
$r_{min}$ and, consequently, the maximum velocity reached by the
jet. This is due to the fact that asymmetries influence the radial
flow focussing effect on the central axis even before the actual
cavity closure. The development of azimuthal instabilities leads
to a decrease of the liquid acceleration towards the axis before
pinch-off and thus reduces the speed of the ejected jet. This is
clearly observed in figures \ref{ExperimentosI} and
\ref{ExperimentosII}, which show the cavity formation and jet
ejection processes when either a brass disc (smooth surface) or a
golf ball (structured surface) impact perpendicularly on a
quiescent pool of water. Despite the fact that both the velocity
and the diameter of the ball are larger than those of the disc,
the maximum jet velocity is larger for the disc case. Indeed,
while the shape of the cavity in figure \ref{ExperimentosI} is
smooth, the cavity interface in figure \ref{ExperimentosII}
clearly exhibits asymmetric modulations already right after the
impact (which -- in addition to the rough surface
structure -- may in part also be due to a rotation of the ball).
Note that the overall shape of the cavity is very similar in both
cases. Consequently, since the self-acceleration of the liquid
towards the axis is lost when the amplitude of azimuthal
disturbances is similar to the radius of the cavity, the maximum
velocity reached during the collapse process decreases when the
cavity interface is not smooth. Note that figures
\ref{ExperimentosI} and \ref{ExperimentosII} are representative of
an exhaustive set of experiments. The analysis of the whole
experimental data has shown that the rough surface systematically
produces lower jet speeds.

The initial amplitude or the precise instant at which such
azimuthal instabilities may develop is not easy to predict. For
instance, \cite{Keim, SchmidtEtAl_NaturePhys_2009} pointed out
that tiny geometrical asymmetries in the initial setup might break
the cylindrical symmetry of the cavity at the pinch-off location.
Moreover, even if the cavity is perfectly axisymmetric, the strong
shear between the gas and the liquid will induce instabilities
that tend to break the cylindrical symmetry of the cavity
[\cite{LeppinenLister2003,BergmannEtAl_PRL_2006}].

\begin{figure}
  \includegraphics[width=0.5\textwidth]{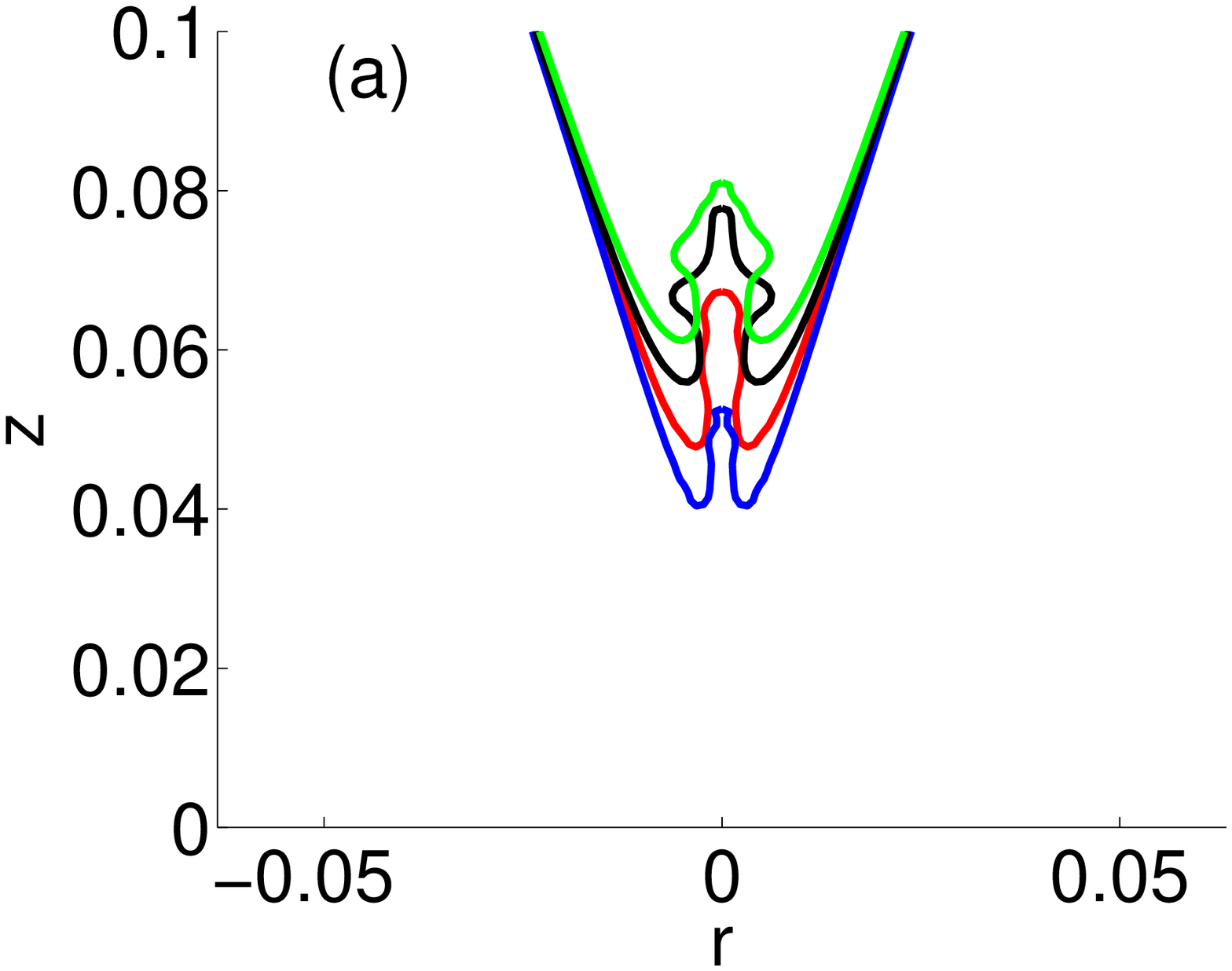}
  \includegraphics[width=0.5\textwidth]{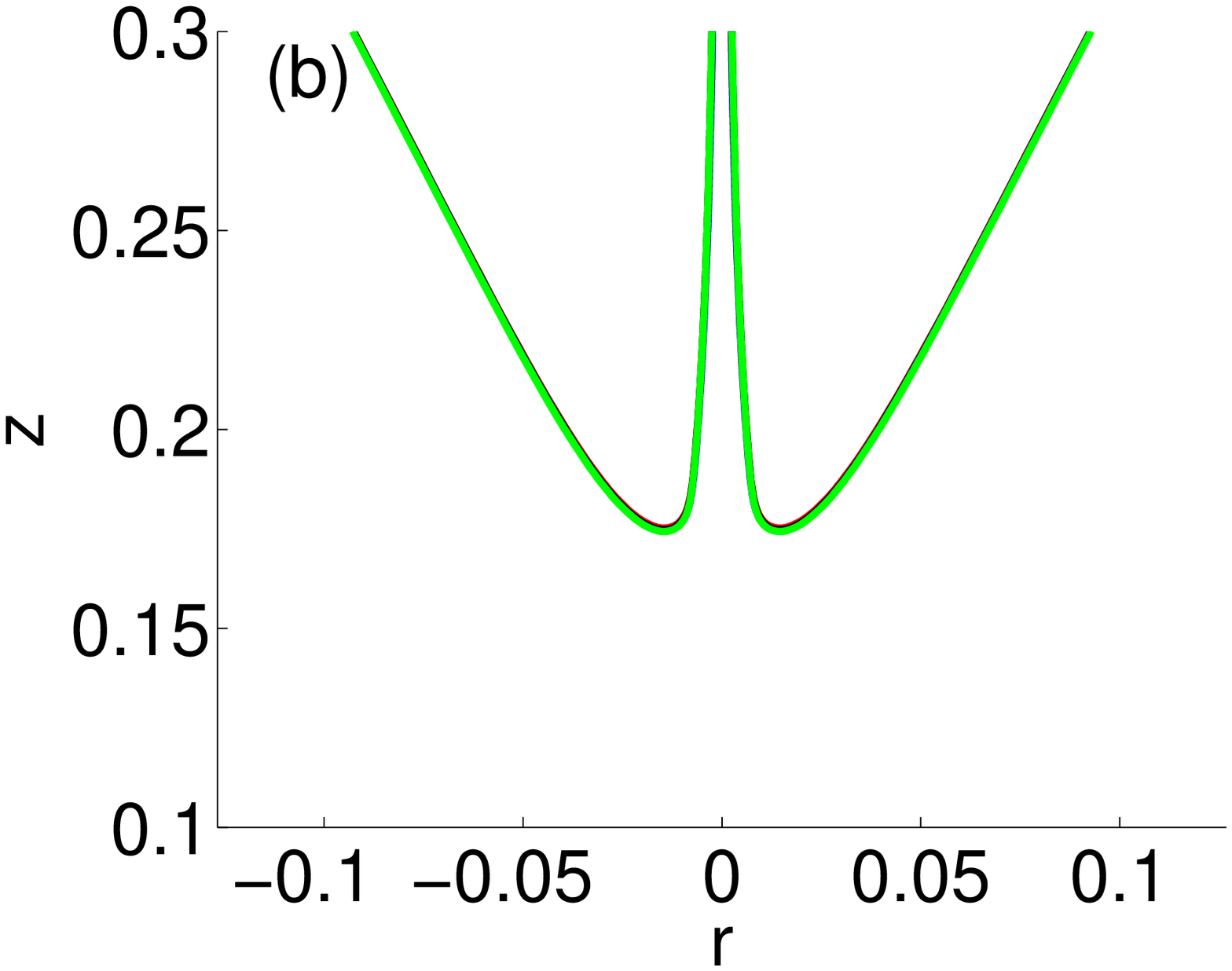}
  \includegraphics[width=0.5\textwidth]{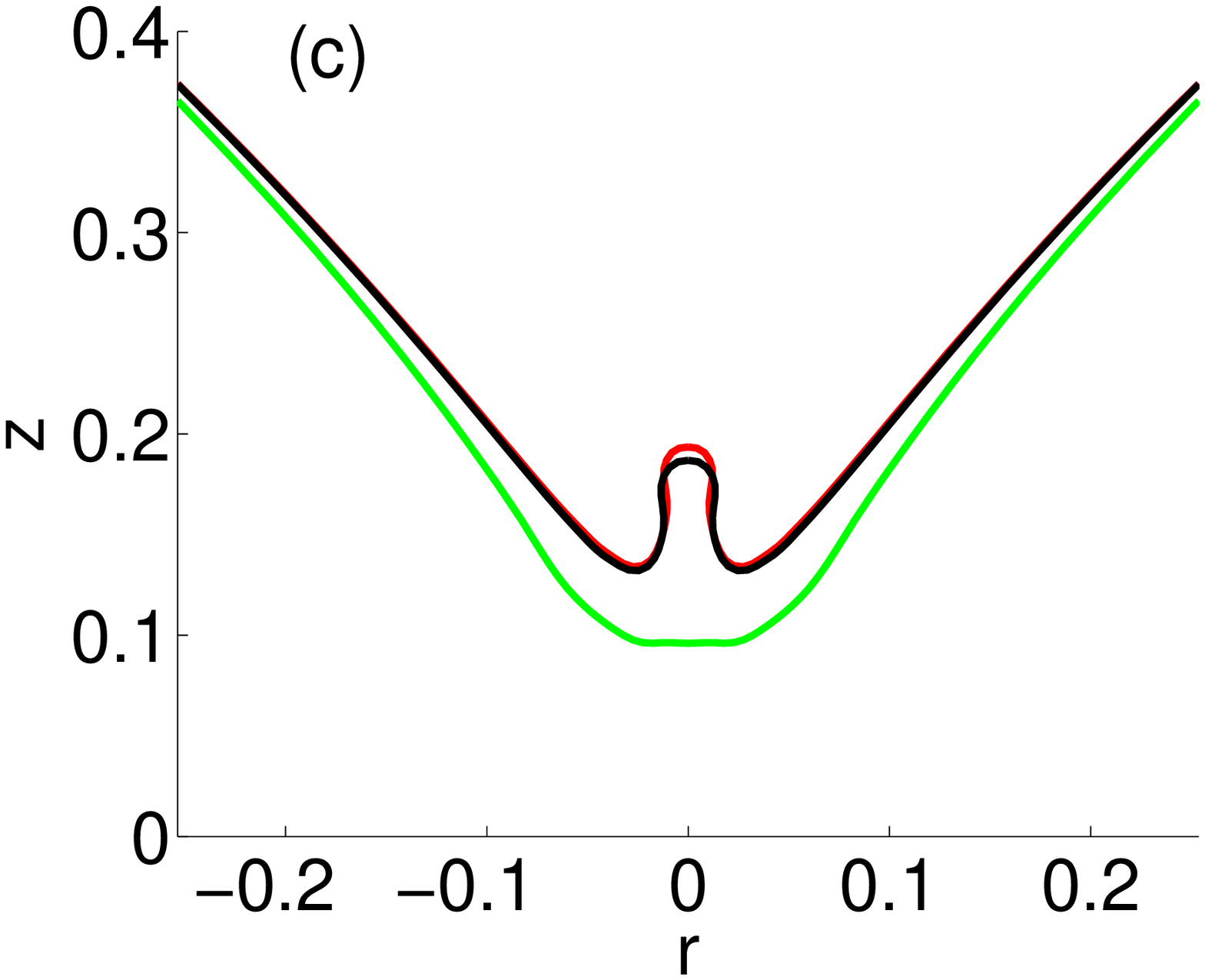}
  \includegraphics[width=0.5\textwidth]{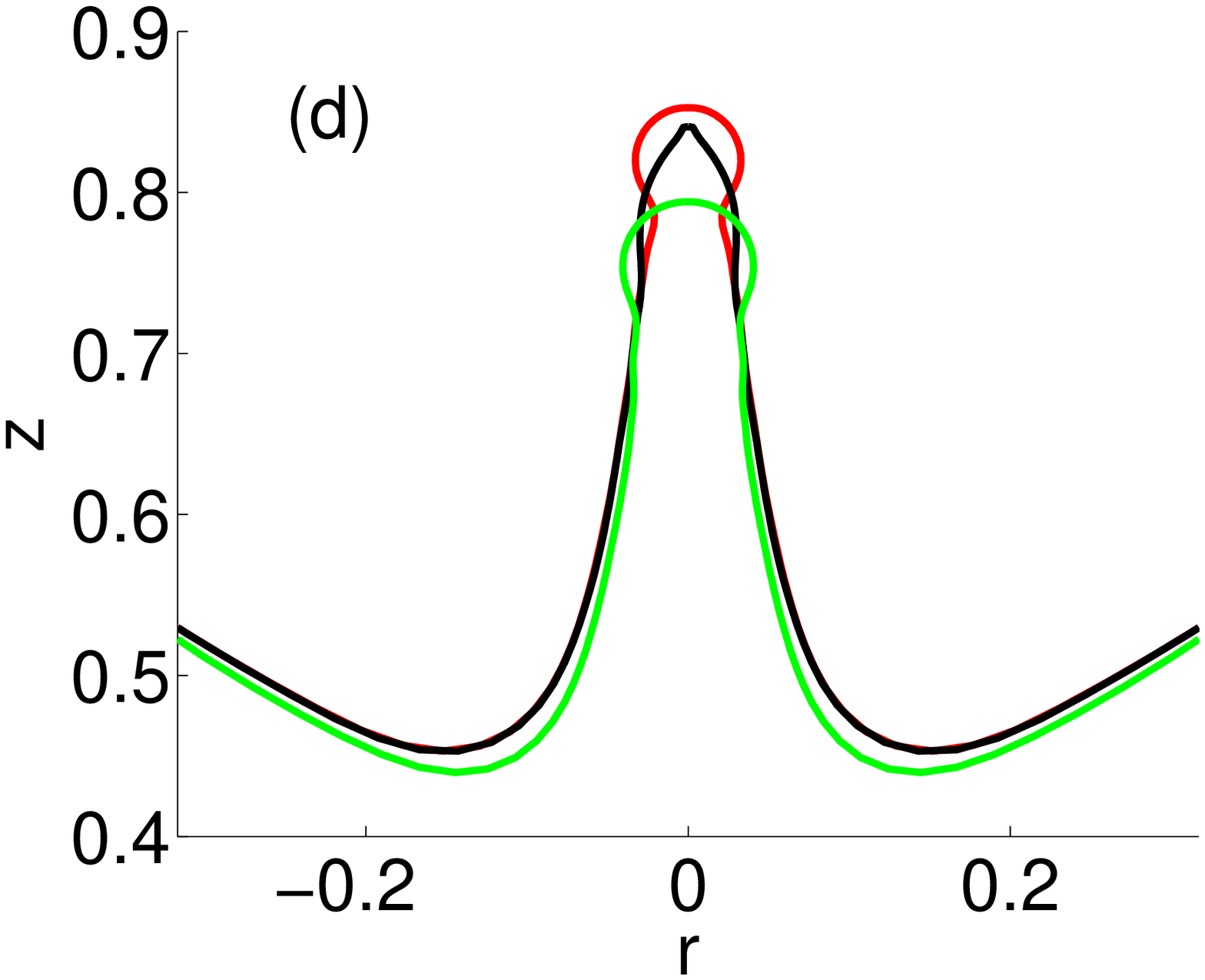}\\
\caption{Top row: Jet shapes for the disc impact at Fr$=5.1$ at two different
instants of time, $t=10^{-4}$(a) and $t=3.2\times 10^{-3}$ (b),
for four different values of the cut-off radius, $r_{min}=0.005$
(blue), $r_{min}=0.01$ (red), $r_{min}=0.02$ (black), and
$r_{min}=0.05$ (green). Bottom row: Jet shapes for the underwater nozzle with different cut-off radii
(colors as in top row) at $t=0.0003$ (c) and $t=0.004$ (d), respectively
(here the simulations are extended beyond the ejection of the
first droplet). It is evident in both cases that the influence of
varying the cut-off is significant only in the very first instants
after pinch-off and at the very tip of the jet.}\label{Jetshapes}
\end{figure}

Therefore, the precise determination of $r_{min}$ is a very
complex and difficult subject which in addition will heavily
depend on the system under study and must therefore remain outside
the scope of this contribution. We have instead decided to vary
$r_{min}$ within reasonable bounds and to analyze carefully the
effect on the subsequent time evolution of the jet. It can be
clearly appreciated in figure \ref{Jetshapes} that differences in
the simulations can be observed in both the jet base and tip
region right after pinch-off occurs. However, as soon as the jet
radius at its base becomes of the order of the maximum value of
$r_{min}$ explored, differences in the jet base region disappear
and only remain appreciable in the jet tip region. Physically,
this means that gas effects and small asymmetries will only be
felt at the highest part of the jet, which represents only a very
small fraction of both the total volume and of the total kinetic
energy of the jet. Note also that, in spite of the jet tip being
the spatial region where the highest velocities are reached, it is
also the least reproducible one from an experimental point of view
since it strongly depends on the precise details of pinch-off.
Thus, regarding experimental reproducibility, our study will be
valid to accurately describe the most robust part of the jet. In
the case of the impacting disc we will set $r_{min}=0.01$ and in
the case of the gas injection needle, the minimum radius will be
fixed to $r_{min}=0.05$.

Finally, note that our axisymmetric approach has been proven to be
in excellent agreement with experiments whenever either the radius
of the collapsing cavity or the radius of the emerging jet, are
larger than the cut-off radius $r_{min}$ for which any of the
effects enumerated above -- gas, azimuthal perturbations -- become relevant [see, for instance,
\cite{BergmannEtAl_PRL_2006,
GekleEtAl_PRL_2008,PoFRocioI,PRL09,PoFrocioII}].


\subsection{Jet ejection process for the disc impact}\label{jetEjectionDisc}

The different stages of the jet formation process have been
illustrated in figure~\ref{Cavities}. After the solid body impacts
against the free surface, an air cavity is generated (a). As a
consequence of the favorable pressure gradient existing from the
bulk of the liquid to the cavity interface, the liquid is
accelerated inwards (b). These radially inward velocities focus
the liquid towards the axis of symmetry, leading to the formation
of two fast and sharp fluid jets shooting up- and downwards, as
depicted in figure~\ref{Cavities}~(c). Here we will mainly focus
on the detailed description of the upwards jet and demonstrate
that the downward jet can be treated in the same way.

From figure \ref{Cavities}, observe that larger Froude numbers
create more slender cavities and also increase the non-dimensional
depth at which the cavity pinches-off.  Furthermore, it can be
appreciated that the jets are extremely thin and that the time
needed for the tip of the jet to reach the free surface is only a
small fraction of the pinch-off time. This latter observation
means that the jets possess a much faster velocity than the
velocity of the impacting solid, a conclusion which was also
extracted from the analysis of the experiments in figures
\ref{ExperimentosI}-\ref{ExperimentosII}. Motivated by this
striking fact, one of the main objectives in this paper will be to
address the following question: what is the relationship between
the impact velocity $V_D$ - or, in dimensionless terms, between
the Froude number - and the liquid velocity within the jet?

\begin{figure}
  \centerline{\includegraphics[width=0.5\textwidth]{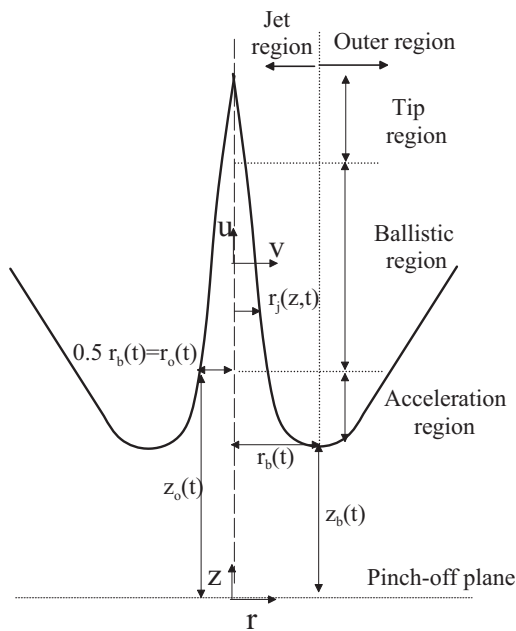}  }
  \caption{Sketch showing the different lengths used to define the jet base and the regions of the jet. The jet base $(r_b, z_b)$ is located where the interface possesses a local minimum. The outer region covers the bulk of the fluid with $r>r_b$ and $z<z_b$. The jet region is subdivided into the acceleration, the ballistic and the tip region. Note that, in the following, $u$ and $v$ will be used to denote axial and radial velocities, respectively}\label{JetGeometry}
\end{figure}

With this purpose in mind, it will prove convenient to define
first the length scale that characterizes the jet width. In
\cite{PRL09} we showed that the time evolution of the jet is a
\emph{local} phenomenon, independent of the stagnation-point type
of flow generated after pinch-off at the location where the cavity
collapses. Therefore, this characteristic length needs to be
related to a \emph{local} instead of a global quantity and,
following \cite{PRL09}, we choose the radial position at which the
interface possesses a local minimum i.e., the radius $r_b(t)$
indicated in figure \ref{JetGeometry}. We shall in the following
call this point the \emph{jet base} and denote its vertical
position by $z_b(t)$.

\begin{figure}
  \centerline{\includegraphics[width=0.5\textwidth]{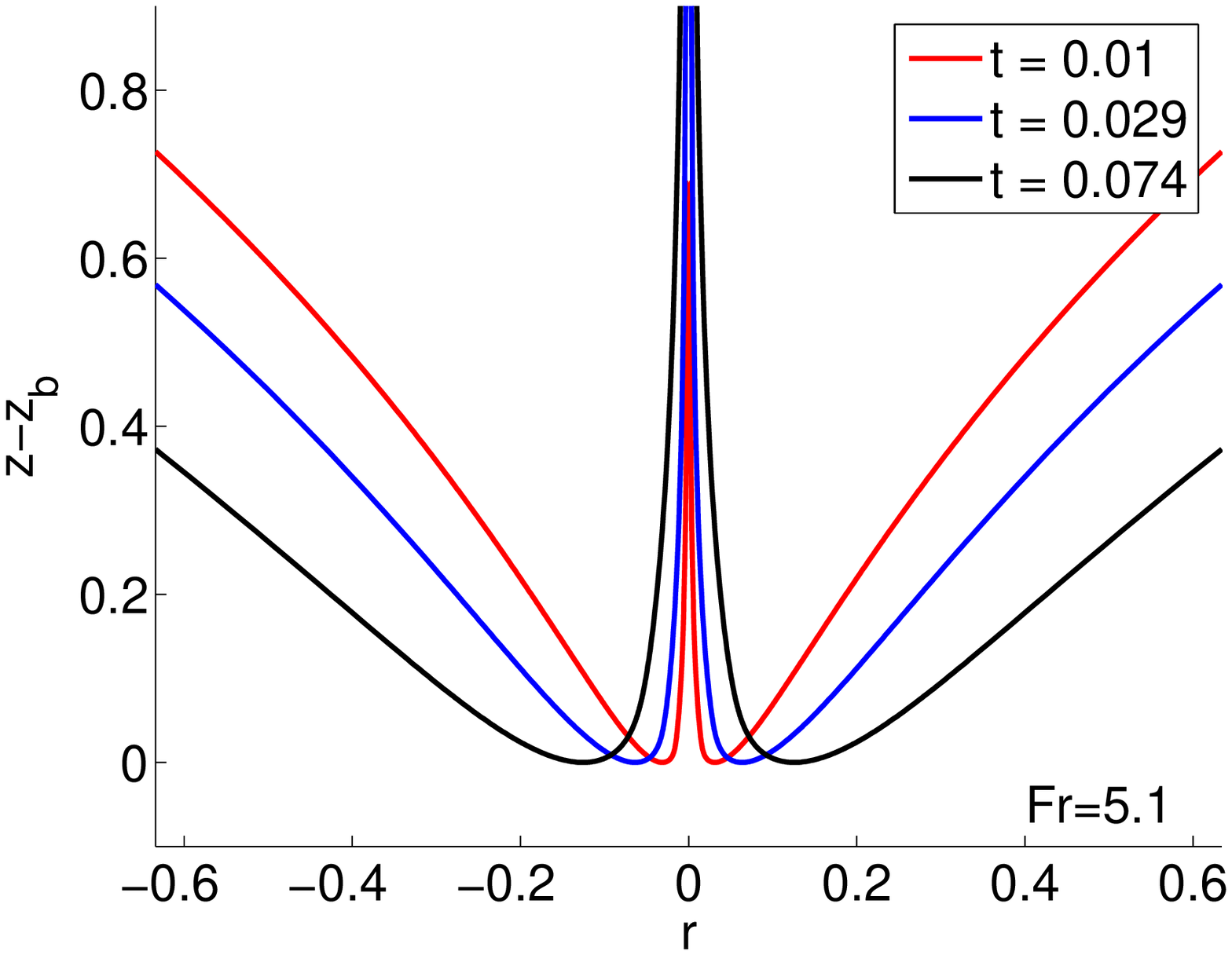}
   \includegraphics[width=0.5\textwidth]{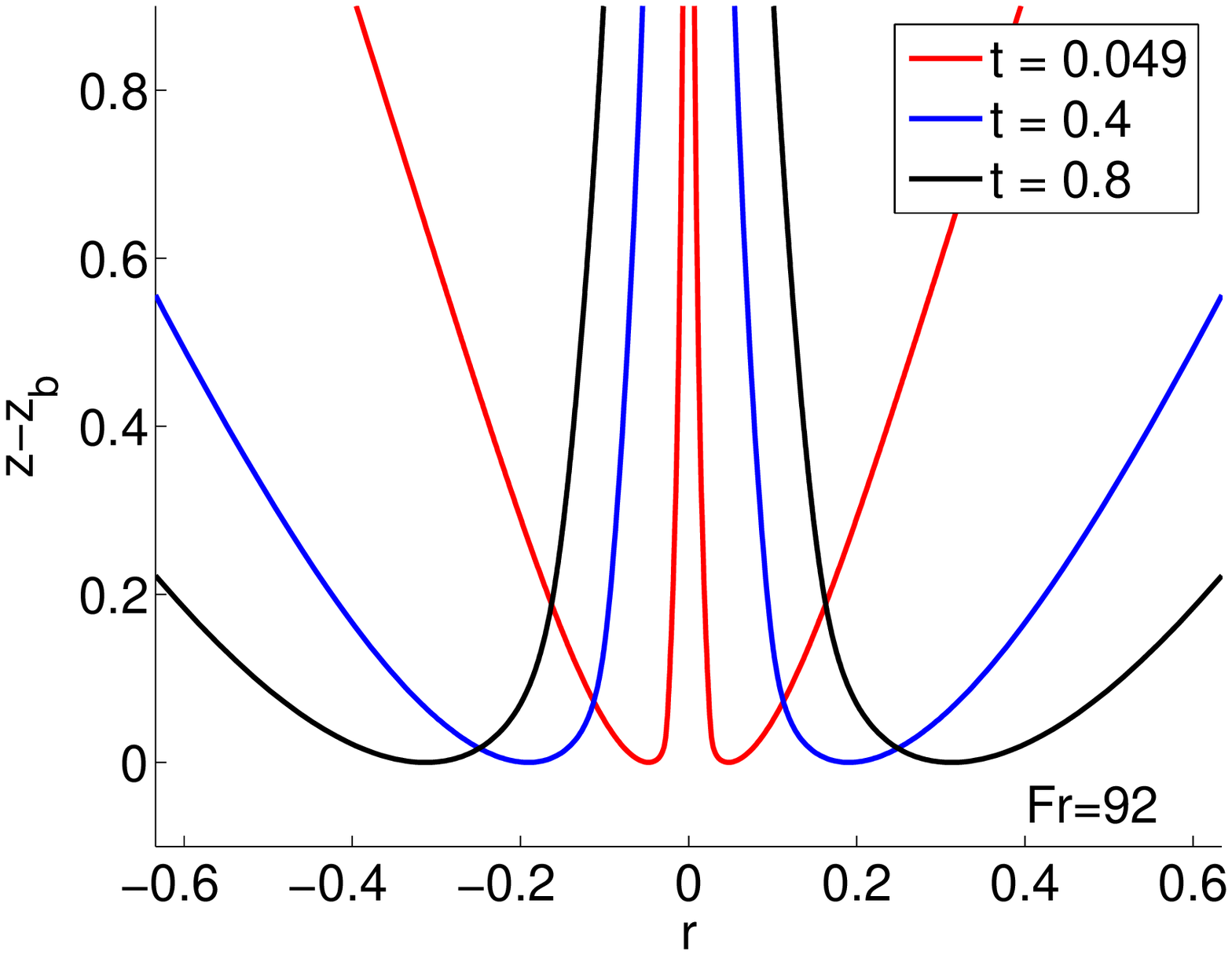}  }
  \caption{Jet shapes translated vertically for different instants of time and different values of the impact Froude number.}\label{JetShape}
\end{figure}

\begin{figure}
  \centerline{\includegraphics[width=0.5\textwidth]{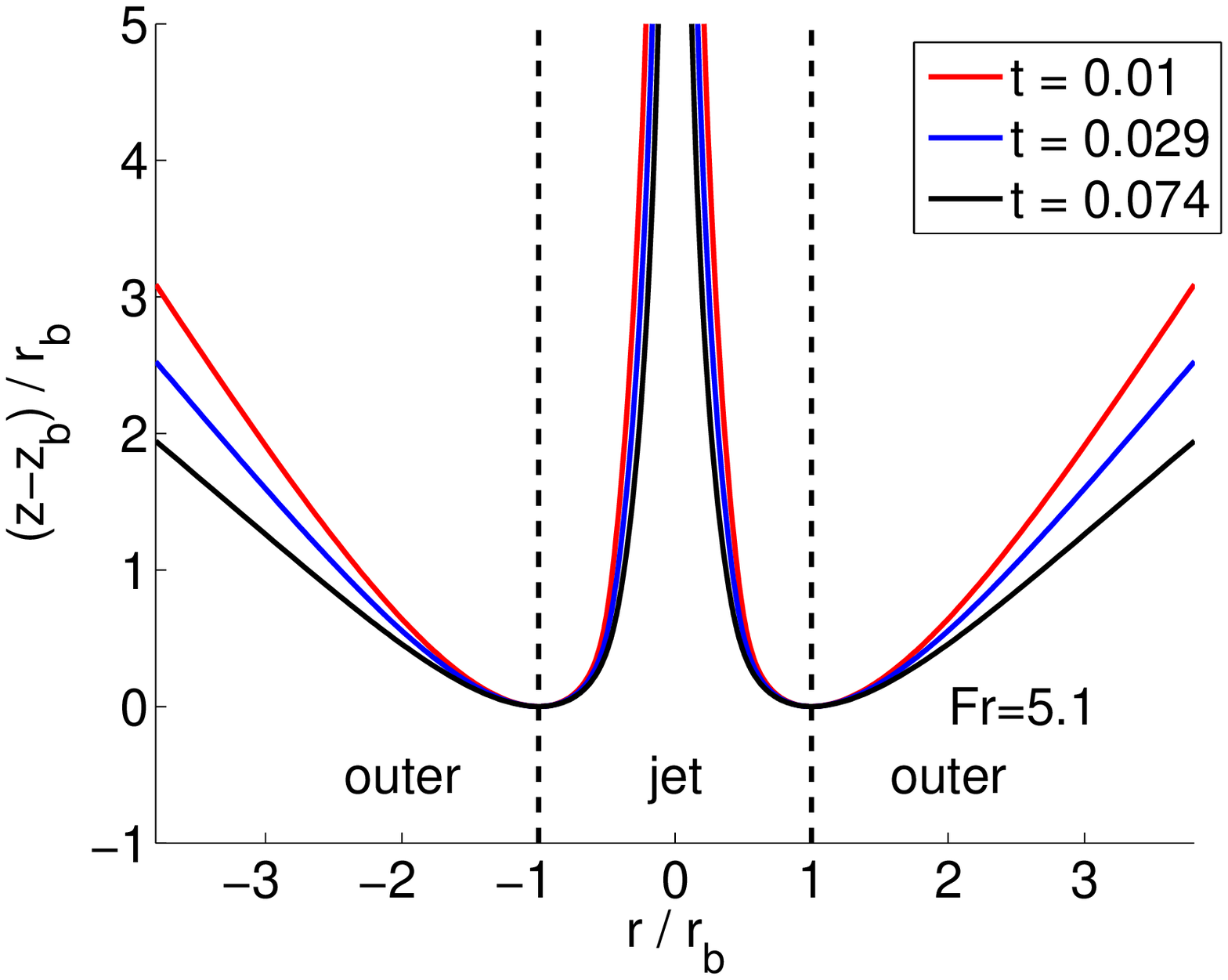}
   \includegraphics[width=0.5\textwidth]{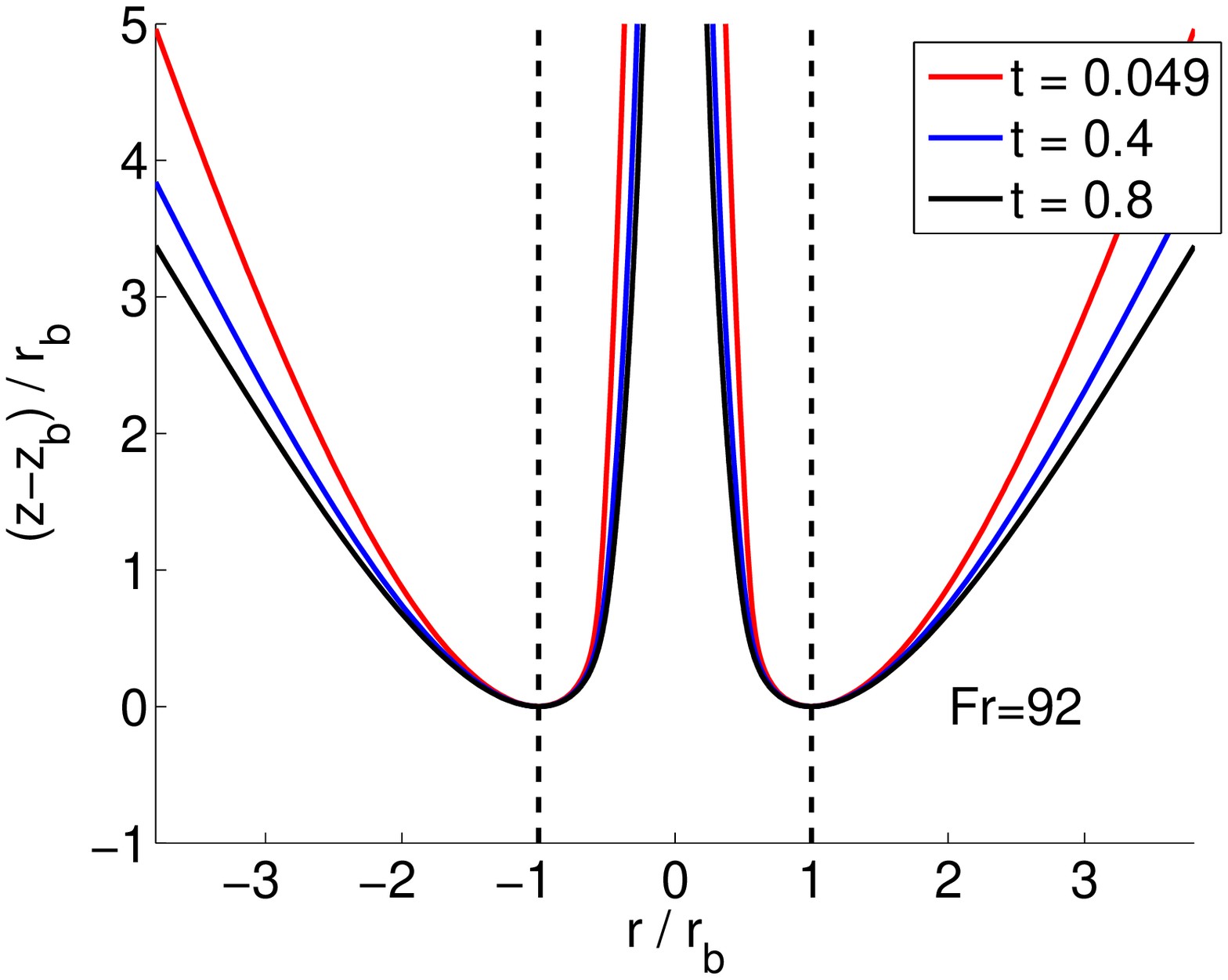}  }
  \caption{Shapes of the jets depicted in figure \ref{JetShape} when distances are normalized using $r_b$ overlay reasonably well indicating that $r_b$ is a good choice for the characteristic local length scale. The definition of the outer and jet regions is also indicated.}\label{JetShapeadim}
\end{figure}

\begin{figure}
  \centerline{\includegraphics[width=0.5\textwidth]{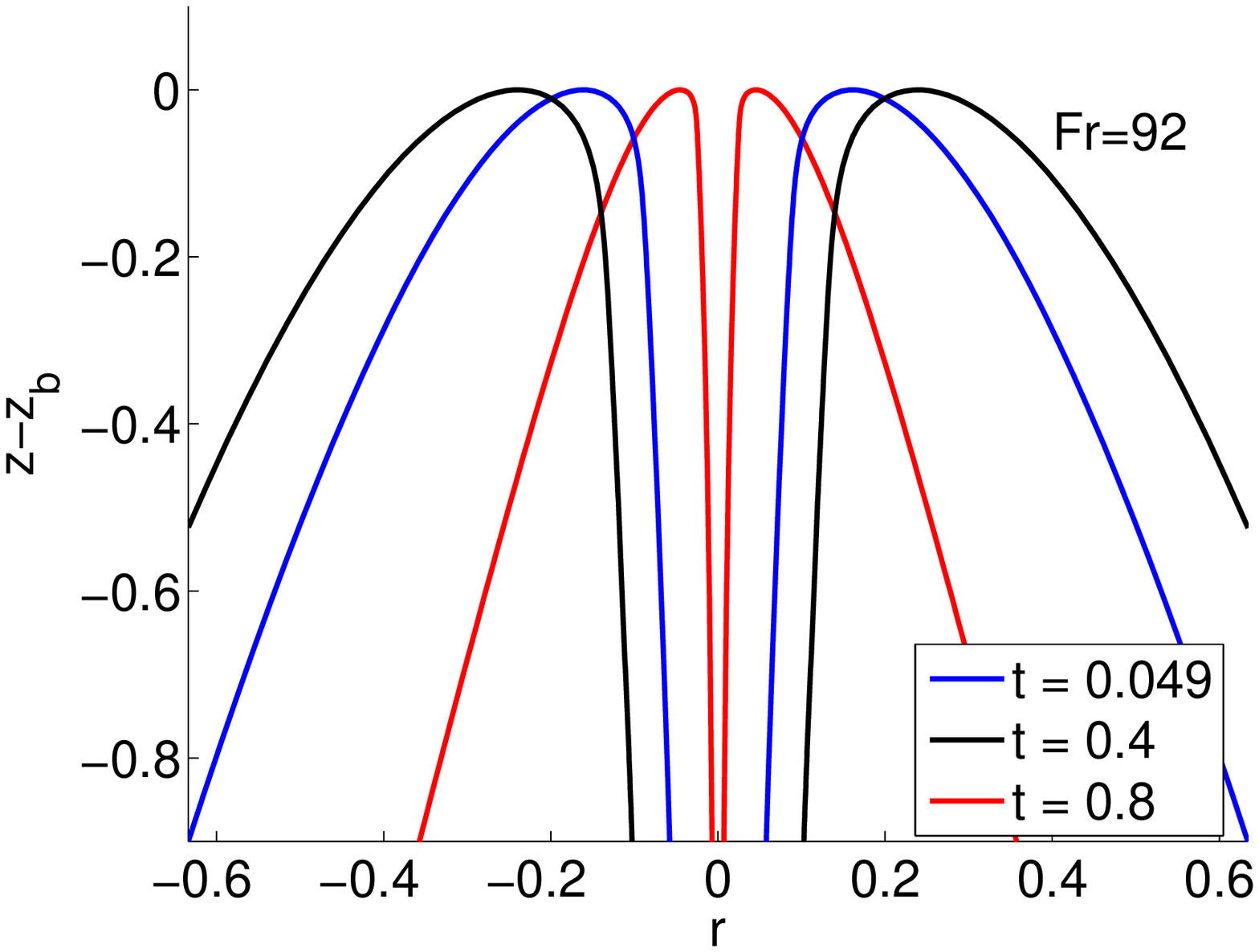}
   \includegraphics[width=0.5\textwidth]{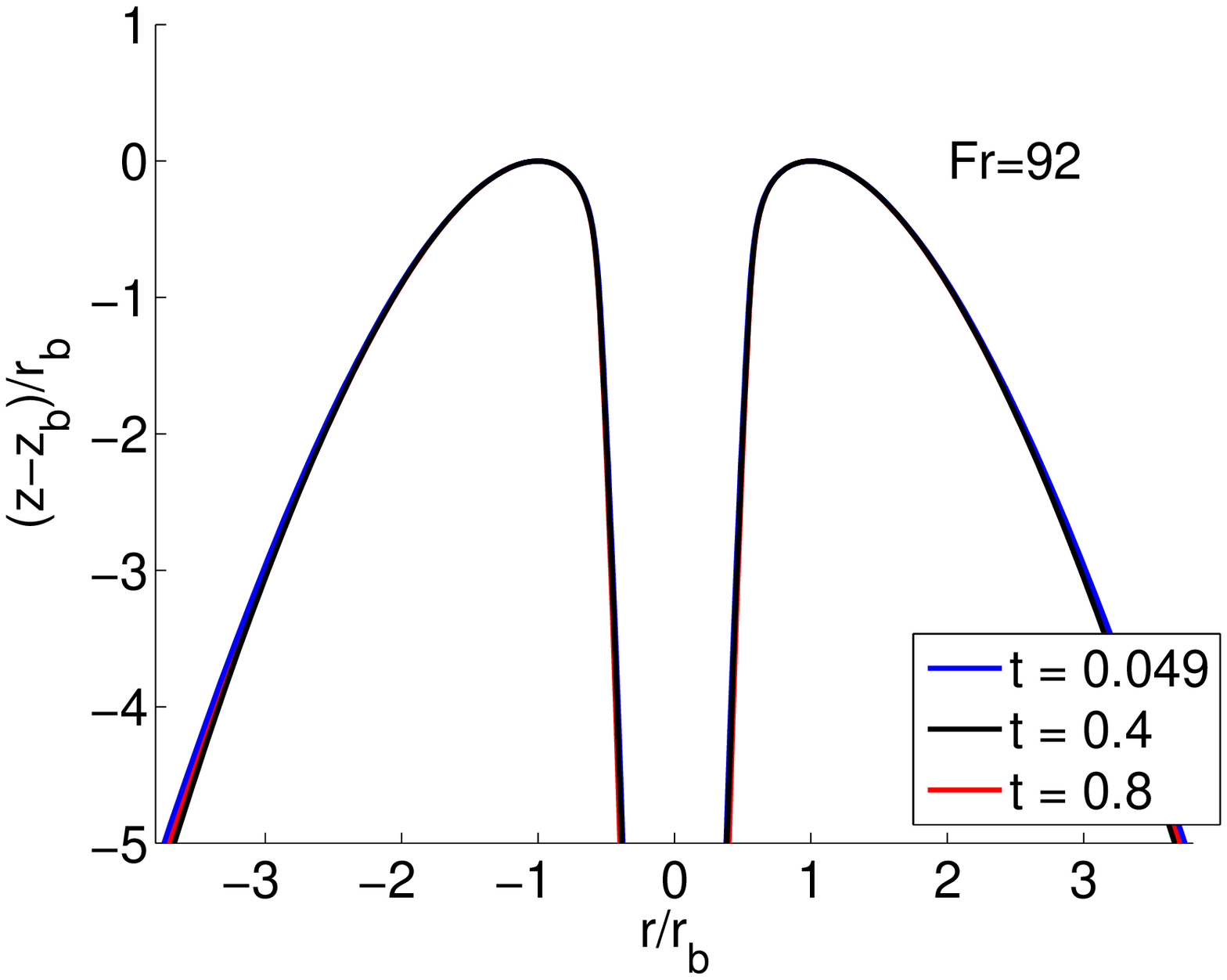}  }
  \caption{Jet shapes for the downward jet at Fr=92 taken at the same times as in figures \ref{JetShape} and \ref{JetShapeadim} show a similarly good rescaling with the corresponding base radius $r_b$.}\label{JetShapeDown}
\end{figure}

To clearly show the spatial region surrounding the jet base, some
of the different jet shapes taken from the time evolutions of
figure \ref{Cavities}, are translated vertically so that they
share a common vertical origin, as depicted in figure
\ref{JetShape}. Note that both the jet base and the jet itself
widen as the time from pinch-off increases. Interestingly enough,
figure \ref{JetShapeadim} shows that jet shapes exhibit some
degree of self-similarity since they nearly collapse onto the same
curve when distances are normalized using $r_b$. This fact
indicates that $r_b$ is not an arbitrary choice, but a relevant
local length that plays a key role in the dynamics of the jet. The same arguments hold for the downward jet as illustrated in figure~\ref{JetShapeDown}.

In order to model the full process of jet ejection and break-up we
divide the liquid flow field into two different regions: the outer
region, defined for $r>r_b$, $z<z_b$ and the jet region, extending
from the jet base to the axis i.e, $r<r_b$ and $z\geqslant z_b$,
as illustrated in figure \ref{JetShapeadim}. The jet region is
further divided into three different axial subregions: the \emph{acceleration region}, the \emph{ballistic
region} and the \emph{tip region} as illustrated in figure
\ref{JetGeometry}.

\begin{figure}
  \centerline{\includegraphics[width=0.5\textwidth]{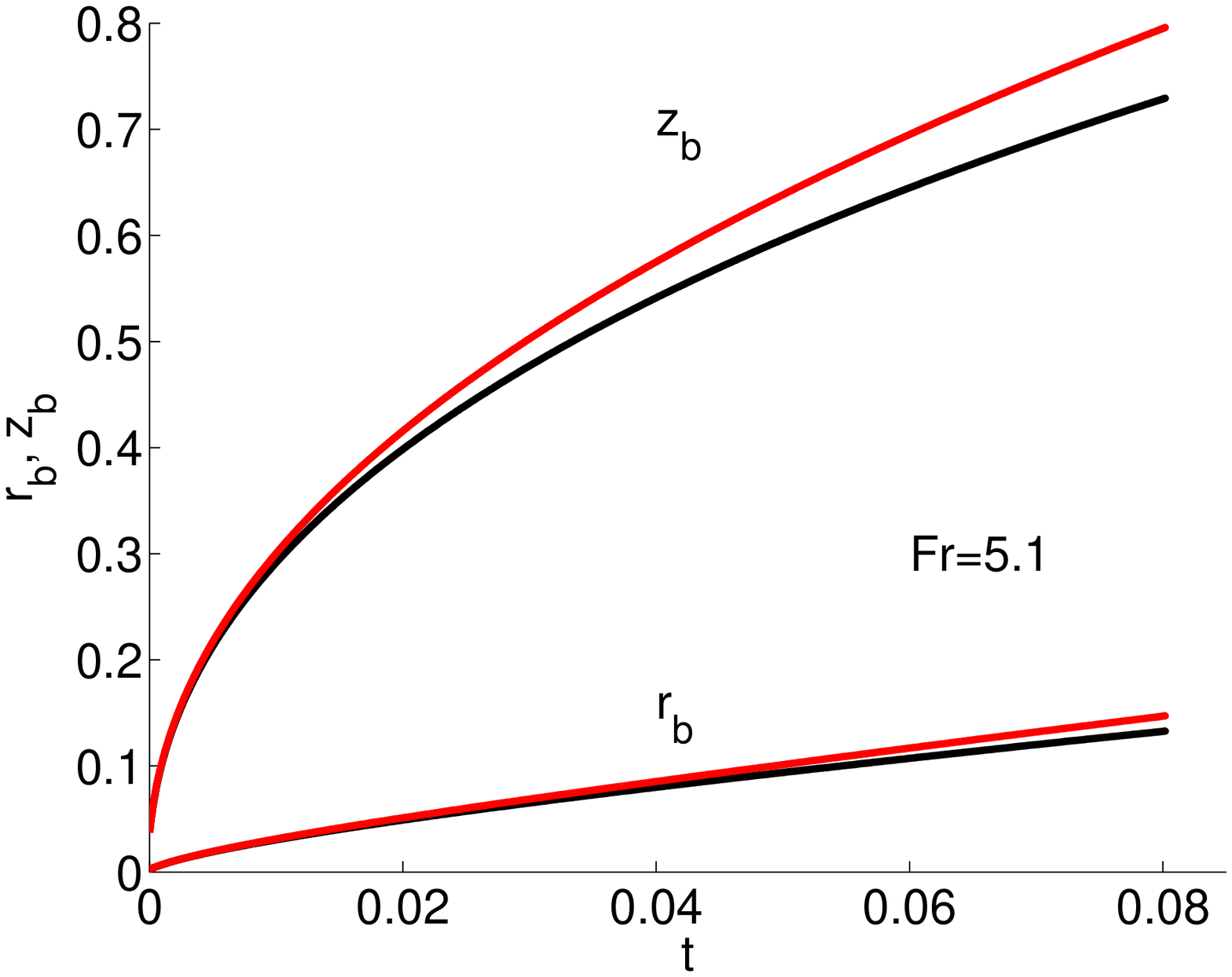}
  \includegraphics[width=0.5\textwidth]{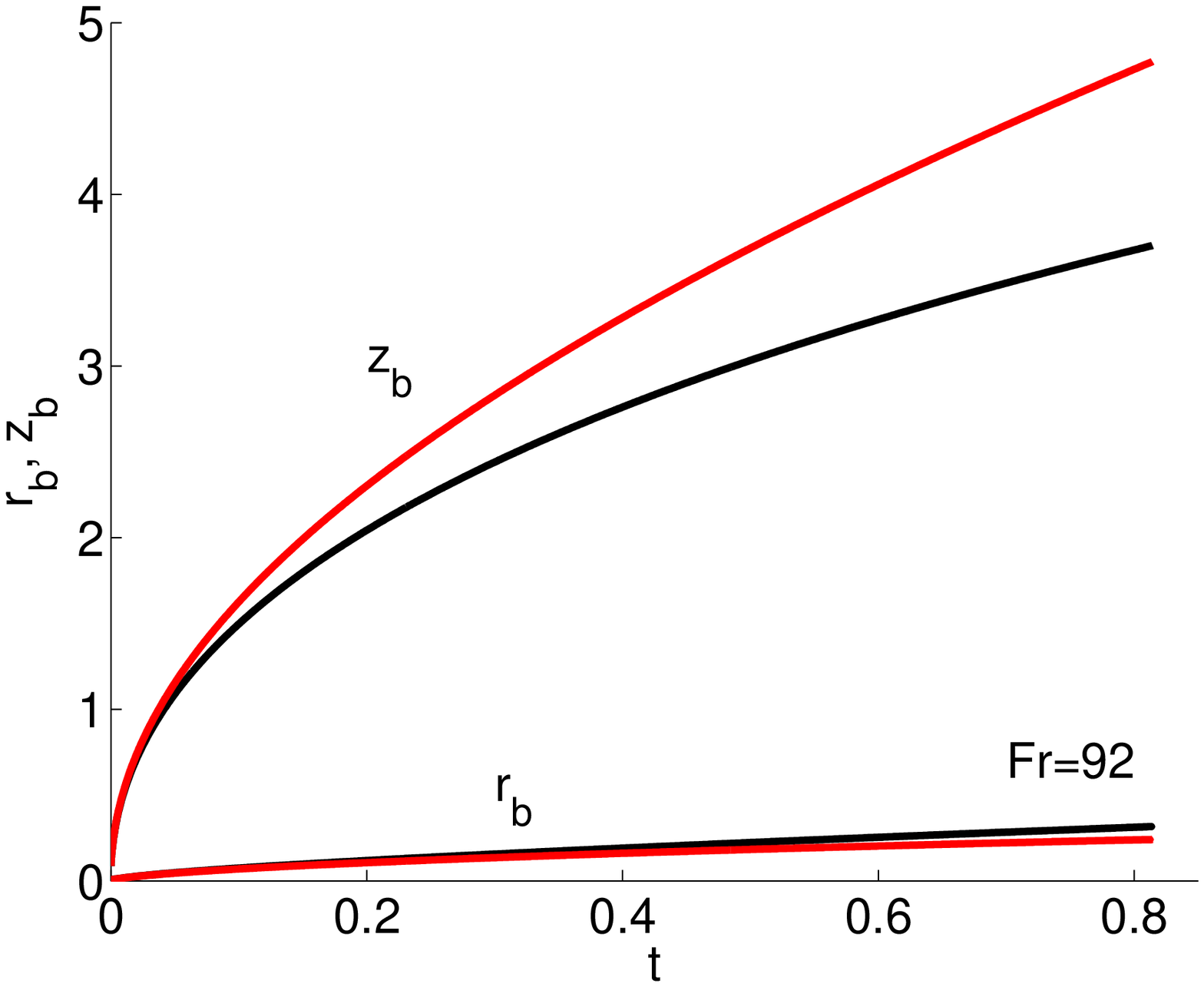}}
  \caption{Time evolution of radial and axial positions of the jet base, $r_b$ and $z_b$ respectively. The upward jet is shown in black and the downward in red (for the downward jet $-z_b$ is shown for convenience). The behavior of both jets is very similar.}\label{rbzb}
\end{figure}

Figure \ref{rbzb} shows that $z_b(t)\gg r_b(t)$. These
comparatively large values of $z_b(t)$ with respect to $r_b(t)$
are caused by the confinement of the jet by the cavity walls,
which inhibits the widening of the base radius. Moreover, the
small values of $r_b$ are responsible for the large axial
velocities within the jet (and, thus, for the large values of
$d\,z_b/d\,t$) since, as it will become clear below, vertical
velocities are inversely proportional to $r_b$.

\begin{figure}
\includegraphics[width=.45\textwidth]{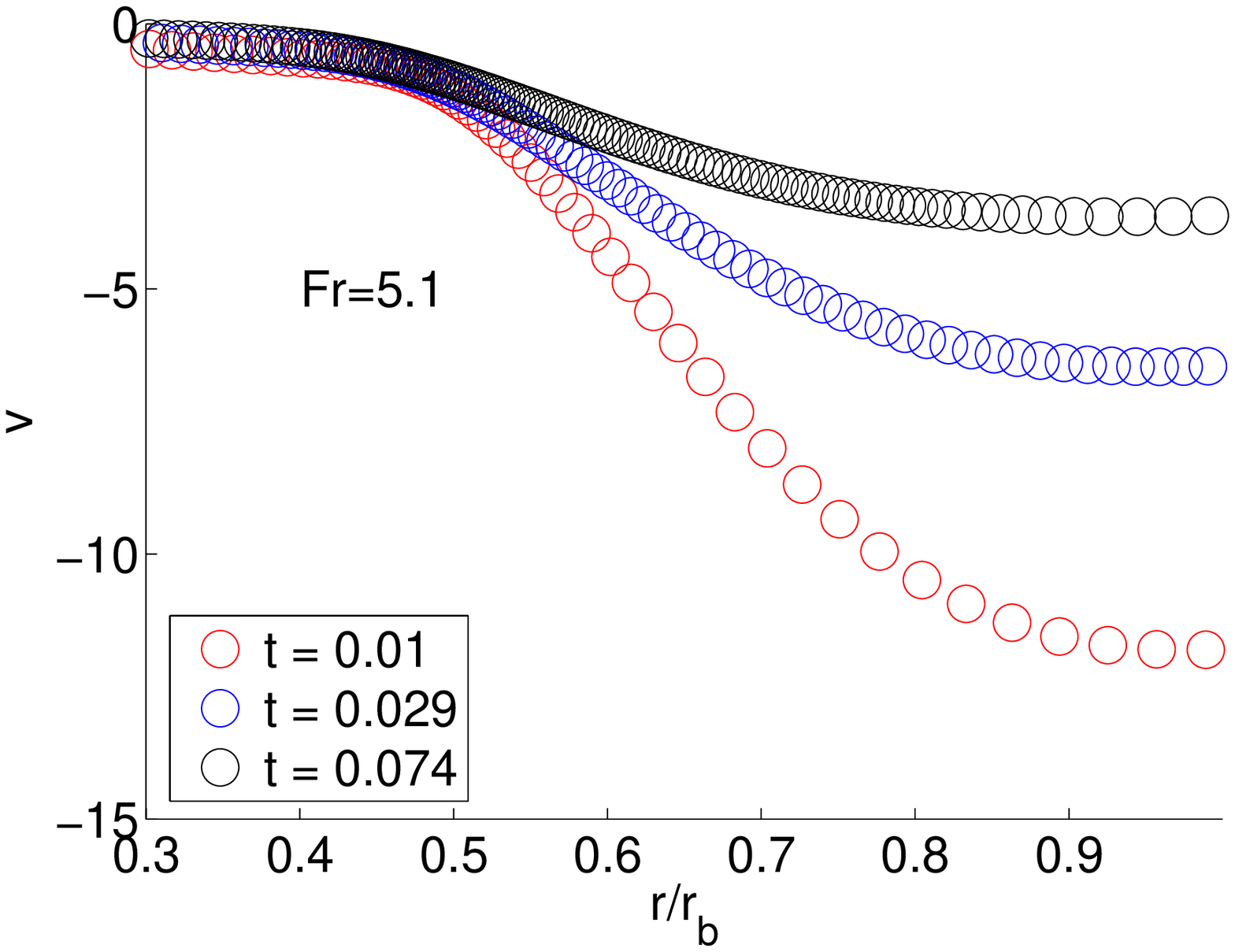}
\includegraphics[width=.45\textwidth]{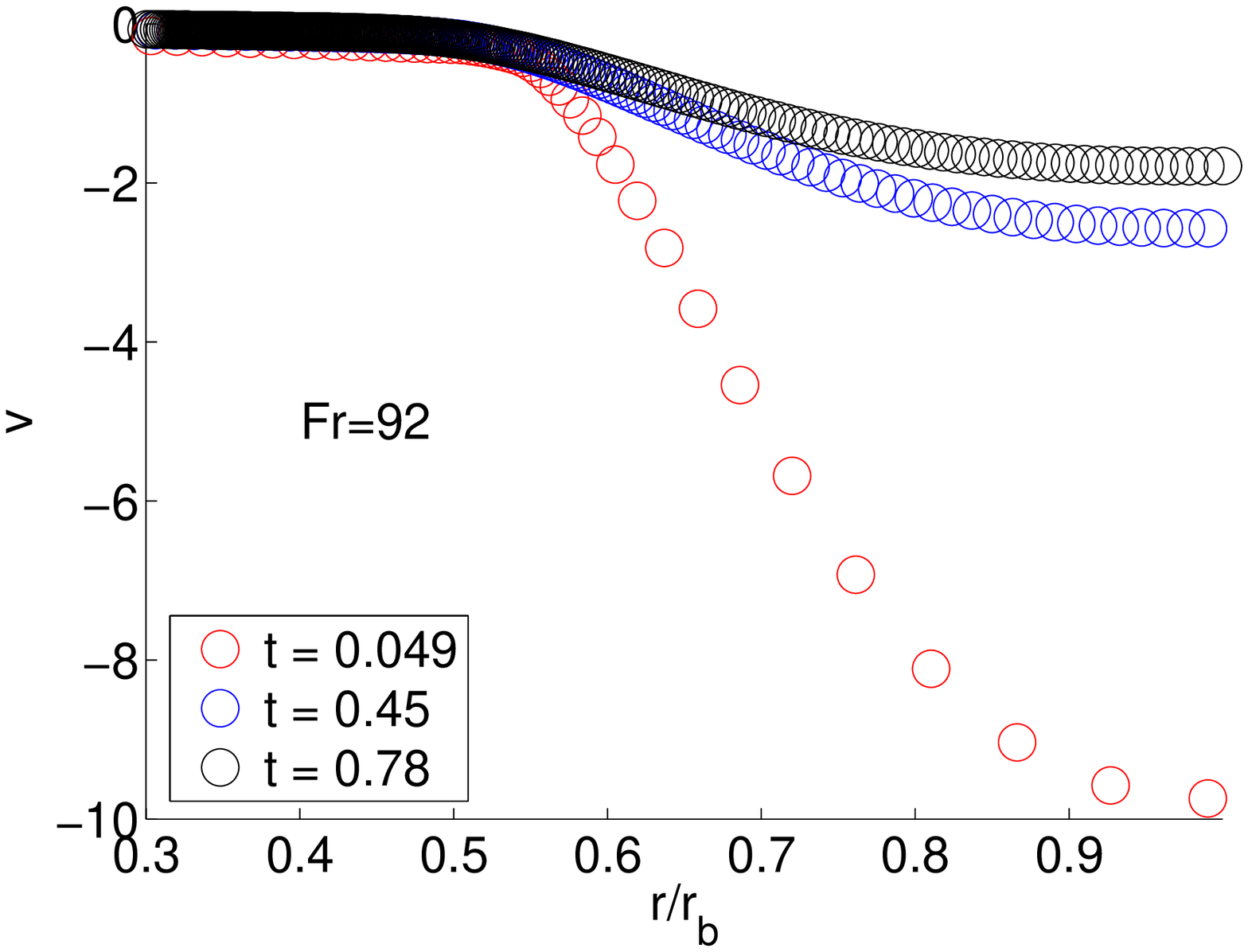}\\
\includegraphics[width=.45\textwidth]{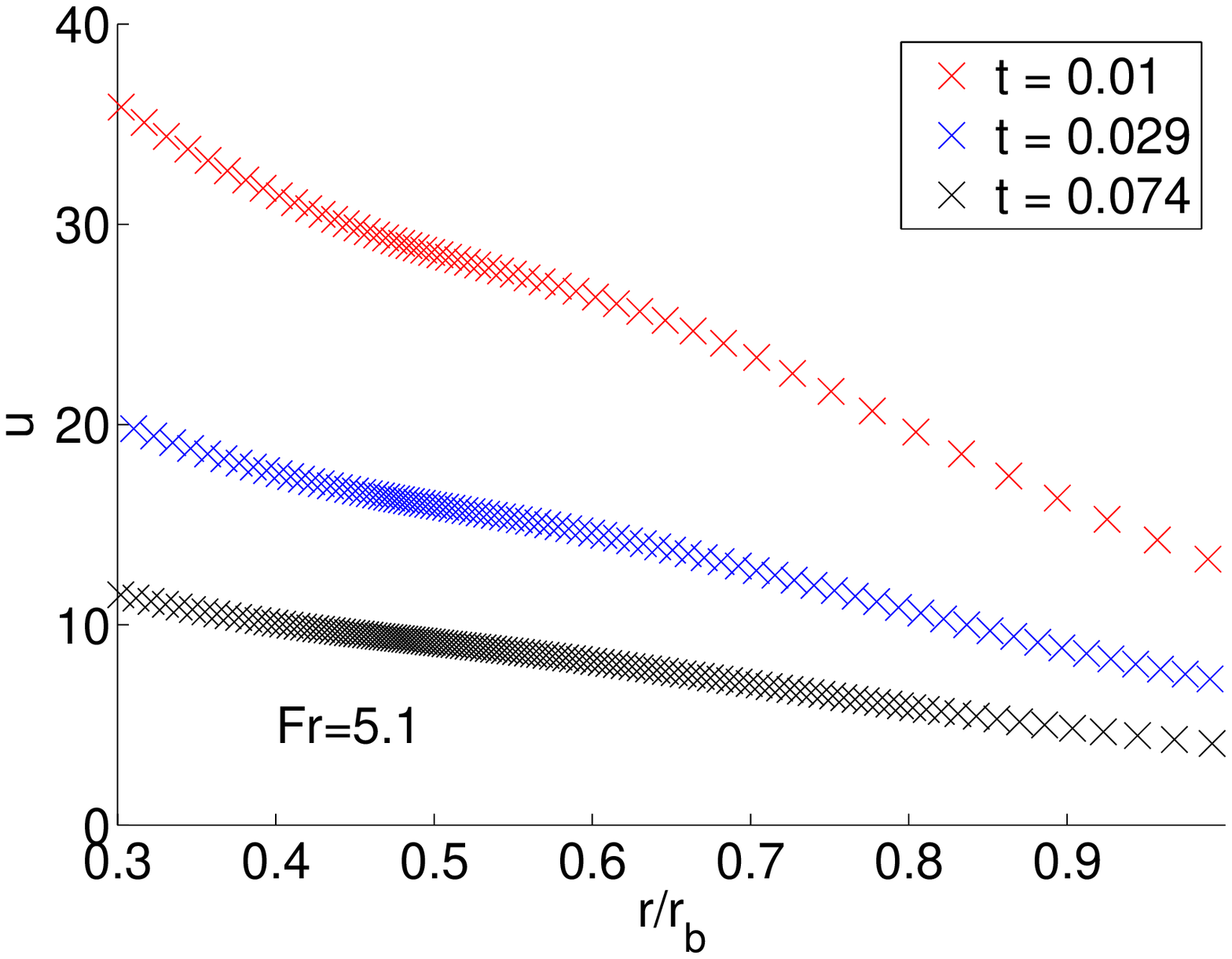}
\includegraphics[width=.45\textwidth]{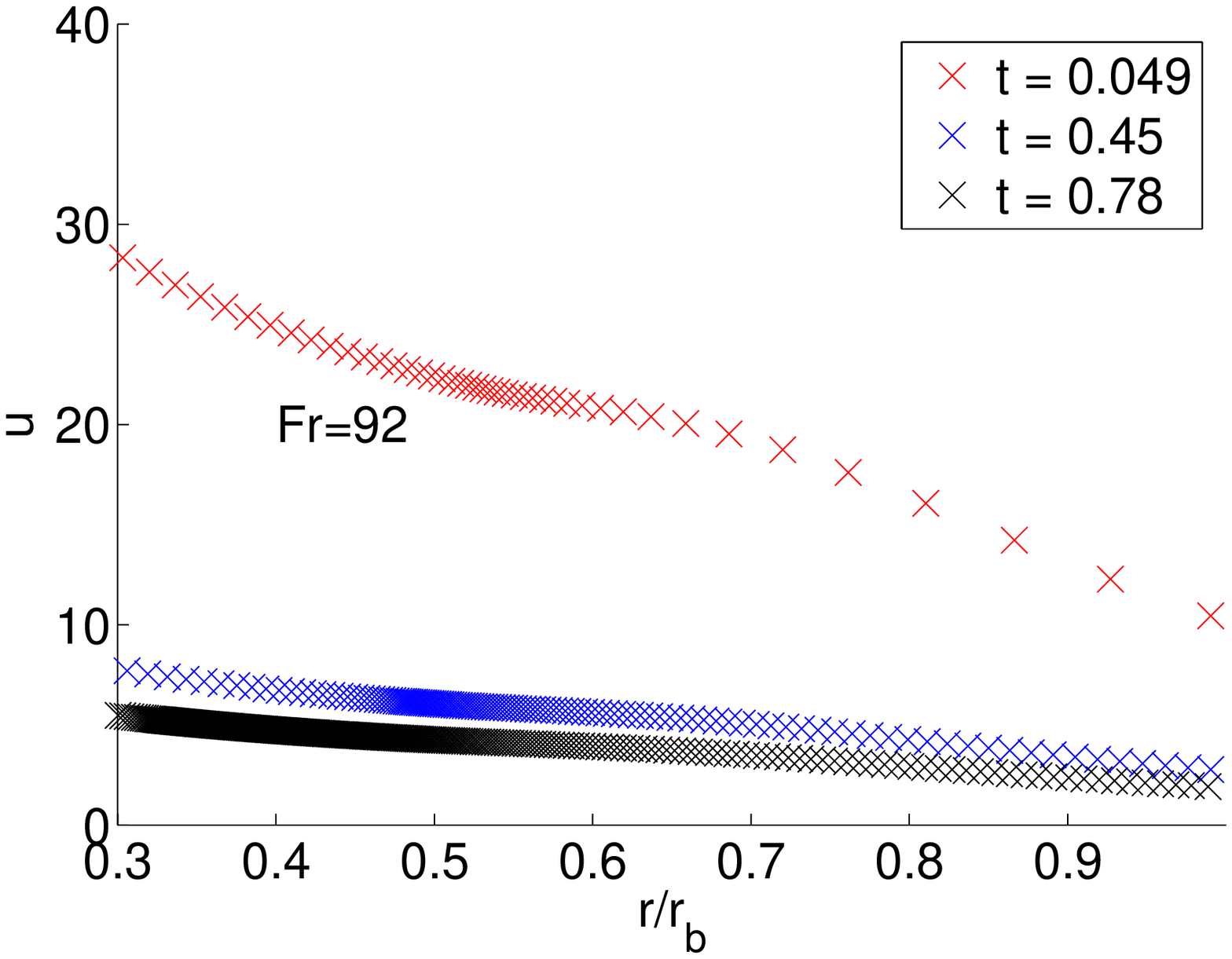}
\caption{Time evolutions of the radial and axial velocities ($v$
and $u$ respectively) of the liquid evaluated at the jet interface
for Fr=5.1 and Fr=92.} \label{vrvz}
\end{figure}

The importance of local processes around the jet base is even more
clearly illustrated in figure \ref{vrvz} where both the axial and
radial velocities evaluated at the jet air/liquid interface, $u$
and $v$ respectively, are represented for different instants of
time. In this figure one can observe that while the axial
velocities are of similar magnitude as the radial velocities at
$r=r_b$, they monotonically increase to much higher values as the
jet radius diminishes. Contrarily, the modulus of the (negative)
radial velocities decays from $\sim O(10)$ at $r=r_b$ to zero at
$r\simeq 0.5\,r_b$ and, therefore, the radial inflow experiences a
strong deceleration in the \emph{small} distance $\sim 0.5r_b$.
Since the liquid is at atmospheric pressure at the free surface of
the jet, the strong radial deceleration provokes an overpressure
below the jet base. Accordingly, a strong favorable
\emph{vertical} pressure gradient is created and, therefore, the
liquid experiences a large upwards acceleration in the vertical
direction, creating the high speed jet ejected into the
atmosphere.

\begin{figure}
  \centerline{\includegraphics[width=0.5\textwidth]{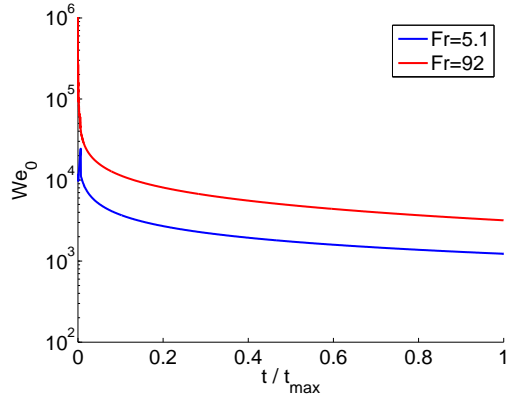}}
  \caption{Time evolution of the local Weber number at the beginning of the ballistic region for two different values of the impact Froude number. The large values demonstrate that surface tension is not relevant during the jet ejection process. To facilitate the comparison between the different Froude cases, times have been normalized by $t_\mathrm{max}$, which is the time when the
downward jet hits the disc and the simulation stops.}
\label{Welocal}
\end{figure}

In the following, we shall define $r_0=0.5r_b$ as the radial
position on the jet interface at which radial velocities become
negligible -$v\approx 0$ for jet radii smaller than $r_0$- and the
corresponding vertical position and velocity, will be denoted in
what follows $z_0$ and $u_0$, respectively. Moreover, we will also
define at this point a local Weber number as
$\mathrm{We}_0=\rho\,U_0^2\,R_0/\sigma=\mathrm{We}\,u_0^2\,r_0$
whose time evolution is depicted in figure \ref{Welocal}. The
large values indicate that surface tension effects can be
neglected in the description of the jet ejection process.

Thus, since the jet interface can be considered to be at constant
atmospheric pressure and surface tension effects are negligible
near the jet base, the only source of axial acceleration is the
axial pressure gradient caused by the radial deceleration of the
flow. Remarkably, this radial deceleration takes place in a very
localized region nearby the jet base. (For radial positions on the
jet smaller than $r_0$ already $v\simeq 0$ as shown
in figure \ref{vrvz}.) Therefore, the source of axial acceleration
(radial deceleration) is no longer active high up into the jet,
but only near the jet base. This key observation is used to define
two of the three different regions within the jet: the \emph{axial
acceleration region} for $r_0<r<r_b$ and $z_b<z<z_0$ and the
\emph{ballistic region} for $r<r_0$, $z\geqslant z_0$. The term
used to name the latter region is based on the fact that, since
$v\simeq 0$ for $r<r_0$ and the pressure at the jet
interface is atmospheric, the momentum equation projected in the
axial direction yields
\begin{equation}
\frac{D\,u}{Dt}=0\, \quad\mathrm{for}\quad
z>z_0\quad\mathrm{with}\quad u\neq u(r)\, ,\label{Momentum1D}
\end{equation}
and $D/Dt$ indicating the material derivative. Equation
(\ref{Momentum1D}) implies that fluid particles are no longer
accelerated upwards and conserve the vertical velocities they
possess at $z=z_0$, which is the axial boundary between the
\emph{axial acceleration region} and the \emph{ballistic region}.
In equation (\ref{Momentum1D}), $u\neq u(r)$ since the radial
velocity gradients of axial velocities are negligible in the
ballistic region (not shown).

As a next step, we would like to scale the radial velocity field
in the vicinity of the jet base which is, as discussed above, the
source of momentum driving the jet ejection process. These
radially inward velocities are originally created by the
difference between the hydrostatic pressure in the bulk of the
liquid and the gas pressure inside the cavity. After pinch-off
however, the radial velocity field feeding the liquid jet is not
appreciably modified by gravity during the time evolution of the
jet since the local Froude number at the beginning of the
ballistic region is
$\mathrm{Fr}_0=U^2_0/(gZ_{\mathrm{surface}})\gg 1$ with
$Z_{\mathrm{surface}}$ the axial distance between the beginning of
the ballistic region and the height of the free surface far from
the impact region (see figure \ref{Cavities}).

\begin{figure}
  \centerline{\includegraphics[width=0.5\textwidth]{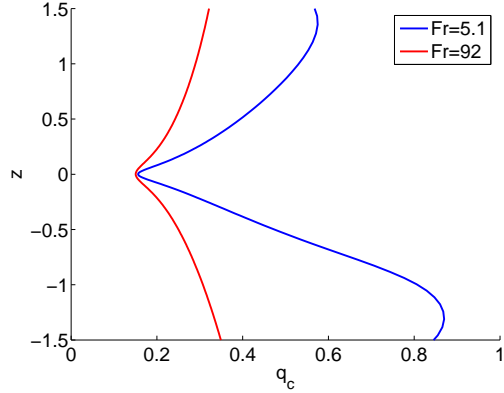}              }
  \caption{The sink strength distribution $q_c(z)$ for two values of the Froude number.
  }\label{qc}
\end{figure}

Therefore, the radial velocities which give rise to the jet
emergence can be characterized by the sink strength distribution
at $t=0$ right before pinch-off occurs:
$q_c(z)=-\,r_c(z)\dot{r}_c(z)$, where $r_c$ and $\dot{r}_c$
indicate the radius of the cavity and its associated radial
velocity, respectively (see \cite{PRL09}). The values of $q_c(z)$
are shown in figure \ref{qc} for several values of the impact
Froude number.

\begin{figure}
  \centerline{\includegraphics[width=0.5\textwidth]{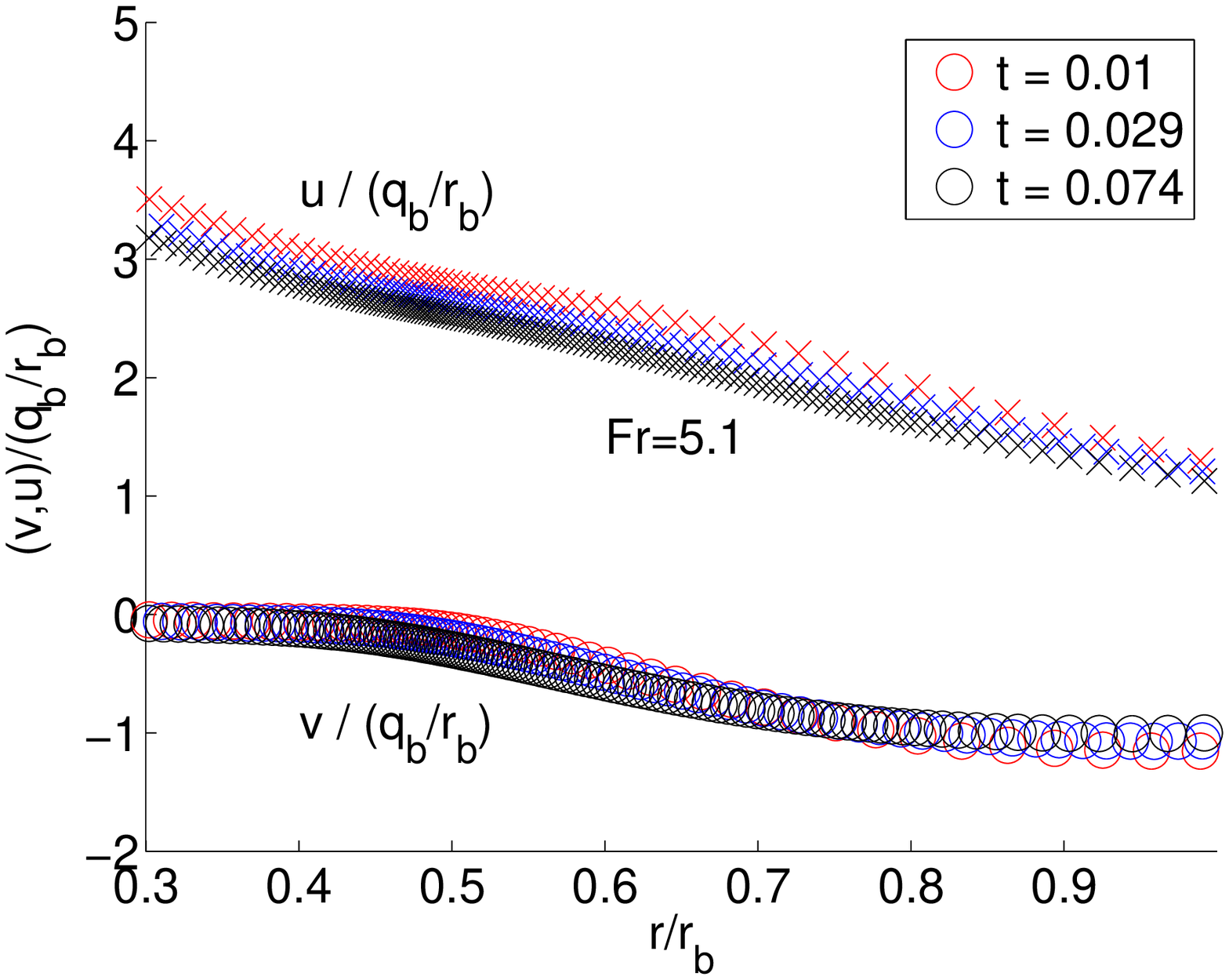}
             \includegraphics[width=0.5\textwidth]{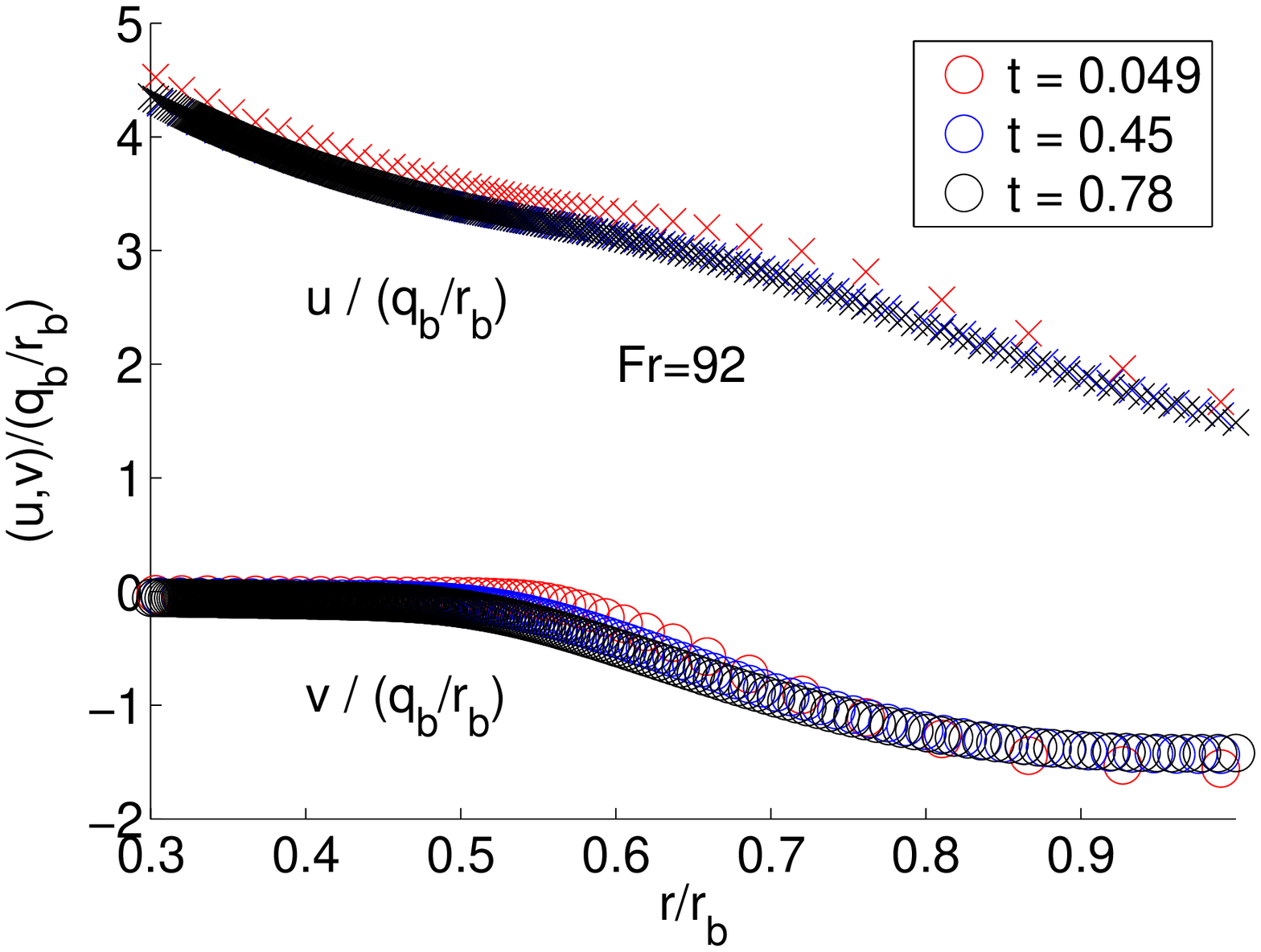}  }
 \caption{Spatial and temporal evolutions of the radial and axial velocities depicted in figure \ref{vrvz} normalized with $q_b(t)/r_b(t)$.
 Observe that $u/(q_b/r_b)$ and $v/(q_b/r_b)$ nearly collapse onto the same curves for each of the two values of the impact Froude number considered, $\mathrm{Fr}=5.1$ (a) and $\mathrm{Fr}=92$ (b).}\label{vrvzscaled}
\end{figure}

In order to demonstrate the intimate relation of the jet ejection
process with the velocity field right before pinch-off, we
normalize the velocities $v$ and $u$ at the jet surface (as shown
in figure \ref{vrvz}) using, as the characteristic scale for
velocities, $q_b(t)/r_b(t)$, where $q_b(t)=q_c(z=z_b(t))$ is the
sink strength at the height of the jet base. The remarkable
result, depicted in figure \ref{vrvzscaled}, is that both rescaled
velocities nearly collapse onto the same master curves for a given
Froude number and thus are almost constant in time for a fixed
value of the rescaled position $r/r_b<1$. This implies that, for a
fixed value of $q_b$, axial velocities are inversely proportional
to $r_b$ i.e., the smaller the jet base radius - or, equivalently,
the more confined is the jet by the cavity walls -, the larger
will be the axial liquid velocities within the jet.

\begin{figure}
  \centerline{\includegraphics[width=0.5\textwidth]{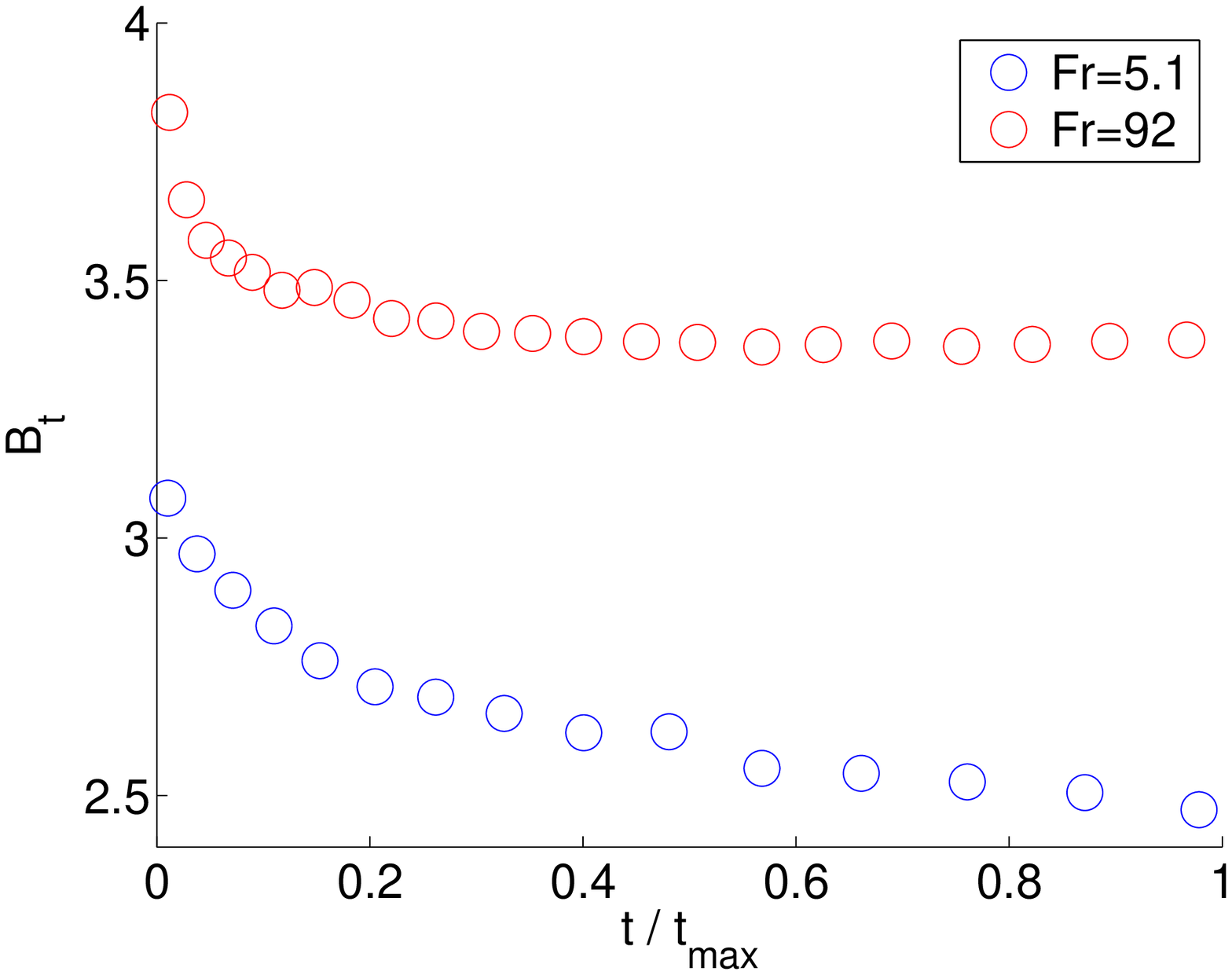}
  \includegraphics[width=0.5\textwidth]{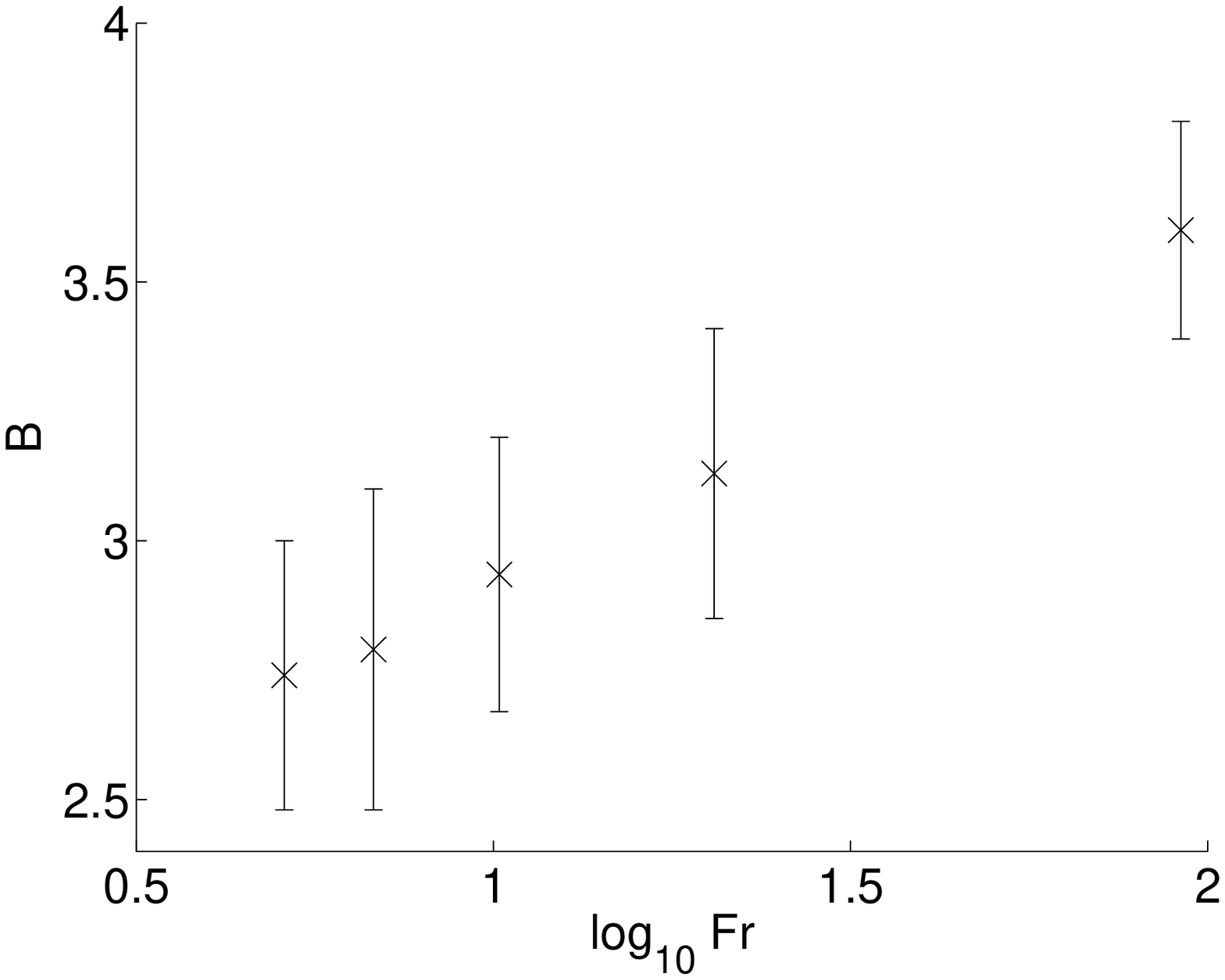}              }
  \caption{(a) The time evolution of $B_t$ demonstrates that $B_t$ is roughly constant in time, but does depend somewhat on the Froude number. (b) Taking the average of $B_t$ over time (with error bars indicating the min/max) for different Froude numbers yields a function $B(\mathrm{Fr})$ which varies only between 2.5 and 3.5 in the range $3\leqslant \mathrm{Fr}\leqslant 92$. As indicated in figure \ref{Welocal}, $t_\mathrm{max}$ is the time when the downward jet hits the disc and the simulation stops.}\label{BFr}
\end{figure}

Of critical importance for our forthcoming discussion is the
rescaled axial velocity evaluated at the boundary of the ballistic
region, $B_t=u_{0}(t)/(q_b(t)/r_b(t))$, whose time evolution is
depicted in figure \ref{BFr} (a). In accordance with the collapse
of the rescaled velocities on a single master curve depicted in
figure \ref{vrvzscaled}, $B_t$ hardly changes with time and, thus,
we can define the function $B(\mathrm{Fr})=u_{0}/(q_b/r_b)$ which
depends also very weakly on the Froude number, as depicted in
figure \ref{BFr} (b).

The result in figure \ref{BFr} possesses the additional remarkable
implication that axial velocities within the jet are larger than
the radial velocities existing at the cavity interface before
pinch-off occurs. This can be seen directly by recalling that
$\left| q_b/r_b\right| = \left|\dot{r}_b\right|$, such that $B$ is
the ratio between the axial velocity $u_0$ with which fluid is
ejected into the jet and the radial inward velocity at the jet
base. Then, during the initial instants of jet formation,
$r_b\simeq r_{min}$, with $r_{min}$ the minimum radius of the
cavity before jet emerges. Therefore, since the maximum radial
velocity before pinch-off occurs is
$\left|\dot{r}_{min}\right|=\left|q_c(z=0)/r_{min}\right|$, the
maximum axial velocity within the jet is given by
$\mathrm{max}(u_0)=B(\mathrm{Fr})\,q_c(z=0)/r_{min}\sim
3\,\dot{r}_{min}$. This means that, essentially, the velocity with
which the jet is ejected is roughly three times larger than the
maximum radial velocity attained before pinch-off!

In addition, provided that $\mathrm{We}_0\gg 1$, fluid particles
conserve their axial velocity within the ballistic region [see
equation (\ref{Momentum1D})] and, consequently, the tip of the jet
\emph{transports} away from the pinch-off location very valuable
information about the largest velocities reached during the cavity
collapse process. The knowledge of the function $B$ could thus
allow an experimentalist to estimate the maximal pinch-off
velocity simply from measurements of the jet tip velocity.

\subsection{Jet ejection after bubble pinch-off from an underwater nozzle}

This section is devoted to the study of Worthington jets which are
ejected after the bubble collapse into a liquid pool
[\cite{Manasseh1,PoFRocioI}]. As depicted in figure
\ref{surfaceProfilesNeedle}, these jets are quite similar to the
ones formed after the impact of a solid body against a free
surface and, thus, we expect that the conclusions of section
\ref{jetEjectionDisc} can be also used for their description.

\begin{figure}
  \centerline{\includegraphics[width=0.5\textwidth]{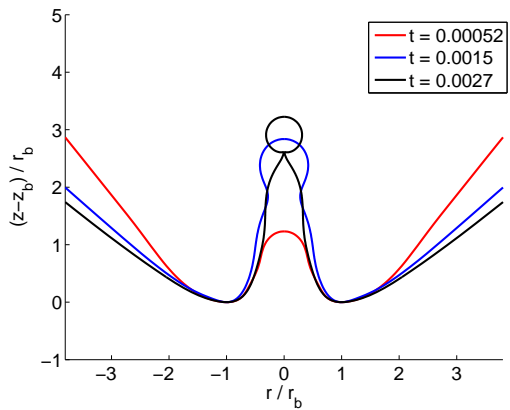}}
  \caption{The time evolution of the shapes of the jets ejected after bubble pinch-off from an underwater nozzle, show good overlap when distances are normalized using $r_b$.}\label{selfSimNeedle}
\end{figure}

\begin{figure}
  \centerline{\includegraphics[width=0.5\textwidth]{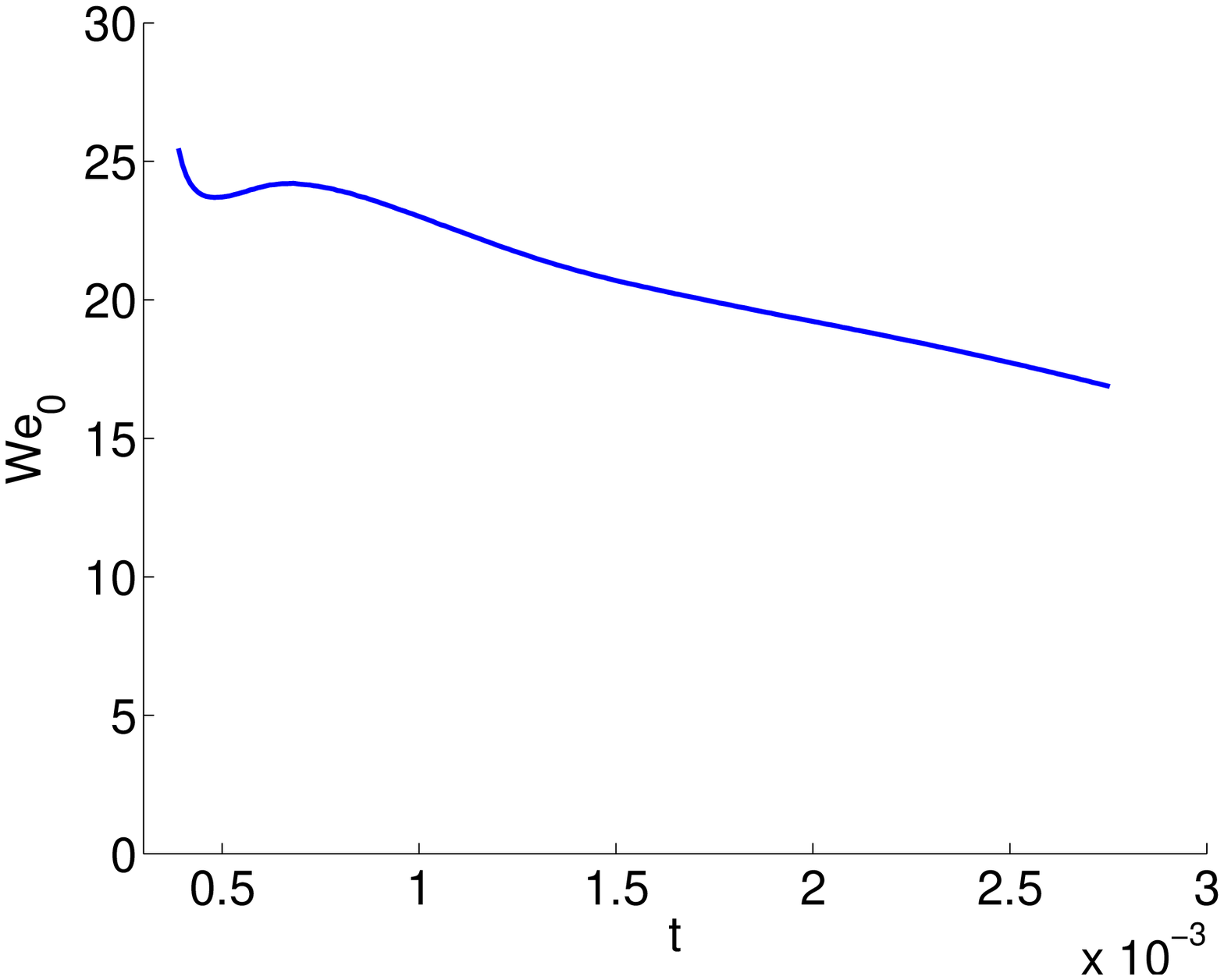}
\includegraphics[width=0.5\textwidth]{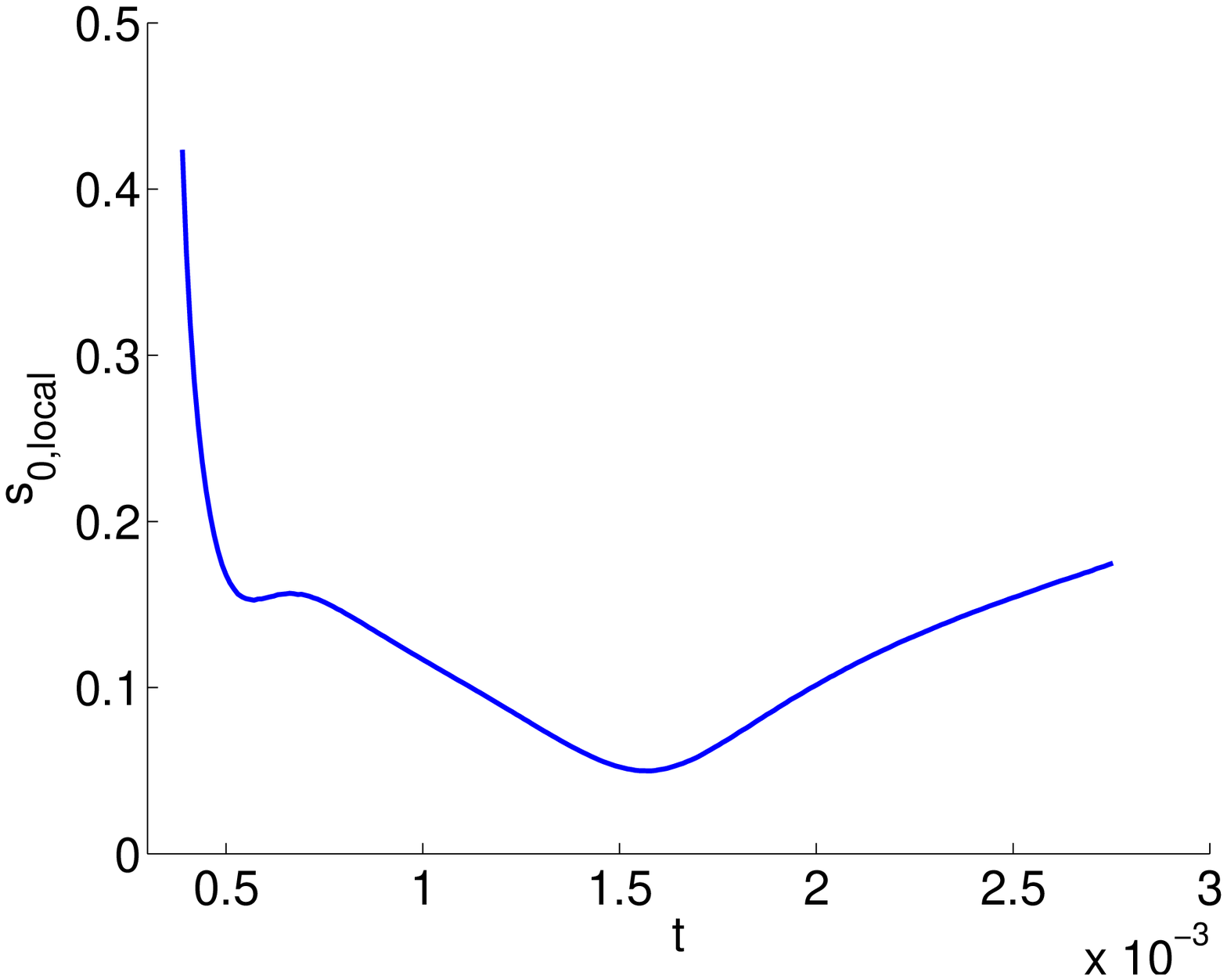} }
  \caption{(a) Time evolution of the local Weber number for the jet depicted in figure \ref{surfaceProfilesNeedle}. (b) Time evolution of the normalized strain rate $s_{0,local}$ at the beginning of the ballistic region for the same case as in (a).}\label{Weo}
\end{figure}

\begin{figure}
  \centerline{\includegraphics[width=0.5\textwidth]{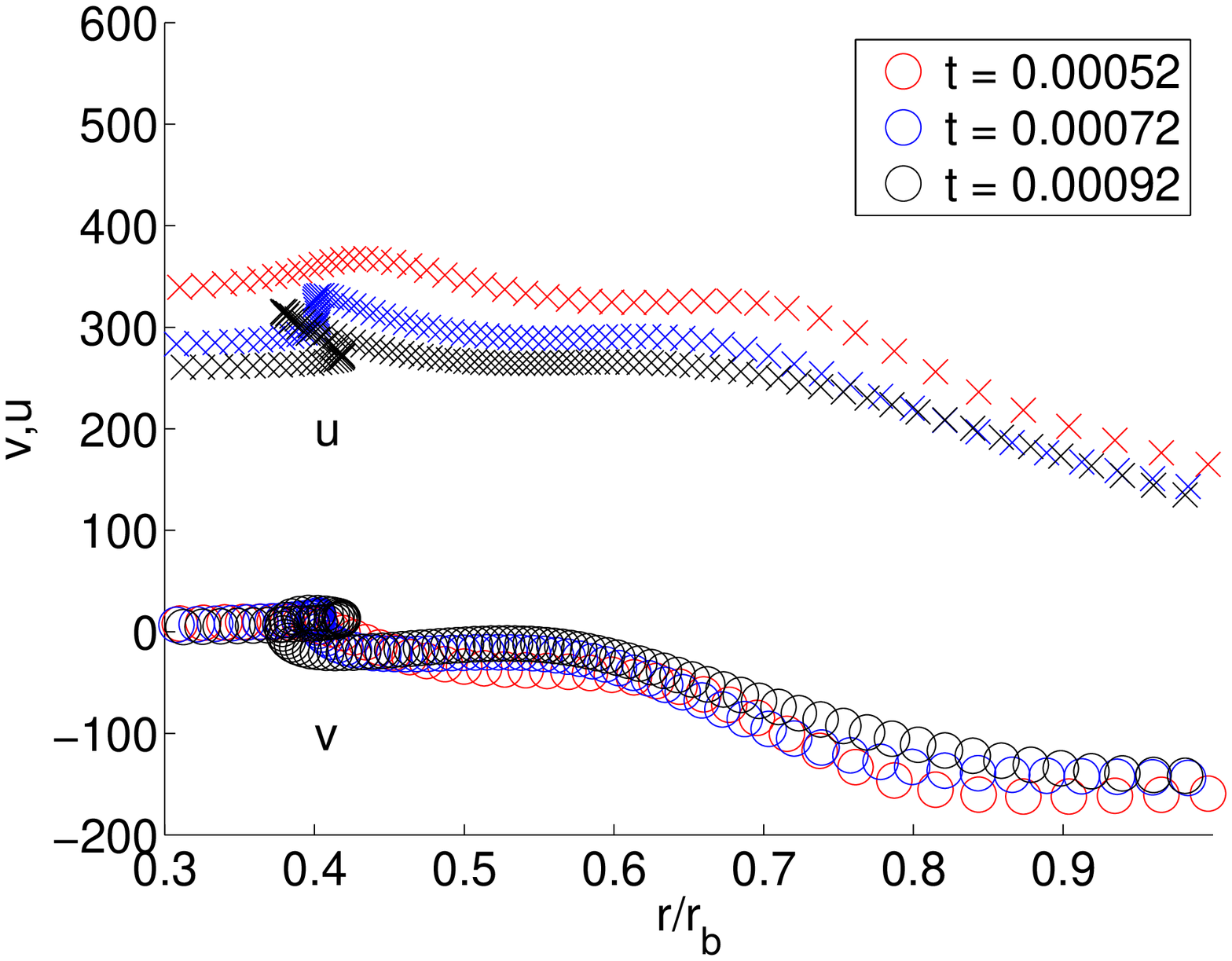}
\includegraphics[width=0.5\textwidth]{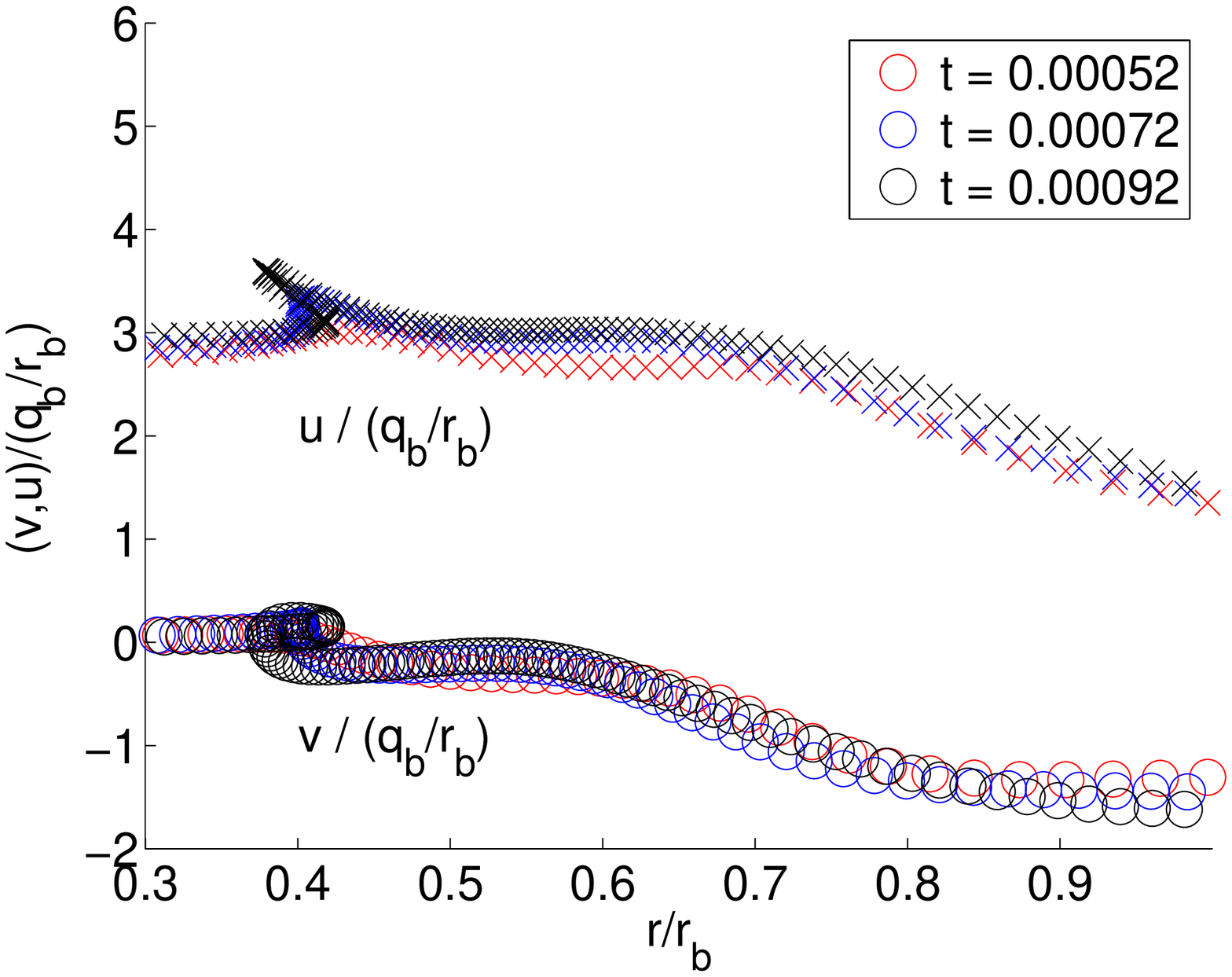}              }
  \caption{(a) Axial and radial velocities evaluated at the jet interface for the case depicted in figure \ref{surfaceProfilesNeedle}. In analogy with figure \ref{vrvzscaled}, both the acceleration and ballistic regions are clearly identified: the modulus of the radial velocities decreases from $r=r_b$ to become negligibly small for $r/r_b\lesssim 0.5$. (b) The same as in (a), but with velocities scaled with $q_b/r_b$. Due to the fact that the Weber number is considerably smaller in this case than for the impacting disc, the jet tip region can be appreciated in this figure as the multivalued part of the curves $u$ and $v$ for $r/r_b\simeq 0.4$.}\label{vrvzNeedle}
\end{figure}

Figure \ref{selfSimNeedle} shows that, similarly to section
\ref{jetEjectionDisc}, the different shapes nearly collapse onto
the same master curve when distances are normalized using $r_b$.
This fact corroborates that $r_b$ is also the correct length scale
to characterize this type of jets. However, differently to the
case of Worthington jets ejected after the impact of a solid body
against a free surface, in which $We_0\gtrsim 10^3$, the local
Weber number evaluated at the beginning of the ballistic region is
$\sim O(10)$ in this case (see figure \ref{Weo}a). As a
consequence of this, the total length at breakup of these jets is
$\sim O(r_b)$ (see figure \ref{selfSimNeedle}), i.e, much shorter
than the length of the Worthington jets in section
\ref{jetEjectionDisc}. Moreover, such comparatively low values of
the local Weber number indicate that surface tension has an effect
in the description of the jet ballistic region. This is clearly
appreciated in figures \ref{selfSimNeedle} and \ref{vrvzNeedle}
where the collapse onto each other of the normalized time
evolutions of the axial and radial velocity components evaluated at
the free surface ($u$ and $v$) is also a bit deteriorated when
compared with the case depicted in figure \ref{vrvzscaled}.
Nevertheless, the two main prerequisites for the model to be
presented in the following sections are also satisfied in this
case: firstly, the acceleration and ballistic regions are clearly
differentiated in figure \ref{vrvzNeedle} and, secondly, the
normalization of the interfacial velocities with $q_b/r_b$ leads
to a reasonable collapse onto a single master curve (see figure
\ref{vrvzNeedle}).

%
%
%
%

\subsection{Jet breakup}\label{sec:breakup}

The growth of capillary perturbations in a cylindrical liquid jet,
firstly quantified by \cite{Rayleigh} [see also
\cite{EggersVillermaux_RepProgPhys_2008}], is based on the
assumption that fluid particles conserve, to first order, their
longitudinal velocity $U$. Rayleigh's analysis shows that, moving
in a frame of reference with the jet velocity $U$ (which in his
case is constant $U\neq U(z,t)$), and no matter how large the
Weber number is, the jet breaks due to the growth of capillary
perturbations of wavelengths larger than the jet perimeter. The
characteristic time needed for such perturbations to disrupt the
jet into drops is the capillary time, $\sim
(\rho\,R^3/\sigma)^{1/2}$, with $R$ the jet radius. Therefore the
jet breakup length, $L_b$, is such that $L_b/R\propto
(\rho\,U^2\,R/\sigma)^{1/2}$ if aerodynamic effects are absent
[\cite{Sterling,JFM05Air}]. Notice that the study of jet breakup
in our case is somewhat related to that considered by Rayleigh
since the fluid particles conserve their velocities, in a first
approach, along the ballistic region of the jet.

Similarly to the case considered by \cite{Rayleigh}, the study of
the capillary breakup of stretched jets will be divided in two:
the calculation of the \emph{basic flow}, which is free of
capillary effects and the analysis of capillary waves propagating
and growing in amplitude at the jet tip region. Viscous effects will be neglected in the analysis.

\subsubsection{Unperturbed flow}\label{unperturbed}

If $\mathrm{We}_0\gg 1$, the time evolution of both the jet radius and the
liquid velocities in the ballistic region can be calculated
neglecting surface tension forces and making use of the
slenderness of the jet. However, differently to the case
considered by \cite{Rayleigh}, in which the jet radius $r_j=1$ and
$u=constant$, here $u$ and $r_j$ are functions of $z$ and $t$. In
effect, if the fluid is assumed to follow a purely vertical motion
inside the ballistic region, the couple of equations that
determine $u$ and $r_j$ are the momentum equation
(\ref{Momentum1D}), which can be also written as
\begin{equation}
\frac{Du}{Dt}=0\rightarrow \frac{\partial u}{\partial
t}+u\frac{\partial u}{\partial z}=0\, , \label{Momentum}
\end{equation}
and the unidirectional version of the continuity equation, namely,
\begin{equation}
\frac{\partial r_j^2}{\partial t}+\frac{\partial
(u\,r_j^2)}{\partial z}=0\rightarrow
\frac{D\,\ln\,r_j^2}{D\,t}=-\frac{\partial\,u}{\partial\,z}\,
\label{Continuity},
\end{equation}
where $D/Dt\equiv \partial/\partial t+u\partial/\partial z$
indicates again the material derivative. From equations
(\ref{Momentum})-(\ref{Continuity}), $u$, $r_j$ and $z_j$ - the
height at which the jet radius is $r_j$ - are completely
determined if the relevant quantities at the beginning of the
ballistic region ($r_0$, $z_0$, and the velocity $u_0$) are known
functions of time. Indeed, equation~(\ref{Momentum}) expresses
that fluid particles conserve the vertical velocity they possess
at the beginning of the ballistic region. Consequently, a particle
ejected from the acceleration into the ballistic region at time
$\tau<t$ will, at time $t$, have attained a height
\begin{equation}
z_j(t)=z_0(\tau) + \left(t-\tau\right)\,u_0(\tau).\label{Z_Jet}
\end{equation}
To obtain the corresponding jet radius $r_j$, equation
(\ref{Continuity}) can be readily integrated to give
\begin{equation}
\begin{split}
r_j^2(z=z_j,t)=r^2_o(\tau)&\frac{u_o(\tau)-d\,z_o/d\tau}{u_o(\tau)-d\,z_o/d\tau-d\,u_o(\tau)/d\tau(t-\tau)}\,
. \label{Radius}
\end{split}
\end{equation}
Introducing the definition of the strain rate at the beginning of
the ballistic region
\begin{equation}
s_o(\tau)=\frac{\partial\,u}{\partial\,z}(z=z_0)=-\frac{\dot{u}_o(\tau)}{u_o(\tau)-\dot{z}_o(\tau)}\,\label{So}
\end{equation}
allows us to rewrite equation (\ref{Radius}) in a more compact
form as
\begin{equation}
r_j^2(z_j,t)=\frac{r^2_o(\tau)}{1+(t-\tau)s_o}. \label{Radius2}
\end{equation}
Note that Rayleigh's original analysis, $u=constant$ and $r_j=1$
(cylindrical jet) may be recovered from equations
(\ref{Momentum})-(\ref{Continuity}) by setting $s_0=0$ and $u_0$,
$z_0$ and $r_0$ constants in time. However, in our case,
$\dot{u}_0<0$ and $u_0=B q_b/r_b>\dot{z}_b\simeq\dot{z}_0$ and,
therefore, by virtue of equation (\ref{So}), these conditions imply
$s_o>0$; consequently, from equation (\ref{Radius2}), the jet is not
cylindrical since it stretches downstream.

\begin{figure}
     \centerline{\includegraphics[width=0.5\textwidth]{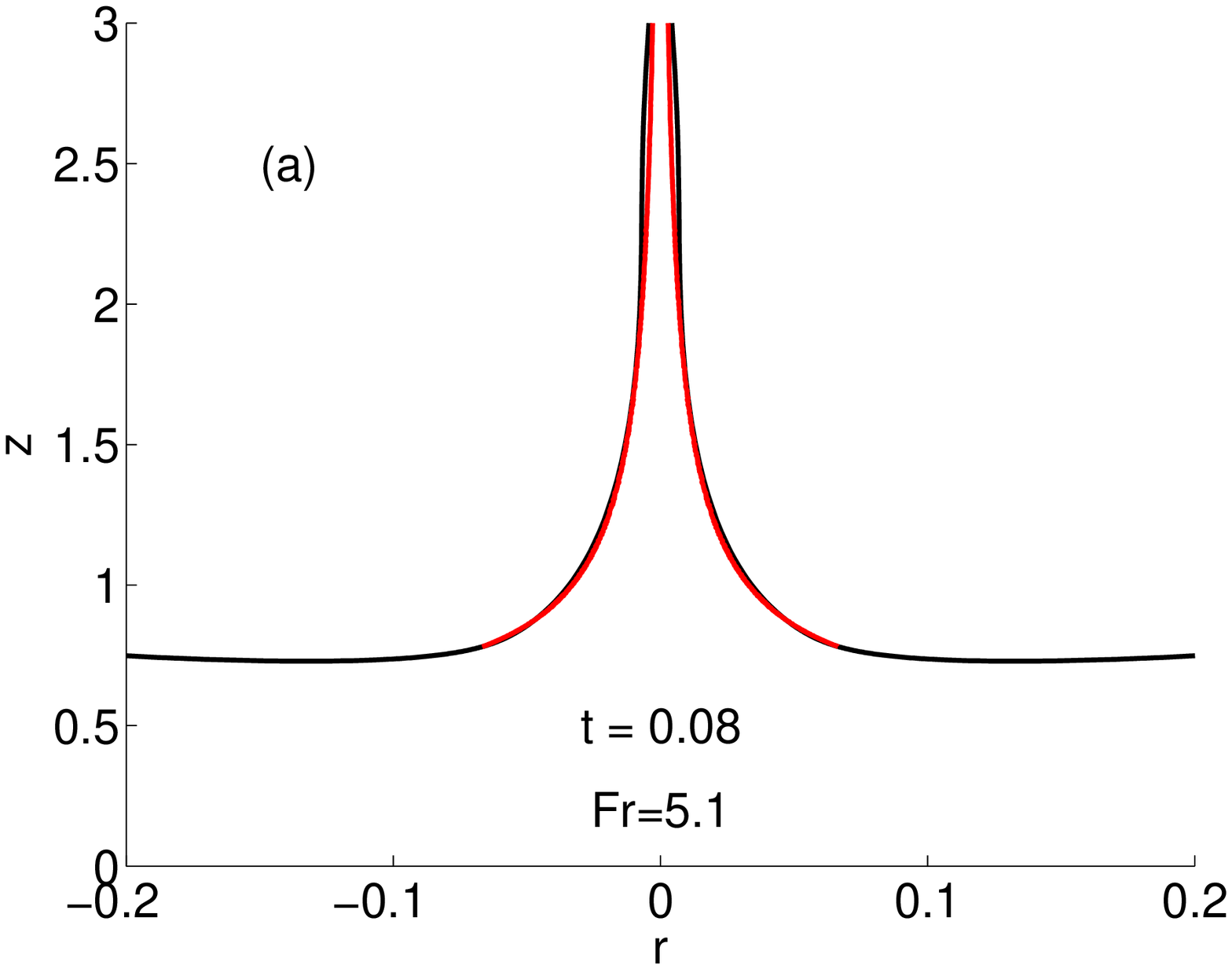}
                 \includegraphics[width=0.5\textwidth]{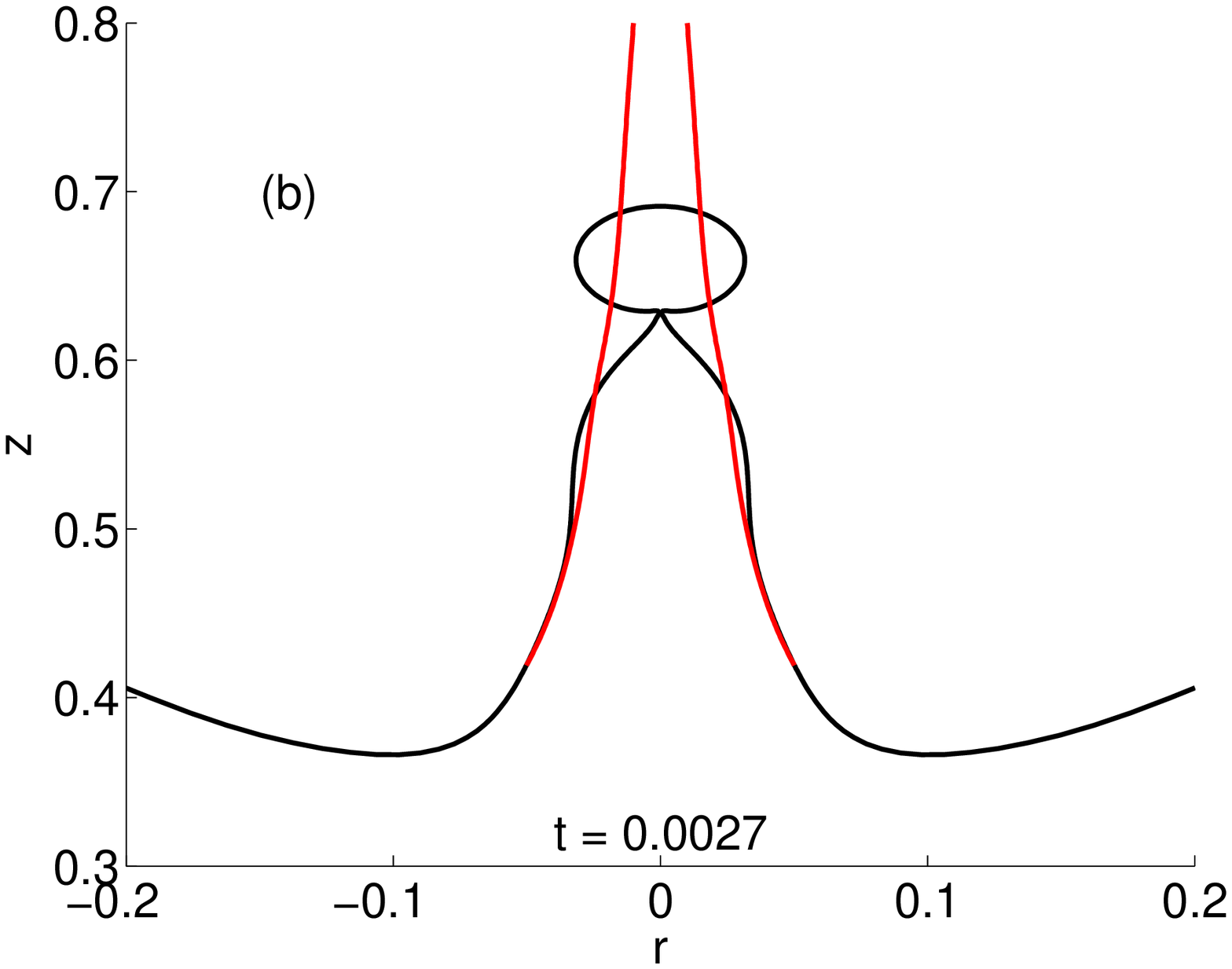}     }
  \caption{Comparison between the numerical jet shape and that obtained from equation (\ref{Radius}) for the disc impact at Fr=5.1 (a) and the jet ejected from the underwater nozzle (b). The black line is the simulation and the red line is the analytical model. The input values of $r_0(t)$, $z_0(t)$ and $u_0(t)$ for the jet stretching model are taken from the simulations. Note that, since surface tension is not included in this model, the tip of the jet requires a separate treatment as described in section \ref{sec:breakupAnalysis}}\label{Jetshape}
\end{figure}

Now, in order to obtain the complete jet shape at time $t$, we
vary $\tau$ between 0 and $t$ and use equations (\ref{Z_Jet}) and
(\ref{Radius}) to compute the corresponding vertical and radial
coordinates of the jet. Note that, clearly, the particle ejected
at $\tau=0$ will end up forming the tip of the jet. The comparison
between the numerical results and those obtained from the
integration of (\ref{Momentum})-(\ref{Continuity}), with the
values of $u_o(\tau)$, $z_o(\tau)$ and $r_o(\tau)$ taken from the
numerical simulations, is depicted in figure \ref{Jetshape}. The
excellent agreement between numerics and the model validates the
approach of considering that fluid particles conserve their axial
velocities within the ballistic region. It should be pointed out,
however, that equations (\ref{Momentum})-(\ref{Continuity}) need
to be corrected at the tip of the jet, where surface tension
effects need to be retained.

\subsubsection{Growth of capillary disturbances}\label{sec:breakupAnalysis}

The linear stability analysis for the type of velocity field given
in section \ref{unperturbed} was firstly accomplished by \cite{Frankel}, who
recovered Rayleigh's original result in the limit of $s_0=0$. It is
our purpose here to extend the analysis on the breakup of stretched
jets of \cite{Frankel} to account for non linear effects and also
for the influence of the tip. It is worthy to mention that, in our
numerical approach, the wavelength of fastest growth rate is
naturally selected by the local flow around the jet tip and,
therefore, a linear stability analysis of the type reported in
\cite{Frankel}, is avoided.

As a first step, the dimensional counterparts of $r_0(\tau)$
[$R_0(\tau)$] and $u_0(\tau)$ [$U_0(\tau)$] are chosen as the
characteristic scales for lengths and velocities, respectively.
Consequently, dimensional analysis indicates that the evolution of
capillary perturbations in the ballistic region for $t>\tau$ will
solely depend on the dimensionless parameters $\mathrm{We}_0$ and
$s_{0,local}=S_0(\tau)\,R_0/U_0=s_0\,r_0/u_0$.

The values of $\mathrm{We}_0$ and $s_{0,local}$ depend
non-trivially on the dimensionless parameters controlling the two
different physical situations analyzed here. Consequently, in
order to study systematically the jet breakup process as a
function of $s_{0,local}$ and $\mathrm{We}_0$ we employ the third
type of simulations of the axial strain type described in section \ref{sec:description} and illustrated in
figure \ref{Geometry}. The real jet breakup process can then be
reproduced provided that the values of the Weber number and the
strain rate at the nozzle exit coincide with those at the
beginning of the ballistic region i.e,
$\mathrm{We}_0=\mathrm{We}_N=\rho\,U^2_N(0)\,R_N/\sigma$ and
$s_{0,local}=\alpha$.

Note that, since the values of the Weber number based on $U_N$ and
$\rho_g$ are always such that
$\mathrm{We}_g=\rho_g\,U^2_N\,R_N/\sigma\ll 1$, the gas dynamics
can be neglected and the only two relevant dimensionless
parameters characterizing the axial strain system of figure \ref{Geometry} are $\mathrm{We}_N$ and $\alpha$.

\begin{figure}
  \centerline{\includegraphics[width=0.75\textwidth]{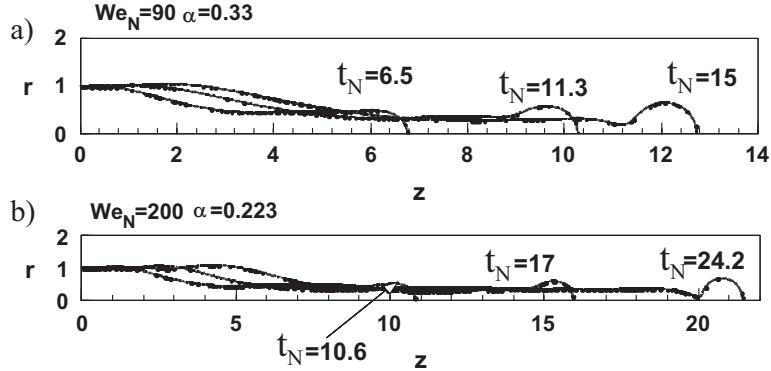}
              }
  \caption{Time evolution of jets calculated for two different values of the pair of variables $(\mathrm{We}_N,\alpha)$ but the \emph{same} value of the product $\mathrm{We}_N\alpha^2$.}\label{formadim}
\end{figure}

\begin{figure}
  \centerline{\includegraphics[width=0.75\textwidth]{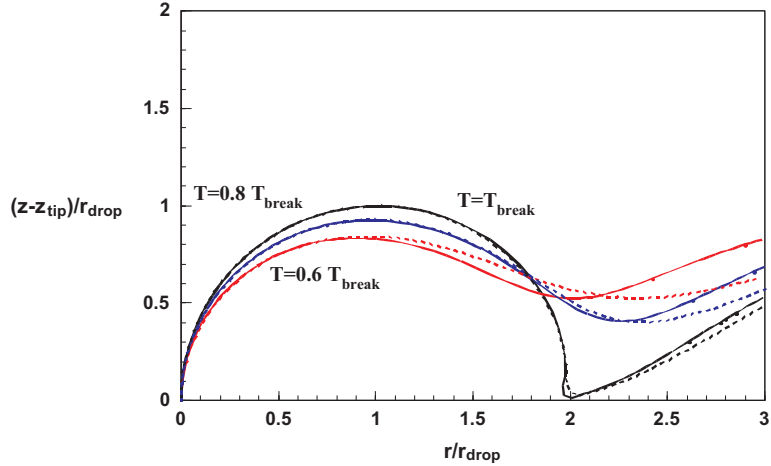} }
  \caption{Jet shapes of figure \ref{surfaceProfilesNeedle} (in continuous lines) at three different instants of time ($0.6T_{break},\,0.8T_{break}$ and $T_{break}$, with $T_{break}$ the breakup time) compared to those obtained from the type of simulations depicted in figure \ref{formadim} (dashed lines) for the same values of $T$. The values of the parameters have been set to $\mathrm{We}_N=25$ and $\alpha=0.4$ which are the initial values of the local Weber number and the dimensionless axial strain rate of the simulations depicted in figure \ref{Weo}. Note that distances are rescaled using the final radius of the drop $R_{drop}$ as the characteristic length scale and that $z_{tip}$ denotes the axial coordinate of the tip of the jet.}\label{Compara}
\end{figure}

The numerical results depicted in figure \ref{formadim} show a
slender liquid thread which breaks many diameters downstream the
nozzle exit. Moreover, it can be observed that the effect of
increasing the Weber number is to increase the breakup time and
the breakup length. Figure \ref{Compara} shows a comparison
between the shapes of the jets formed after bubble collapse
depicted in figure \ref{selfSimNeedle} and those obtained from the
simulations of the type illustrated in figure \ref{formadim} with
$\mathrm{We}_N=\mathrm{We}_0(\tau=0)$ and
$\alpha=s_{0,local}(\tau=0)$. The excellent agreement between both
type of numerical results corroborates the fact that tip breakup
of Worthington jets can be reproduced by means of the simulations
considered in this section if the values of the parameter
$\mathrm{We}_N$ and $\alpha$ coincide with the initial values of
$\mathrm{We}_0$ and $s_{0,local}$.

However, the numerical code used in this section is unstable for
$\mathrm{We}_N\gtrsim O(10^3)$. Consequently, the third type of
simulations cannot reproduce, at first sight, the breakup of jets
ejected by an impacting disc since, in this case,
$\mathrm{We}_0\gtrsim 10^3$ as depicted in figure \ref{Welocal}.
Thus, is it nevertheless possible to describe the breakup process
of jets with such high values of $\mathrm{We}_0$ using the
numerical simulations of the axial strain type illustrated in figure \ref{Geometry}? The answer to this question
is affirmative if we realize that, in a frame of reference moving
at the tip velocity, the parametrical dependence on the velocity
$U_0(\tau=0)$ disappears. Consequently, since both the local flow
field and the jet radius still depends in this frame of reference
on $S_0(\tau=0)$ (see equation (\ref{So})), dimensional analysis
indicates that jet breakup can be described in terms of the
dimensionless variables $T\,S_o$ (or, analogously, $t_N\alpha$)
and
$\mathrm{We}_S=\rho\,S^2_0\,R^3_0/\sigma=\mathrm{We}\,r^3_0\,s^2_0=\mathrm{We}_0\,s^2_{o,local}$
(or, analogously, $\mathrm{We}_N\alpha^2$).

\begin{figure}
     \centerline{\includegraphics[width=1\textwidth]{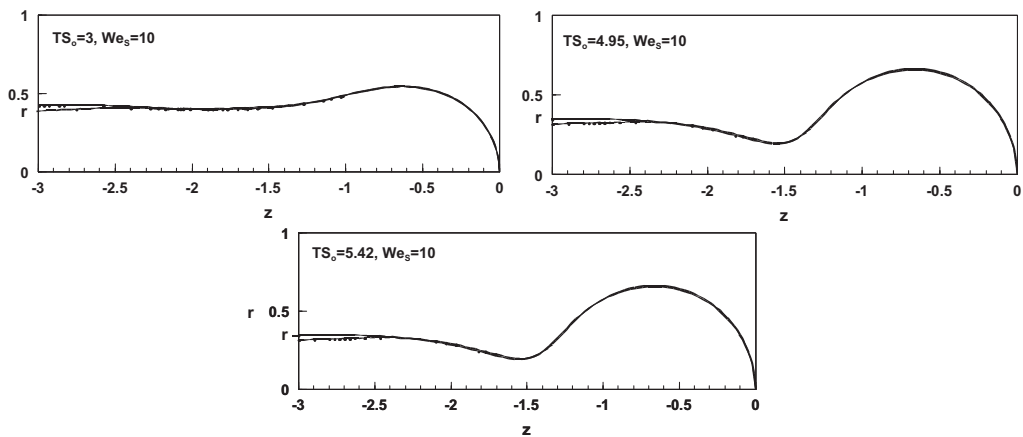}         }
  \caption{Translated jet shapes corresponding to the conditions $\mathrm{We}_N=90$, $\alpha=0.33$ and $\mathrm{We}_N=200$ and $\alpha=0.223$, depicted in figures \ref{formadim} (a) and (b), respectively. Since $\mathrm{We}_S=\mathrm{We}_N\alpha^2=10$ in both cases, the time evolution near the tip region is identical in the normalized temporal variable $TS_0$.}\label{formaadim}
\end{figure}

\begin{figure}
     \centerline{\includegraphics[width=1\textwidth]{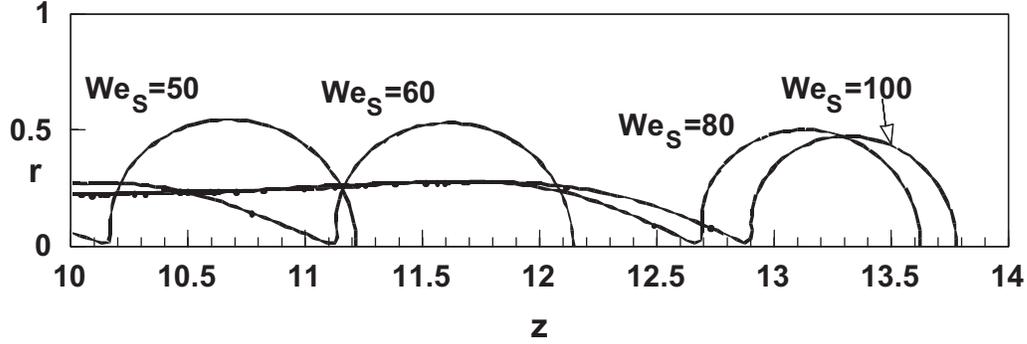}         }
  \caption{Tip region at the instant of breakup for different values of $\mathrm{We}_S$. Observe that the drops generated are nearly spherical and that their volume decreases for increasing values of $\mathrm{We}_S$.}\label{tipbreak}
\end{figure}

\begin{figure}
  \centerline{\includegraphics[width=0.5\textwidth]{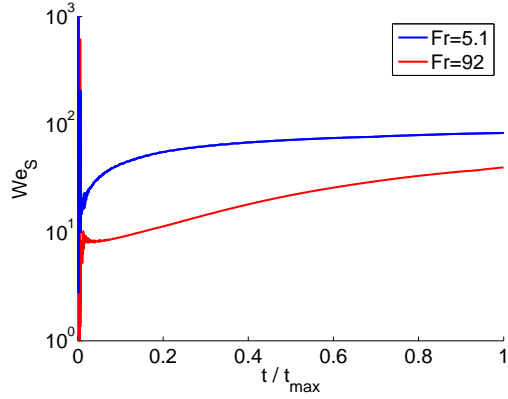}}
 \caption{Time evolution of the normalized Weber number $\mathrm{We}_S=\mathrm{We}_0\,s^2_{0,local}$ for the impacting disc. As indicated in figure \ref{Welocal}, $t_\mathrm{max}$ is the time when the downward
jet hits the disc and the simulation stops.}\label{Wes}
\end{figure}

\begin{figure}
     \centerline{\includegraphics[width=1\textwidth]{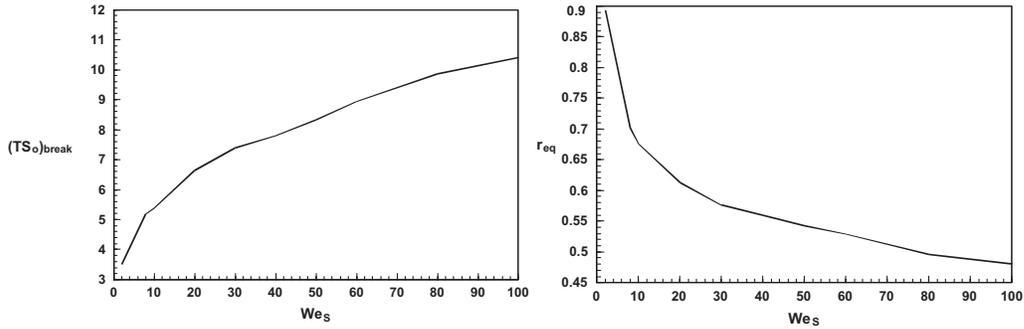}         }
 \caption{Numerical results of the dimensionless breakup time, $(TS_0)_{break}$ (a), and the dimensionless equivalent radius, $r_{eq}$ (b), with $4/3\pi\,r^3_{eq}\,R^3_0$ the volume of the drops formed.}\label{tbreakvol}
 \end{figure}

To check this, the results depicted in figure \ref{formadim},
which correspond to different values of $\mathrm{We}_N$ and
$\alpha$ but to the same value of $\mathrm{We}_N\alpha^2$, are
represented in figure \ref{formaadim}. Remarkably, the different
jet shapes superimpose onto each other for the same values of the
dimensionless time $T\,S_0$, what indicates that the breakup
process depends solely on $\mathrm{We}_S$ (or, equivalently, on
$\mathrm{We}_N\alpha^2$) and on the dimensionless time $T\,S_0$
(or, equivalently, on $t_N\alpha$) for sufficiently large values
of $\mathrm{We}_0$. Figure \ref{tipbreak} illustrates that the
volume of the nearly spherical drops formed at breakup, decreases
for increasing values of $\mathrm{We}_S$. In figure \ref{tipbreak}
note also that the range of values of $\mathrm{We}_S$ investigated
is realistic even for the impacting disc, as depicted in
\ref{Wes}. Consequently, even though $\mathrm{We}_0$ in some
situations such as the disc may be very high, the important
parameter which is $\mathrm{We}_S=\mathrm{We}_0s_{0,local}^2$ can
be matched to the simulations in this section. The dimensionless
breakup time and the dimensionless size of the drops in figure
\ref{tbreakvol} behave as $r_{drop}\propto \mathrm{We}^{-1/7}_S$
and $(TS_o)_{break}\propto \mathrm{We}^{2/7}_S$, respectively. A
detailed corresponding theory will be the subject of a forthcoming
contribution [\cite{Stretchedjets}]. We emphasize that figure
\ref{tbreakvol} describes a universal relation for the breakup of
Worthington jets at high Weber numbers which allows one to obtain
the breakup time and volume of the first ejected droplet knowing
merely the value of $\mathrm{We}_S$ defined at the beginning of
jet formation.

Finally, note that, in order for $\mathrm{We}_N<O(10^3)$, the
computations have been performed choosing $1\ll \mathrm{We}_N\ll
\mathrm{We}_0$ and, correspondingly, $\alpha>s_{0,local}$. The
condition $\mathrm{We}_N\gg 1$ is essential since, if
$\mathrm{We}_N$ was not sufficiently large, the jet breakup
process of real Worthington jets would depend on the liquid
velocity $U_N(0)$ and, thus, on $\mathrm{We}_N$ and $\alpha$
separately.


%
%
%
%
%
%
%
%

\section{Modeling the jet ejection and breakup processes}\label{sec:model}

Here we aim to develop a model to explain, in simple terms, the
jet ejection and breakup processes. Our model will be based on the
main conclusions of the previous section which are: (i) both $r_b$
and $z_b$ are \emph{local} quantities which, therefore, do not
depend on the large scales of the flow, (ii) the velocity field
within the jet can be characterized solely in terms of the sink
strength intensity at pinch-off, $q_c(z)$ and (iii) the flow field
within the jet can be divided in three parts: the
\emph{acceleration region}, the \emph{ballistic region} and the
\emph{jet tip region}.

This section is structured as follows: in subsection
\ref{sec:baseModel} $r_b(t)$ and $z_b(t)$ are calculated in terms
of \emph{only} $q_c(z)$ using the theory developed in
\cite{PRL09}. Then, in section \ref{sec:modelBreakup} the axial
velocity and the jet shape within the ballistic portion of the jet
are calculated through equations
(\ref{Momentum})-(\ref{Continuity}) using, as initial conditions,
$r_0=0.5\,r_b$, $u_0=B(\mathrm{Fr})\,q_b/r_b$ and
$z_0(\tau)=z_b(\tau)+0.5r_b$.

\subsection{Reviewing the model for $r_b(t)$ and $z_b(t)$}\label{sec:baseModel}

In this subsection we will very briefly review our model for jet
formation as presented in \cite{PRL09} and show its applications
to predict the flow fields as well as the dynamics of the jet base
for the impacting disc at Fr=5.1 and Fr=92 and the Worthington
jets created after bubble pinch-off from an underwater nozzle.

\begin{figure}
  \centerline{\includegraphics[width=0.5\textwidth]{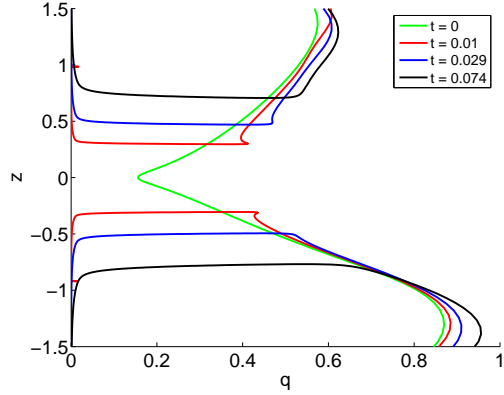} }
  \caption{The sink distribution $q_c(z)$ at the moment of pinch-off (green) for the impacting disc at Fr=5.1 is the essential ingredient to our jet formation model. The sink distributions at later times (red, blue and black curves) are almost unchanged with respect to the sink distribution at pinch-off, confirming our model assumption that $q_c(z)$ is valid even for jet formation when two additional effects are accounted for: the accumulation of sinks around the base and the hole between the up- and the downward jet.}\label{sinksJet}
\end{figure}

The starting point of our model is the description of the cavity
collapse using a line of sinks on the axis of symmetry as depicted
in figure \ref{qc}. After pinch-off most of this distribution
remains intact with two notable exceptions: a hole is created
between the bases of the up- and downward jet and sinks accumulate
around the jet base [\cite{PRL09}]. These effects are illustrated
in figure \ref{sinksJet}. Based on this observation we derived in
\cite{PRL09} an analytical expression for the flow potential
$\phi$ at an arbitrary point in the outer region (note that by
construction the model is not valid inside the jet itself):
\begin{equation}
2\phi = \underbrace{-q_b\int_{-\infty}^{\infty}
\frac{dz'}{\sqrt{r^2+(z-z')^2}}}_{\mathrm{collapsing\;cavity}} +
\underbrace{q_b\int_{-z_b}^{z_b} \frac{dz'}{\sqrt{ r^2 +
(z-z')^2}}}_{\mathrm{hole}} +
\underbrace{\frac{C\,q_b\,r_b}{\sqrt{r^2 +
(z-(z_b+C_{\mathrm{sink}}r_b))^2}}}_{\mathrm{point\;sink}}
\label{eqn:potentialJet}
\end{equation}
with the order one constants $C$ and $C_{\mathrm{sink}}$ chosen appropriately.

\begin{figure}
\includegraphics[width=.45\textwidth]{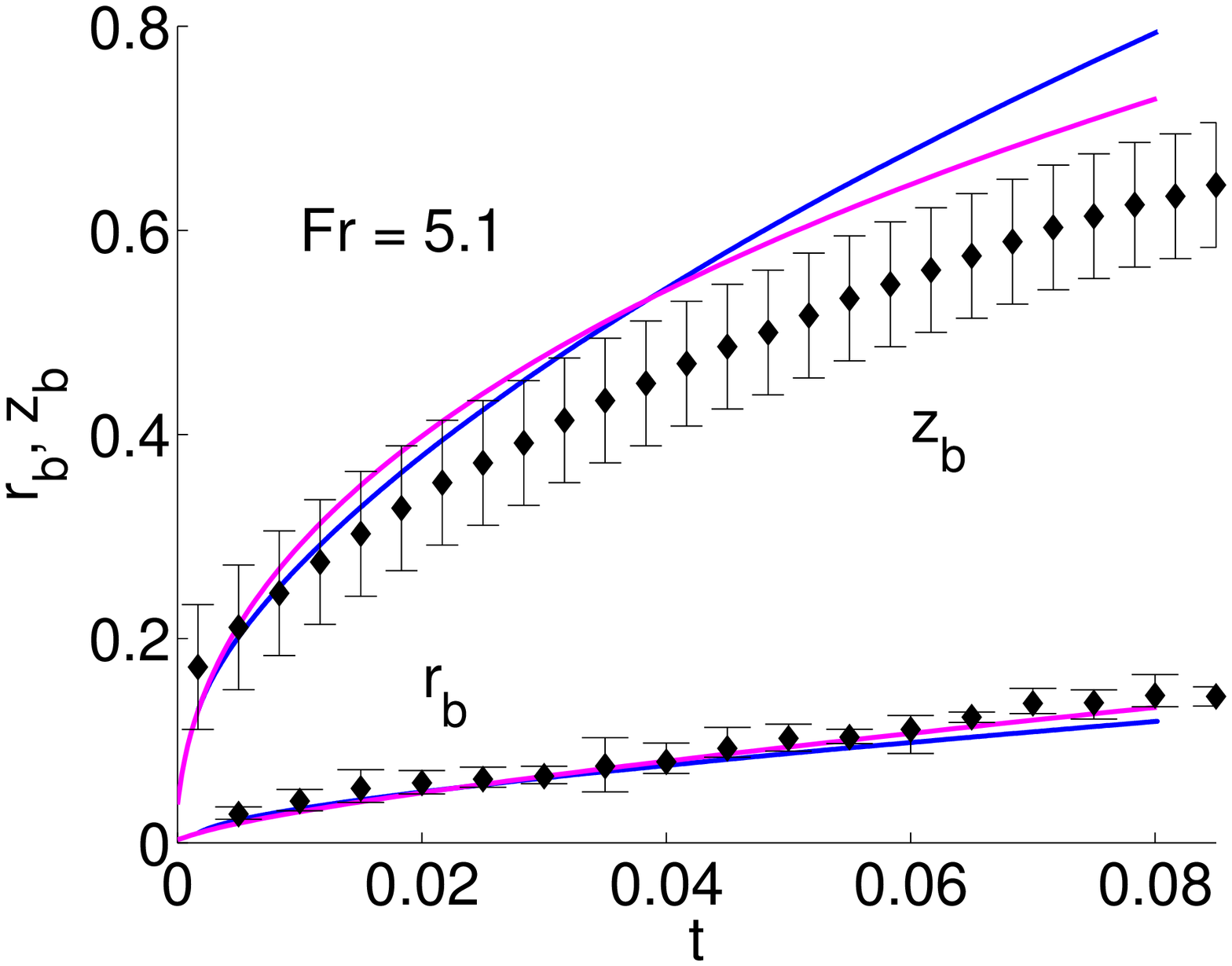}
\includegraphics[width=.45\textwidth]{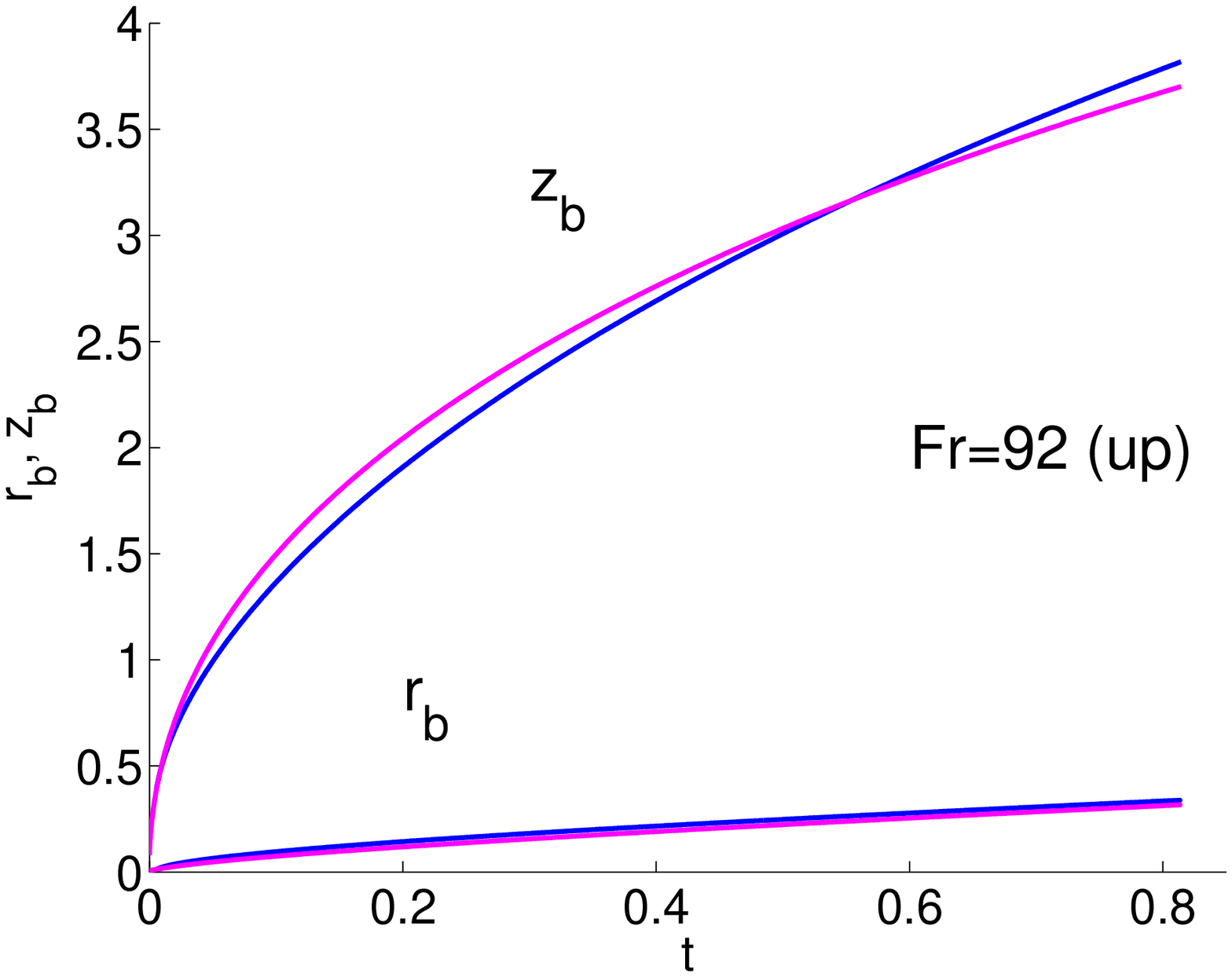}\\
\includegraphics[width=.45\textwidth]{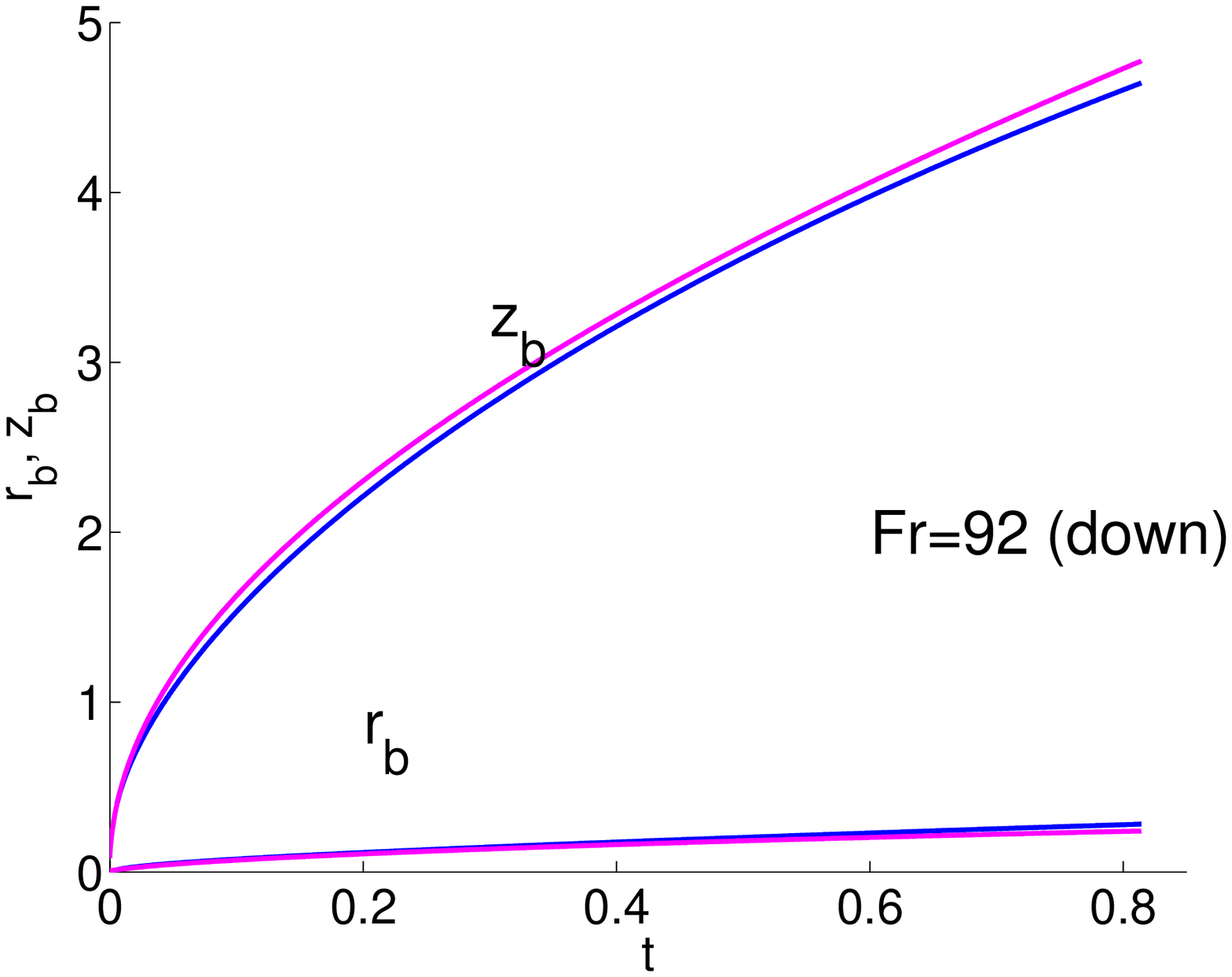}
\includegraphics[width=.45\textwidth]{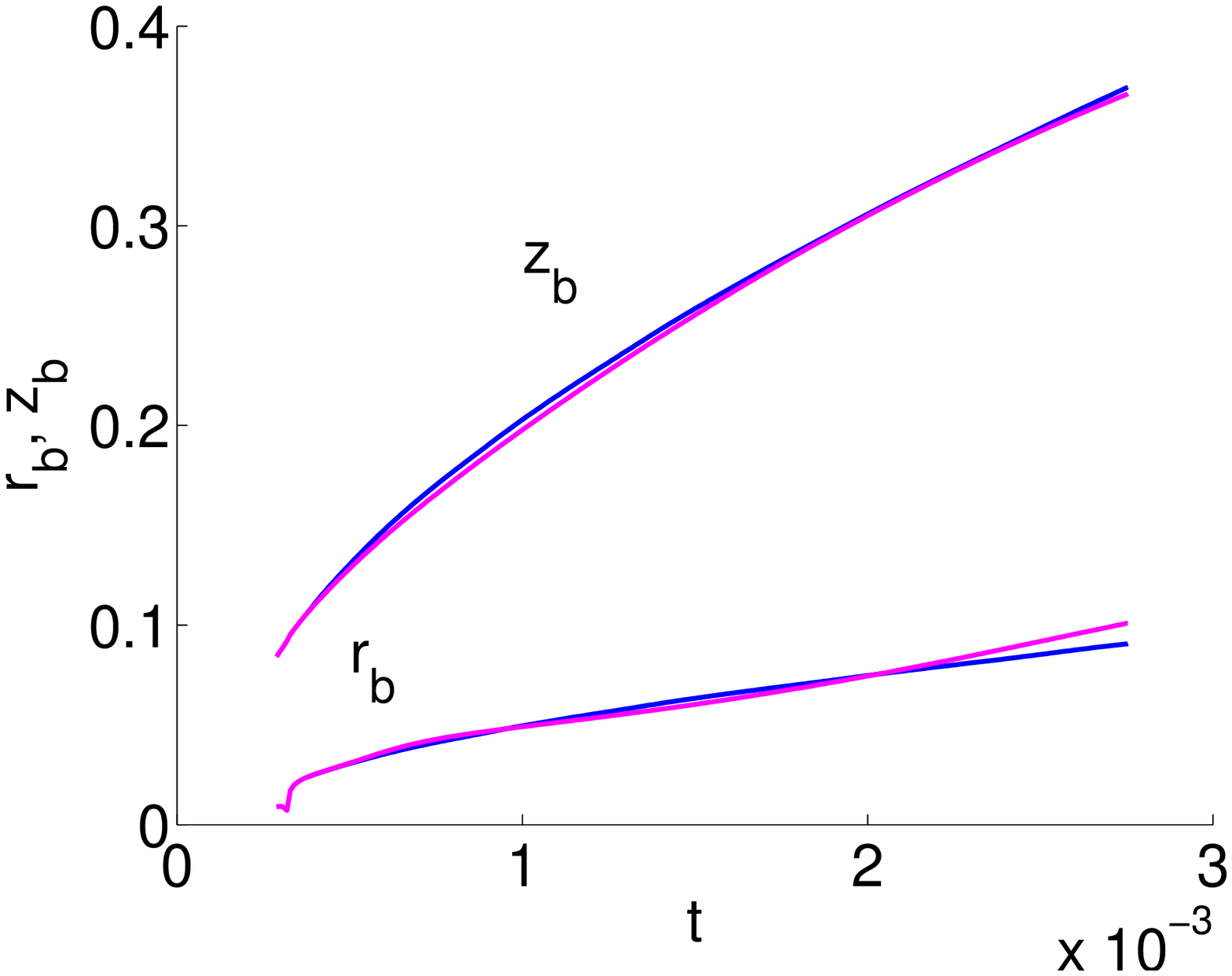}
\caption{Time evolution of the jet base radial and axial
positions, $r_b$ and $z_b$ respectively, taken from the simulation
(magenta lines) and the analytical model (blue lines). (a)
Impacting disc with Fr=5.1 for the upward jet (here black diamonds
represent experimental data) with $C=4.55$ and
$C_{\mathrm{sink}}=0.63$. (b) Impacting disc with Fr=92 (upward
jet) with $C=7.8$ and $C_{\mathrm{sink}}=0.63$. (c) Impacting disc
with Fr=92 (downward jet) with $C=6.66$ and
$C_{\mathrm{sink}}=0.55$. (d) Upwards jet from the underwater
nozzle with $C=4.9$ and $C_{\mathrm{sink}}=0.76$.}\label{JetBase}
\end{figure}

As shown in \cite{PRL09} this model can be used to predict the
temporal evolution of the jet base, i.e. the widening and upwards
motion of the jet base. Here we will restrict ourselves to show
the result of this procedure for the different systems studied in
this work, which are depicted in figure \ref{JetBase}.

\begin{figure}
     \centerline{\includegraphics[width=0.5\textwidth]{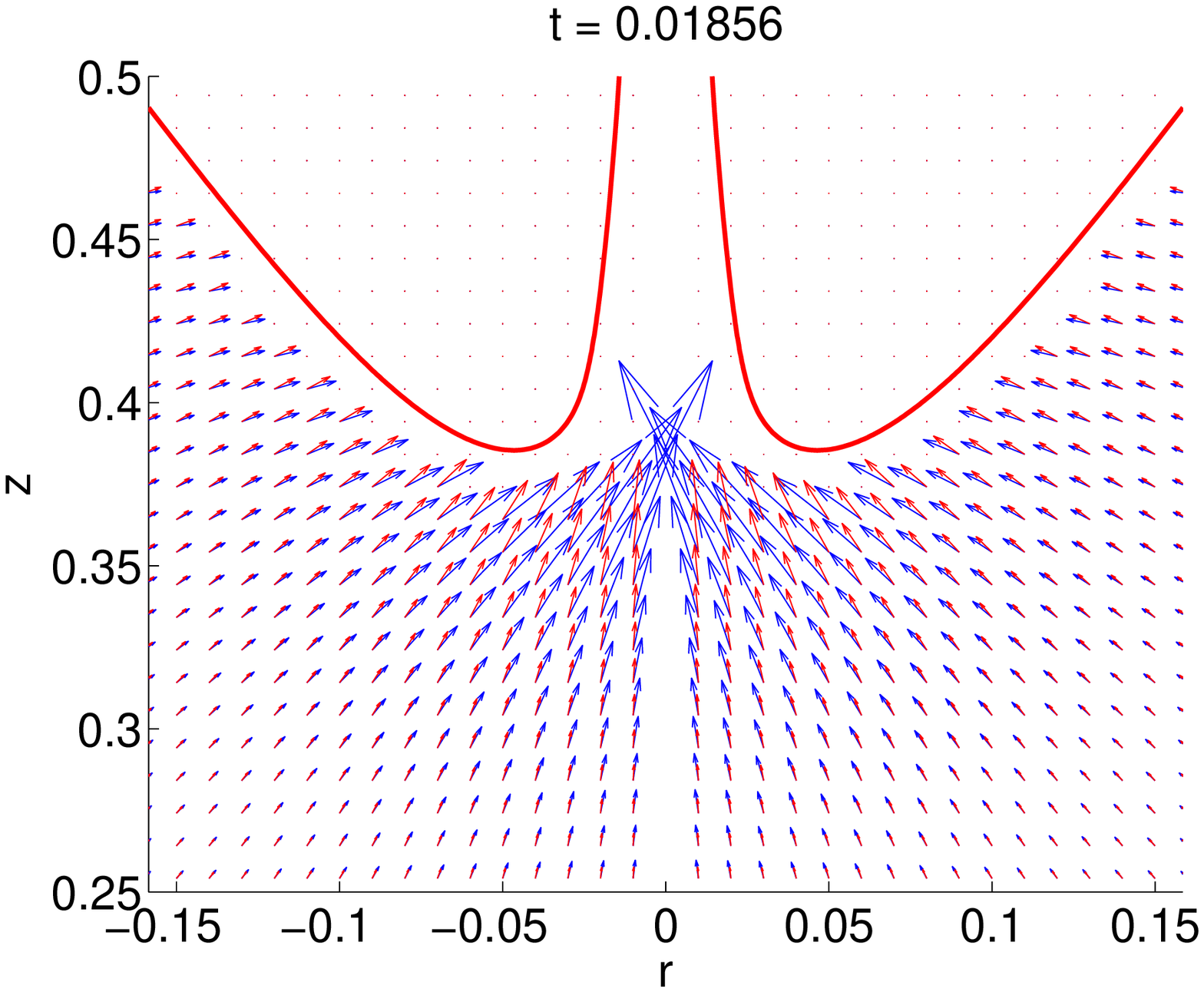}
                 \includegraphics[width=0.5\textwidth]{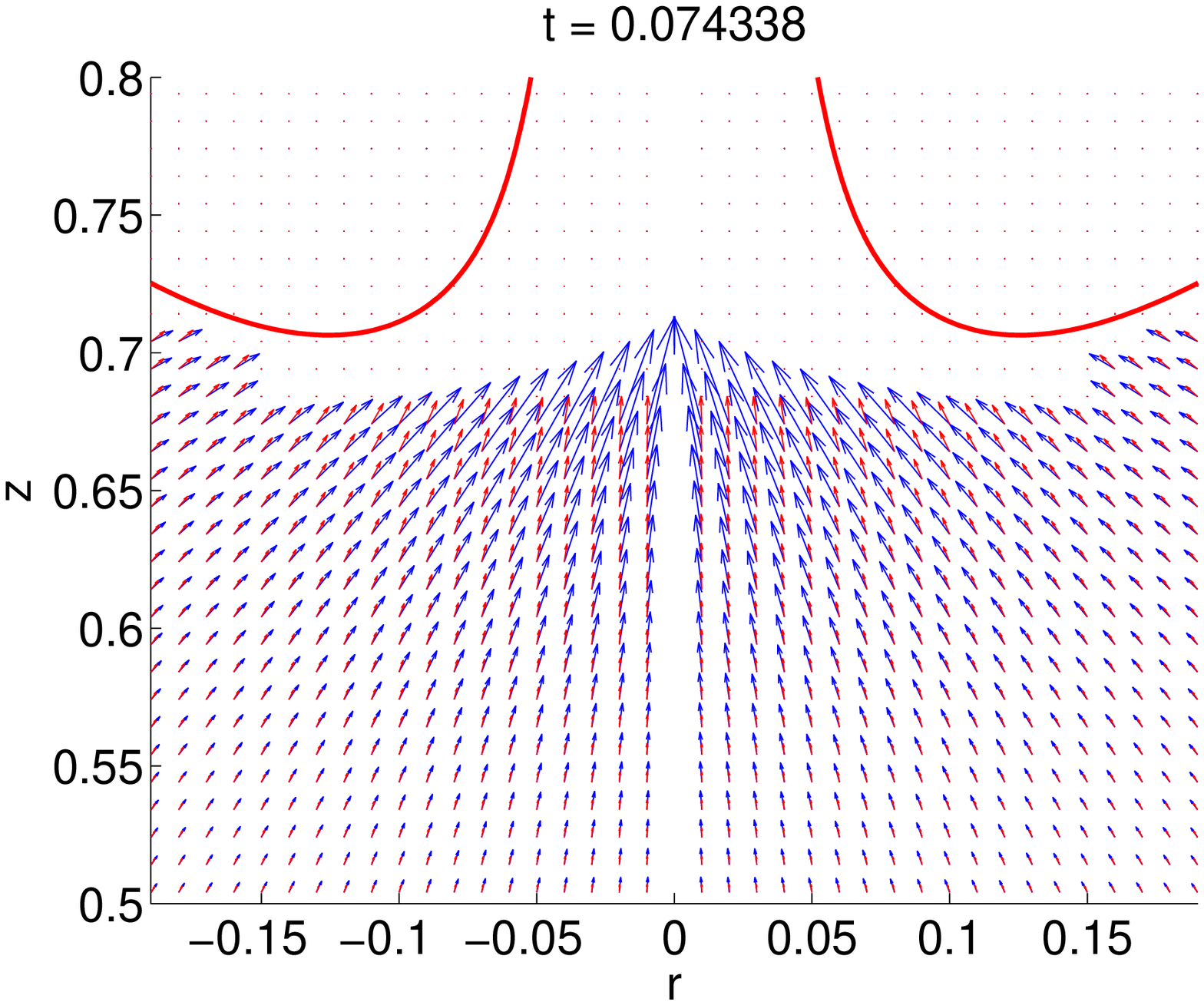}     }
  \caption{The flow field obtained from the model with constants $C=4.55$ and $C_{\mathrm{sink}}=0.63$ for the disc impacting at Fr=5.1 (blue arrows) shows very good agreement with the numerically calculated flow field (red arrows). The region inside the jet and very close around the base is excluded since the model is not perfectly reliable there (due to the assumption of the observation point far from the base, see \cite{PRL09}).} \label{FlowField}
\end{figure}

In fact, as shown in figure \ref{FlowField}, equation
(\ref{eqn:potentialJet}) can also be used to predict the entire
flow field in the outer region. Figures
\ref{JetBase}-\ref{FlowField} illustrate the rather good agreement
between theory and numerics, which we find in all cases studied.

\subsection{Modeling the jet breakup and drop ejection processes} \label{sec:modelBreakup}

We will now take the model of the previous section one step
further by combining its results with the analysis described in
section \ref{unperturbed}. This will allow us to predict not only
the flow field in the outer region, but also the flow inside the
jet and thus the jet shape as a function of time.

\begin{figure}
     \centerline{\includegraphics[width=0.5\textwidth]{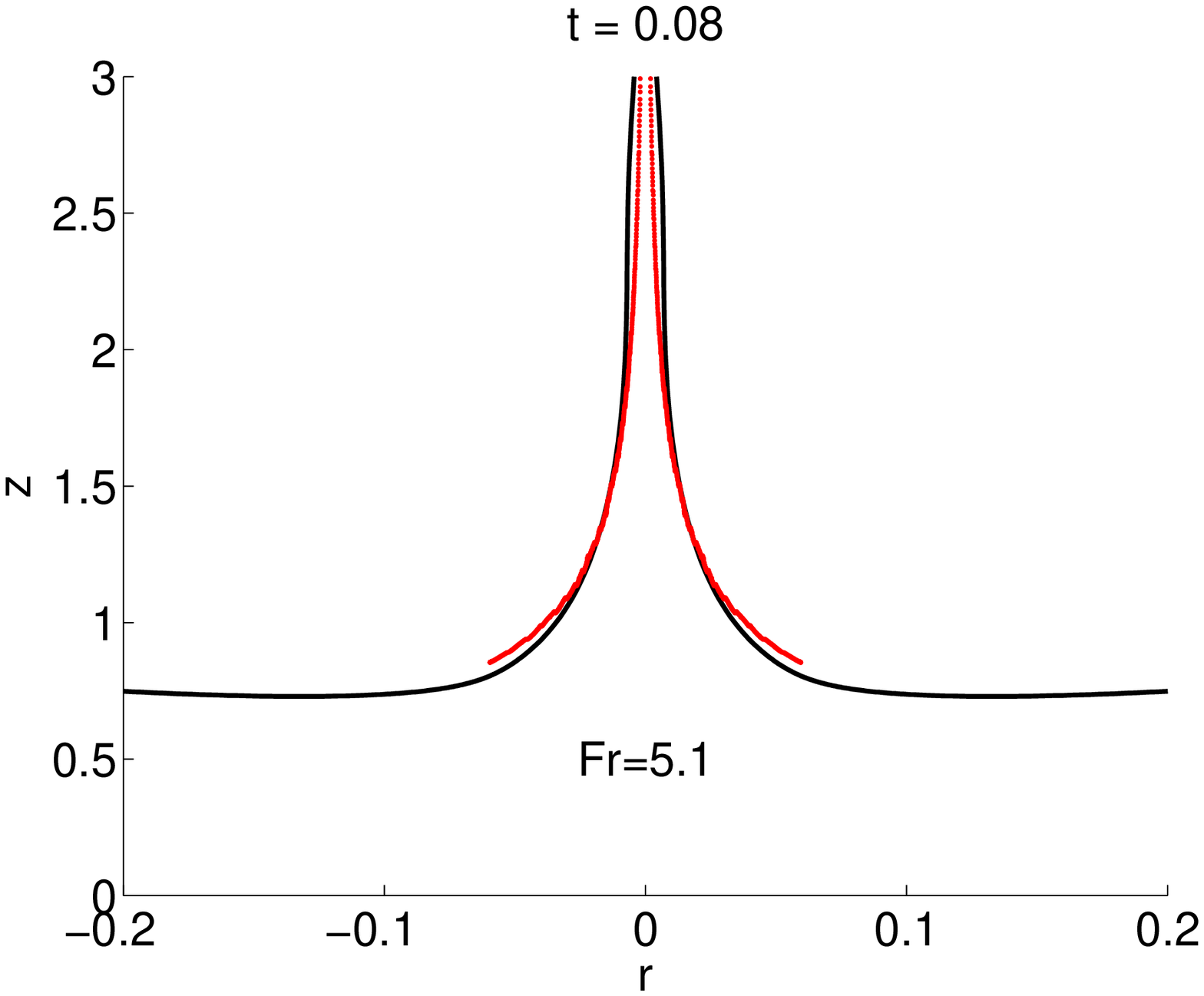}
                 \includegraphics[width=0.5\textwidth]{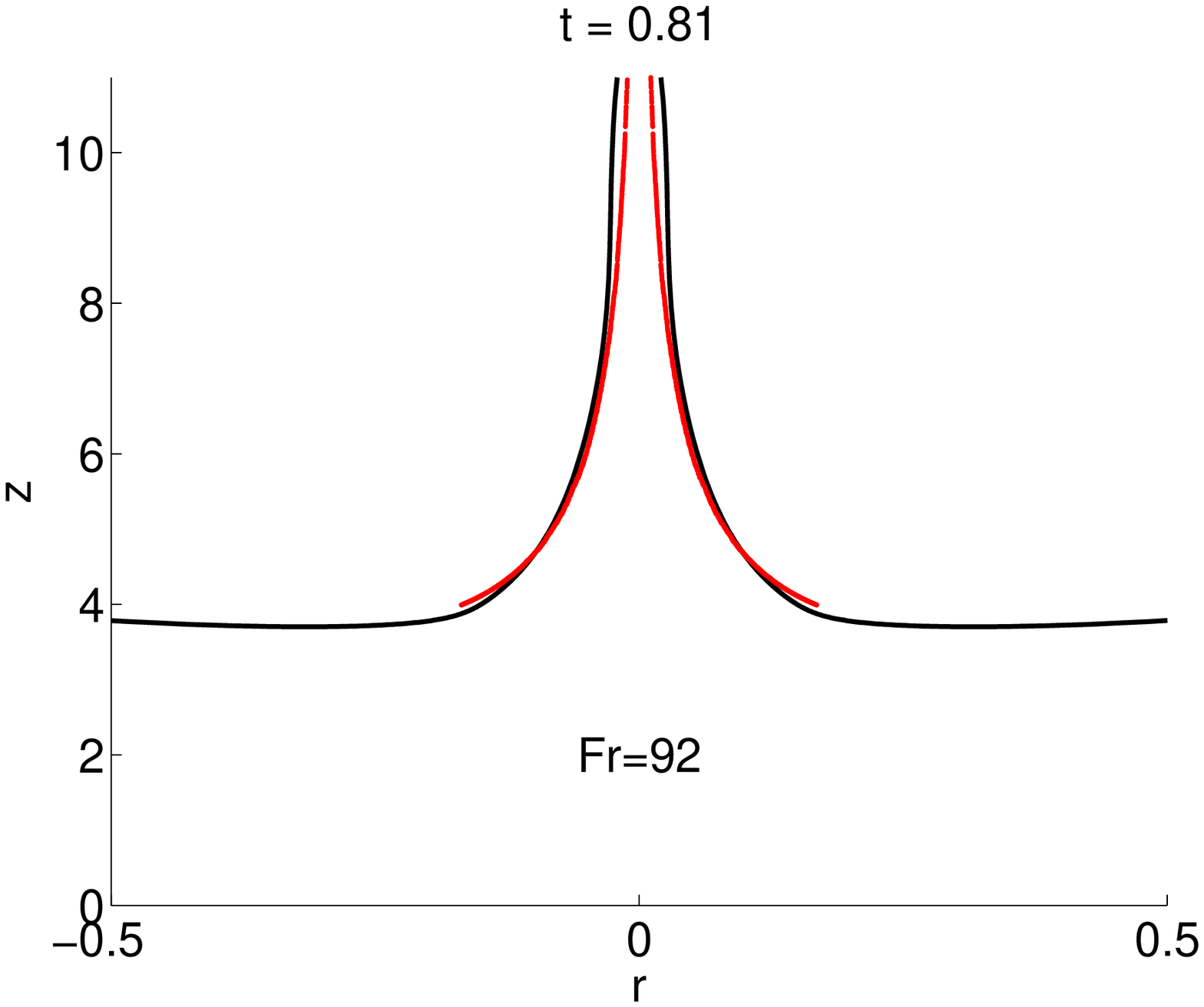}     }
  \caption{Comparison between the jet shape calculated using the boundary integral code (black line) and the one obtained integrating equations (\ref{Momentum})-(\ref{Continuity}) using the values of $r_b$ and $z_b$ given by the model described in section \ref{sec:baseModel} and shown in figure \ref{JetBase} (red line).}\label{Jetsmodel}
\end{figure}

Once $r_b$ and $z_b$ are obtained through the model in section
\ref{sec:baseModel}, the axial velocity at the beginning of the
ballistic region can be calculated as a function of known
quantities as $u_0\simeq B(\mathrm{Fr})q_b/r_b$, with
$B(\mathrm{Fr})$ the function depicted in figure \ref{BFr}.
Therefore, both the flow field and the jet shape within the
ballistic region can be computed from the integration of equations
(\ref{Momentum})-(\ref{Continuity}) using, as initial conditions,
$r_0(\tau)=0.5\,r_b(\tau)$,
$u_0(\tau)=B(\mathrm{Fr})q_b(\tau)/r_b(\tau)$ and
$z_0(\tau)=z_b(\tau)+0.5r_b$. The comparison between the jet shape
calculated numerically and that obtained from the model is
depicted in figure \ref{Jetsmodel} and good agreement is found.

The capillary breakup process of the jet can be also modeled
making use of our numerical results in section
\ref{sec:breakupAnalysis} since, through equation (\ref{So}), both
$\mathrm{We}_0$ and $s_0$ can be easily expressed as a function of
$u_0(\tau)=B\,q_b/r_b$, $z_b(\tau)$ and $r_b(\tau)$, with the
latter two functions given by the model as described above.
Consequently, both the ejection and breakup process of the jet can
be modeled with the only inputs of $q_c(z)$ and $r_{min}$, i.e.
quantities defined \emph{before} pinch-off.

Moreover, the trajectory of the ejected drops can also be modeled
using the results of the previous sections. Indeed, the nearly
spherical drops ejected from the tip of the jet follow a ballistic
trajectory which can be calculated from Newton's second law as
\begin{equation}
\begin{split}
\rho\frac{4\pi\,R^3_D\,r^3_{eq}}{3}\frac{d\,U_{drop}}{d\,T}=-\frac{1}{2}\rho_g\,U^2_{drop}\,c_d\,\pi\,r^2_{eq}R^2_D-\rho\frac{4\pi\,R^3_D\,r^3_{eq}}{3}\,g\, \label{BallisticDrop}
\end{split}
\end{equation}
with $U_{drop}$ and $c_d$ indicating the drop velocity and the drag coefficient, respectively. The drag term needs to be included since the relative variation of the drop velocity associated to aerodynamic effects, $\Delta\,U_{drop}/U_{drop}$, can be estimated from equation (\ref{BallisticDrop}), yielding
\begin{equation}
\frac{\Delta\,U_{drop}}{U_{drop}}\sim-\frac{\rho_g}{\rho}\frac{U_{drop}\,c_d\,\Delta\,t_{flight}}{R_D}\,
\end{equation}
with $\Delta\,t_{flight}$ the flight time of the drop. Therefore,
since $\Delta\,t_{flight}\sim U_{drop}/g$ and $U_{drop}\simeq
U_0\simeq V_D\,B(\mathrm{Fr})\,q_b/r_b$, the relative variation of
drop velocity associated to aerodynamic forces is given by
\begin{equation}
\frac{\Delta\,U_{drop}}{U_{drop}}\sim -\frac{\rho_g}{\rho}\frac{U^2_{drop}}{g\,R_D}\sim -\frac{\rho_g}{\rho}\frac{q^2_b}{r^3_b}\,\mathrm{Fr}\sim O(1).
\end{equation}

It needs to be pointed out that the validity of equation
(\ref{BallisticDrop}) rests on the assumption that drops conserve
their spherical shape along their trajectory and, thus, are not
atomized due to aerodynamic effects. This will be the case
whenever the aerodynamic Weber number
$\mathrm{We}_a=\rho_g\,U^2_{drop}\,R_D\,r_{drop}/\sigma\sim
\rho_g/\rho \mathrm{We}_0\lesssim 6$
[\cite{Hanson,VillermauxFrag,VillermauxNatPhys}]. Therefore,
except for the very initial instants after cavity pinch-off, in
which the gas Weber number could be larger than 6 - as can be
inferred from figure \ref{Welocal} -, equation
(\ref{BallisticDrop}) is valid to calculate the drop velocity.

We are now able to calculate the trajectories of the drops ejected
from the jet tip. Indeed, once the constants $C$ and
$C_{\mathrm{sink}}$ of the jet base model in section
\ref{sec:baseModel} are properly chosen, they determine the values
of $\mathrm{We}_0$ and $s_{0,local}$, which are the only inputs
for the model and simulations described in section
\ref{sec:breakup}. With this knowledge, the numerical results of
the type illustrated in figures \ref{tipbreak} and \ref{tbreakvol}
(which only depend on $\mathrm{We}_S$ and $TS_0$) allow one to calculate the
size, velocity and ejection time of the first ejected droplet, the
only inputs needed for the integration of equation
\ref{BallisticDrop}.


{\section{Conclusions}\label{sec:conclusion}

Using detailed boundary-integral simulations together with analytical modeling, we have studied the formation and breakup of the high-speed Worthington jets ejected either after the impact of a solid object on a liquid surface or after the pinch-off of a gas bubble from an underwater nozzle. To describe the phenomenon as a whole we divided the flow structure in two parts separated by the jet base ($r_b$, $z_b$): the \emph{outer region} for $r>r_b$, $z<z_b$ and the jet region, extending from the jet base to the axis i.e, $r<r_b$ and $z\geqslant z_b$. The jet region is further subdivided into the three subregions: The \emph{axial acceleration region}, where the radial inflow induced by the cavity collapse is decelerated radially and accelerated axially, the \emph{ballistic region}, where fluid particles are no longer accelerated vertically and, thus, conserve the axial momentum they possess at the end of the acceleration region and the \emph{jet tip region}, which is where the jet breakup process occurs.

We first show that the flow in the \emph{outer region} is well described by the analytical model presented in \cite{PRL09}. This model further provides a set of equations for the time evolution of the jet base $r_b(t)$ and $z_b(t)$. As depicted in figures \ref{JetBase} and \ref{FlowField}, the analytical predictions are in remarkable agreement with numerical simulations for the up- and downwards jets of the disc impact as well as the upwards jet created after the bubble pinch-off from an underwater nozzle. The model uses as its only input parameters the minimum radius of the cavity $r_{min}$ and the sink strength $q_c(z)$, both taken at the moment of pinch-off.

The axial \emph{acceleration region}, of characteristic length $O(r_b)\ll z_b$ is where the fluid is decelerated in the radial direction which causes an overpressure that accelerates the fluid vertically. This is thus a very narrow region, localized nearby the jet base, of crucial importance for the jet ejection process since it is where the fluid particles transform their radial
momentum into axial momentum. We have found the remarkable result
that both radial ($v$) and axial ($u$) velocities, when normalized
with $q_c(z=z_b)/r_b=q_b/r_b$, nearly collapse onto the same
master curves for both the disc and the nozzle. Therefore, the values of
the rescaled velocities $(u,v)/(q_b/r_b)$ are almost constant in
time for a fixed value of the rescaled position $r/r_b<1$. We have
also found that $v/(q_b/r_b)\simeq 0$ for that part of jet surface whose radius
is smaller than $r_0=0.5\,r_b$. Therefore, since the source of axial
acceleration - radial deceleration of the fluid - is no longer
active when $r<r_0$ the corresponding vertical position $z_0$ constitutes the upper boundary of the acceleration region. In addition, we have found that the normalized axial velocity at $z_0$, $u_0/(q_b/r_b)=B$ is a function which depends very weakly on time and on the Froude number for the impacting disc case (see figure \ref{BFr}).

In the slender \emph{ballistic region} the axial pressure gradients are negligible since $v\simeq 0$ and the Weber number evaluated at the beginning of the ballistic region ($\mathrm{We}_0=\mathrm{We}\,u^2_0\,r_0$) is much larger than unity. Therefore, we have developed a 1D model assuming that, in a first approach, fluid particles conserve their vertical velocities along the ballistic portion of the jet. This model allows us to calculate both the velocity field and the jet shape from equations (\ref{Momentum})-(\ref{Radius2}). The only input parameters are the radius, vertical position, and axial velocity at the beginning of the ballistic region. For the impacting disc, these values $r_0(t)$, $z_0(t)$, and $u_0(t)$, respectively, can be obtained directly from the analytical model of the outer region together with the function $B(\mathrm{Fr})\simeq constant$ describing the acceleration region. For the underwater nozzle, the input parameters are provided directly by the numerical simulation. The results of this new model for the jet shape are in remarkable agreement with numerical simulations, as depicted in figures \ref{Jetshape} and \ref{Jetsmodel}.

Finally, we have analyzed the \emph{tip break-up region} of the stretched
jet. The main result is that the jet capillary breakup can be
described as a function of two dimensionless parameters: the local Weber number
$\mathrm{We}_0$ and the strain rate evaluated at the beginning of the
ballistic region, $s_0=\partial u/\partial z(z=z_0)$. Both quantities can again be obtained either from the numerical simulations or from the models of the outer and acceleration regions. In order to study systematically the jet breakup process as a function of these two values we have simulated the injection of a liquid into the atmosphere from a nozzle of constant radius (see figure \ref{Geometry}). The real jet breakup process can then be reproduced provided that the values of the Weber number and the strain rate at the nozzle exit coincide with those at the beginning of the ballistic region, as shown in figure \ref{Compara}.

We have found that the tip breakup in our physical situations is not triggered by the growth of perturbations coming from an external source of noise. Instead, the jet breaks up due to the capillary deceleration of the liquid at the tip, which produces a corrugation to the jet shape. Moreover, for sufficiently large values of $\mathrm{We}_0$, the time evolution of the tip of the jet does not depend on $\mathrm{We}_0$ and $s_0$ separately, but can be described in terms of the dimensionless parameter $\mathrm{We}_S=\mathrm{We}\,r^3_0\,s^2_0$ and the rescaled time $TS_0$. This universal description allows us thus to obtain the size of the droplet ejected from the tip (cf.~figure~\ref{tbreakvol}) if $\mathrm{We}_0$ and $s_0$ are known from either simulations, measurements, or analytical models such as the one described in \cite{PRL09}.

In summary, our description of Worthington jets created by the impact of a solid object on a liquid surface allows us to predict the jet base dynamics, the jet shape, and even the ejection of drops from the tip of the jet based \emph{only} on the knowledge of the minimum radius of the cavity before the jet emerges and the sink distribution at pinch-off.}

\begin{acknowledgements}
We gratefully acknowledge many helpful discussions with Devaraj
van der Meer and Detlef Lohse. We further thank Johanna Bos for
providing the experimental data for the jet base, Arjan van der
Bos for the photograph in figure \ref{discExp} and Francisco del
Campo-Cort\'es for the photographs in figures
\ref{ExperimentosI}-\ref{ExperimentosII}. JMG thanks financial
support by the Spanish Ministry of Education under Project
DPI2008-06624- C03-01. SG's contribution is part of the program of
the Stichting FOM, which is financially supported by NWO.
\end{acknowledgements}

\bibliographystyle{jfm}

\end{document}